\documentclass{jfm}
\usepackage{natbib}
\usepackage{upmath}
\usepackage[british]{babel}
\usepackage{amsmath,bm}
\usepackage{amssymb}
\usepackage{amsbsy}
\usepackage{amscd}
\usepackage{amstext}
\usepackage{tabularx}
\usepackage{float}
\usepackage{makeidx}
\usepackage{amsmath}
\usepackage{subfigure}
\usepackage{afterpage}
\usepackage[T1]{fontenc}
\usepackage[latin1]{inputenc}
\usepackage{multirow}
\usepackage{color}
\usepackage{url}
\usepackage{mathrsfs}

\usepackage{graphicx} 
\usepackage{graphics,epsfig}
\usepackage{amsfonts}
\usepackage{psfrag}

\ifCUPmtlplainloaded \else
  \checkfont{eurm10}
  \iffontfound
    \IfFileExists{upmath.sty}
      {\typeout{^^JFound AMS Euler Roman fonts on the system,
                   using the 'upmath' package.^^J}%
       \usepackage{upmath}}
      {\typeout{^^JFound AMS Euler Roman fonts on the system, but you
                   dont seem to have the}%
       \typeout{'upmath' package installed. JFM.cls can take advantage
                 of these fonts,^^Jif you use 'upmath' package.^^J}%
      }
  \else
  \fi
\fi

\ifCUPmtlplainloaded \else
  \checkfont{msam10}
  \iffontfound
    \IfFileExists{amssymb.sty}
      {\typeout{^^JFound AMS Symbol fonts on the system, using the
                'amssymb' package.^^J}%
       \usepackage{amssymb}%
         \let\leq=\leqslant
         \let\geq=\geqslant
      }{}
  \fi
\fi

\ifCUPmtlplainloaded \else
  \IfFileExists{amsbsy.sty}
    {\typeout{^^JFound the 'amsbsy' package on the system, using it.^^J}%
     \usepackage{amsbsy}}
    {\providecommand\boldsymbol[1]{\mbox{\boldmath $##1$}}}
\fi

\providecommand\bcdot{\boldsymbol{\cdot}}

\newsavebox{\astrutbox}
\sbox{\astrutbox}{\rule[-5pt]{0pt}{20pt}}

\def\Ub{{\U_b}}

\title[Numerical analyses of wire-plate electrohydrodynamic flows]
{Numerical analyses of wire-plate electrohydrodynamic flows}

\author[X. He, P. A. V\'azquez, M. Zhang]%
{
Xuerao He$^1$, Pedro A. V\'azquez$^2$ and Mengqi Zhang$^1$\thanks{Email address for correspondence: mpezmq@nus.edu.sg}
}

\affiliation{
$^1$ Department of Mechanical Engineering, National University of Singapore, 9 Engineering Drive 1, 117575 Singapore  \\
$^2$ Departamento de F\'isica Aplicada III, Universidad de Sevilla, ESI, Camino de los Descubrimientos s/n, 41092 Sevilla, Spain
  \\[\affilskip]
}

\date{\today}

\graphicspath{{figures/}}
\newlength\savewidth

\def\Ub{\mathbf{U}}

\def\Eb{\mathbf{E}}
\def\ub{\mathbf{u}}
\def\eb{\mathbf{e}}
\def\Lh{\mathbf{L}}

\begin{document}
\maketitle

\begin{abstract}
We present numerical analyses of 2-D electrohydrodynamic (EHD) flows of a dielectric liquid between a wire electrode and two plate electrodes with a Poiseuille flow, using direct numerical simulation and global stability analysis. Both conduction and injection mechanisms for charge generation are considered. In this work, we focused on the intensity of the cross-flow and studied the EHD flows without a cross-flow, with a weak cross-flow and with a strong cross-flow. (1)In the case without a cross-flow, we investigated its nonlinear flow structures and linear dynamics. We found that the flow in the conduction regime is steady, whereas the flow in the injection regime is oscillatory, which can be explained by a global stability analysis. (2)The EHD flow with a weak cross-flow is closely related to the flow phenomena in electrostatic precipitator (ESP). Our analyses indicate that increasing the cross-flow intensity or the electric Reynolds number leads to a less stable flow. Based on these results, we infer that one should adopt a relatively low voltage and weak cross-flow in the wire-plate EHD flow to avoid flow instability, which may hold practical implications for ESP. (3)The case of strong cross-flow is examined to study the EHD effect on the wake flow. By comparing the conventional cylindrical wake with the EHD wake in linear and nonlinear regimes, we found that the EHD effect brings forward the vortex shedding in wake flows. Besides, the EHD effect reduces the drag coefficient when the cross-flow is weak, but increases it when it is strong.
\end{abstract}

\section{Introduction}
Electrohydrodynamics (EHD) is an interdisciplinary subject that studies  the coupling of electric forces and fluid motion \citep{castellanos1998electrohydrodynamics}. Techniques based on EHD have broad application prospects in many engineering fields. With the process of urbanization, air pollution has become a critical issue worldwide. The air pollutants can be broadly divided into two categories, particulate matter and waste gases \citep{kulkarni2015studies}. ESP remains one of the most popular devices for the waste-gas treatment. The ESP is widely used in coal-fired power plants, accounting for nearly $ 80\% $ factories in China in the past few decades \citep{teng2020effect}. The ESP in general has the advantages of low energy consumption, large flue gas treatment capacity, and high-efficiency \citep{jaworek2019hybrid}. 

For a theoretical study, the geometry of the ESP device can be idealised as uniformly distributed wire electrodes placed in a channel between two plate electrodes. When a sufficiently high voltage is applied to the wire electrodes, corona discharge occurs, and the generated ions move towards the plate electrodes. As the dusty gas enters and passes through the channel, the neutral particles become charged via ion attachment, moving towards the collecting plates under the action of Coulomb force. Subsequently, the particles are deposited at the collecting plates and get collected \citep{yamamoto1981electrohydrodynamics,zhao2008numerical}. Although they are widely used, the ESP devices encounter problems such as particle re-entrainment and insufficient capacity to remove small particles \citep{teng2020effect}. In particular, in recent years, the emission of $ \rm{PM}_{2.5} $ (particles with diameters less than $ 2.5 \rm{\mu m} $) endangers the environment and public health (especially in developing countries), leading to more stringent $ \rm{PM}_{2.5} $-emission limits in the past decade worldwide \citep{mep2011gb}. Therefore, in order to meet the stricter standard of flue gas emission, it is necessary to further study and improve the performance of ESP. In this regard, we are particularly interested in the fluid dynamics in ESP.

Physically, the flow in the ESP can be divided into two components, the primary flow which carries the particles or dusts, and the secondary flow generated by the electric field. It has been shown that the EHD secondary flow has a significant effect on the transport of particles, especially small particles \citep{yamamoto2013,zehtabiyan2018numerical}; however, the underlying flow mechanism has not been fully elucidated. Therefore, it is important to study the complex interaction between the ion motion and the fluid dynamics in the wire-plate EHD flow. In this work, we will consider a similar but simplified problem related to the fluid dynamics in ESP, that is, the EHD flow in a wire-plate geometry, which is a more general and idealised configuration than the ESP configuration, to focus on the essential flow dynamics. 
We will adopt the direct numerical simulation method and the global linear stability analysis. In particular, we are keen to compare our computations with the experimental results reported in \cite{mccluskey}. As a theoretical study, our parametric study will be extensive, i.e., we will not confine ourselves in exploring the parametric range for the ESP flow solely, but explore a large parameter space. In the following, we summarize the research on EHD flow in a wire-plate configuration.


\subsection{EHD flow in a wire-plate configuration}


\cite{yabe1978ehd} firstly studied the corona wind in the 2-D wire-plate system experimentally and theoretically. They quantitatively proved that the corona wind is generated by the Coulomb force acting upon ions. \cite{yamamoto1981electrohydrodynamics} analyzed the interaction between the primary flow and the EHD secondary flow in both one-wire and two-wire configurations. Their results showed that the collection of dust particles could be influenced by the EHD flow. The experiment of the flow in a wire-plate ESP was performed by \cite{blanchard2001effect} to investigate the effect of the turbulent motion on the distribution of charged particles. In addition, a 2-D numerical model was built, in which the typical velocity of EHD secondary flow was observed of the order of $ 1 m/s $, agreeing with the experimental results. \cite{zhao2008numerical} numerically investigated the EHD flow in a single wire-plate ESP. The interaction between the electrostatic field and airflow was analyzed and a complete airflow regime map was obtained for a wide range of control parameters. The standard $ k-\epsilon $ model has been used in the numerical simulations of turbulent flows in the wire-plate ESP by \cite{chun2007numerical}. The influence of the control parameters on EHD flow patterns was analyzed qualitatively. The results of turbulent structures were presented, but there was no detailed analysis of the turbulent flows therein.  \cite{feng2018characterization} examined the EHD flow pattern and vortex structure quantitatively in the wire-plate ESP using Qkubo-Weiss index. The relationship between EHD flow pattern and pressure drop, turbulence intensity and flow vortex index was explored. \cite{guo2019influence} numerically studied the EHD effect in the ESP. Their results indicated that the EHD has an important influence on the particle deposition pattern, especially when the gas flow velocity is low. 

The working fluid in the above studies was gas. There is also research on the flow of a dielectric liquid in the wire-plate EHD configuration. \cite{mccluskey} conducted an experiment to examine the wake behind a wire in a laminar cross-flow with and without charge injection into the liquid. Their results showed that the wake could be modified by the injected charges when the voltage was high. \cite{fernandes2014electrohydrodynamic} studied the wire-plate EHD flow without cross-flow both numerically and experimentally. The Onsager effect, also named the Onsager-Wien effect, referring to the electric-field enhanced ion dissociation was considered for ion transport \citep{onsager1934deviations,vazquez2019depth}. They showed that the numerical simulation slightly overestimates the flow velocity measured experimentally, but the gap narrows by adopting a proper truncated series for the Onsager function. \cite{barz2018electrokinetic} conducted numerical simulations of confined cylinder wake flow subjected to a DC (direct current) electric field, and a small electrokinetic velocity was applied at the cylinder surface. They studied the influence of electrokinetic manipulation on the flow past a cylinder under different electrode arrangements. The results showed that the vertical electrostatic force component influences the characteristics of the lift coefficient, and the drag coefficient is affected by the horizontal force component. Besides, \cite{wang2021numerical} performed a numerical investigation of the wire-plate EHD-Poiseuille flow of a dielectric liquid. A detailed map of flow patterns at different hydrodynamic Reynolds numbers ($ Re $, quantifying the ratio of inertia to viscosity) and electric Reynolds numbers (based on the ionic transit velocity) was displayed.  

After reviewing the literature, one can realize that the induced EHD flow structure can interact with the primary Poiseuille flow and this interaction is important in determining, e.g., how the ESP flow behaves. In hydrodynamic community, such interaction between large-scale flow structures can be analyzed using stability analysis. This analysis has not been applied to the EHD flow in a wire-plate geometry. In the following, we will summarize the stability analyses on the EHD flows between two plane plates, which has been studied extensively.

\subsection{Stability analysis of EHD flow}

From the perspective of the stability analysis, the uniform electric field configuration was often adopted due to their simplicity. \cite{schneider1970electrohydrodynamic} firstly used the linear stability analysis to predict the start of flow motion of a dielectric liquid between two parallel electrodes subjected to unipolar injection. Meanwhile, \cite{atten1972electrohydrodynamic} analyzed the modal stability of electroconvection, showing that in the weak injection case, that is, when the charge injection intensity parameter $ C \ll 1, $ the criterion for the flow stability depends on the injection strength. In addition, the critical  electric Rayleigh number $ T $ (quantifying the ratio between the Coulomb force and the viscous force) was calculated in the case of space charge limited (SCL, when $ C \to \infty $) injection. \cite{atten1974electrohydrodynamic} investigated the stability of EHD flow subjected to SCL injection during the transient regime adopting a quasistationary approach. It was found that the experimental criterion is lower than the theoretical one under the transient condition. \cite{Zhang2015} studied EHD flows between two parallel plates with and without cross-flow under strong injection case by using the modal and non-modal linear stability analysis theories, the latter of which can depict the non-normality of the linearized Navier-Stokes (NS) operator in EHD flows. They found that, in the hydrostatic EHD flow, the transient energy growth caused by the non-normality of the linear operator is limited. Subsequently, \cite{Zhang2016Weakly} carried out a detailed weakly nonlinear stability analysis of the 2-D EHD in the SCL regime with and without cross-flow by adopting a multi-scale expansion method. Furthermore, direct numerical simulation (DNS) has been employed to study the bifurcation of the EHD flow near and beyond the linear critical threshold $ T_c $, such as the work of \cite{chicon1997numerical}, \cite{wu2013onset} and so on. In particular, \cite{wu2013onset} studied the critical bifurcation of EHD flow in a 2-D finite container without considering the charge diffusion effect. Deterministic and stochastic bifurcations in EHD flow of a dielectric liquid between two parallel plates were investigated by \cite{feng2021deterministic}. They tried to reduce the discrepancy of the linear instability criteria between experiment and theory by considering stochastic boundary conditions.

It is noted that the works on the stability analysis of the EHD flow that we reviewed above all pertain to the configuration of two parallel walls. The linear stability analysis of EHD flow in a blade-plane configuration has been performed by \cite{perez1995dynamics} based on the parallel-flow approximation. The neutral stability curves were generated in the wavenumber-Grashof number (denoting the ratio of the Coulomb force to viscous force) plane. \cite{perez2009numerical} numerically investigated the EHD flow between a blade injector and a flat plate by DNS. They studied the transition from laminar to chaotic flow with a varying electric Rayleigh number $ T $. The EHD flow between two eccentric cylinders was numerically investigated by \cite{huang2020numerical}. The detailed bifurcation diagrams corresponding to different flow regimes were presented in terms of $ T $. Their results indicated that the parameter eccentricity (the distance between the centers of outer and inner cylinders over the difference in the radii of two cylinders) influences the bifurcation. After reviewing the literature, we realise that more work should be conducted for the linear stability of EHD flow with a non-uniform electric field to understand its fluid dynamics, especially in the wire-plate configuration. 

\subsection{The current work}

From the literature review above, one can see that all the previous studies on the wire-plate EHD flow focused on distinguishing different flow patterns in a fully developed phase. The global instability mechanism in this flow, however, has not been studied. Elucidating the instability mechanism is clearly important as it concerns the dynamics of the large-scale flow structures in the flow. Based on this information, one will be able to infer how the flow becomes unstable and transitions to turbulence. Some studies have indicated that turbulence is deleterious for the ESP performance \citep{leonard1983experimental,badran2022evaluating}. Thus, understanding and controlling the flow instability and the turbulence generation in the ESP flow is important. 

In order to distill its global instability mechanism, we will in this work perform a global stability analysis for the 2-D EHD flow in a dielectric fluid layer subjected to a Poiseuille flow in a wire-plate configuration. The flow is not turbulent but is about to become unstable. Both conduction (dissociation/recombination) and injection mechanisms for the charge generation will be considered, and the enhanced dissociation by the electric field will also be included in the model. Since this is a theoretical study, we will explore a large parameter space to gain a global view of the fluid dynamics in this EHD wire-plate flow. In particular, we have chosen to particularly vary the velocity of the cross-flow to understand its effect. According to the strength of the cross-flow, the flow can be categorised into 3 regimes, i.e., no cross-flow, a weak cross-flow and a strong cross-flow. They correspond to three interesting flow phenomena. Namely, the no cross-flow case studies electroconvection due to charge injection from the wire electrode. The weak cross-flow case resembles the flow pattern in ESP and may be helpful for understanding the fluid dynamics in the latter. Note that an analogy of the flow with particles in ESP and fluids in EHD has been made by \cite{atten1987electrohydrodynamic}. The strong cross-flow investigates the EHD effect on the cylindrical wake flow.

The remaining part of this paper is organized as follows. In section \ref{problemformulation}, we describe the physical problem, the governing equations with boundary conditions, and the framework of DNS and the global linear stability analysis. The numerical methods are introduced in section \ref{numerical}. We then report our results on the three flow regimes in section \ref{results}.  The conclusion is drawn in the last section. The three appendices explain the validation of our nonlinear simulations and linear analyses. In particular, the nonlinear results are compared with the previous experimental work in the wire-plate EHD flow \citep{mccluskey} and cylindrical wake flow \citep{verhelst2004visco}.

\section{Problem formulation and numerical method}\label{problemformulation}

\subsection{Mathematical modelling}

\begin{figure}
	\centering
	\includegraphics[width = 0.8\textwidth]{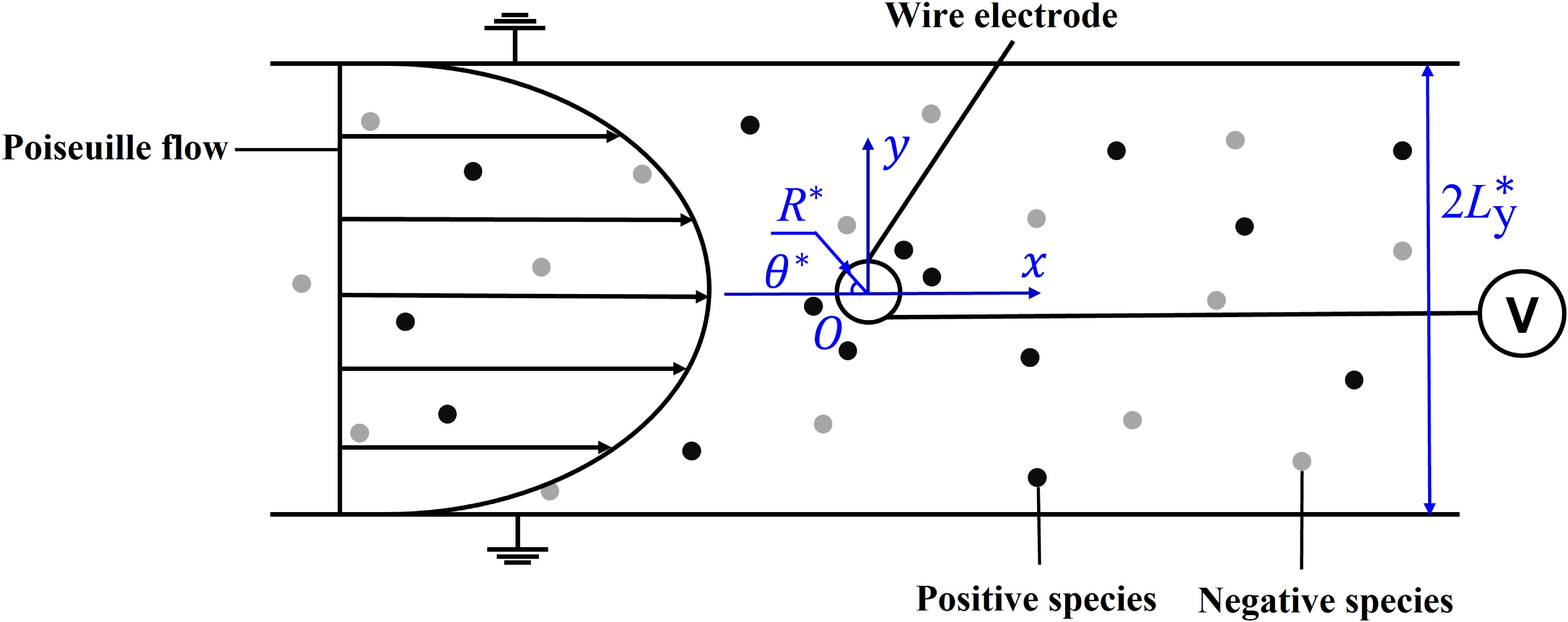}
	\caption{Sketch of the wire-plate EHD-Poiseuille flow problem.}\label{fig.sketch}
\end{figure}

As shown in figure \ref{fig.sketch}, a metallic wire located between two parallel plates is immersed in an incompressible dielectric liquid subjected to a Poiseuille flow. The streamwise direction is $x$ and the wall-normal direction is $y$. The radius of the wire is $ R^* $ and the distance between the two plate electrodes is $ 2L_y^* $. In this paper, dimensional variables and parameters are denoted with superscripts $ ^* $. A constant electric potential $ \phi^*_0 $ is applied to the wire electrode, while the two plate electrodes at plane $  y = \pm L_y^* $ are grounded. 

It has been observed in the experimental works that conduction and injection are two major mechanisms of the charge generation in dielectric liquids \citep{daaboul2017study,sun2020experimental}. In the conduction scenario, ions are generated in the liquid as a result of the dissociation-recombination process of a solute or impurity present in the liquid. In dielectric liquids, only a very small part of the solute is dissociated and a dynamic equilibrium is established between the ionic species and the neutral solute. In the injection mechanism, charges are produced by the electrochemical reaction between the interface of the electrode and the dielectric liquid \citep{atten1996electrohydrodynamic}. Both ion-generation mechanisms will be considered in this work; especially, the main reason of considering the conduction mechanism lies in the conductivity of the liquid. In order to illustrate this point, we will compare the typical time scales relevant in our flow system. There are four time scales, namely, the transit drift time scale, $ \tau_K^* $, the travel time of the ions at the electric drift velocity; the convective time scale, $ \tau_c^* $, the travel time at the velocity of the liquid; the diffusion time scale, $ \tau_D^* $, the time of ionic diffusion; and finally, the ohmic time scale, $ \tau_\sigma^* $, for the charges recombination. For the convenience of explanation, we take the parameters in the typical experiment of \cite{mccluskey} as an example. In the experiment, the typical electric potential is $ \phi^*_0=5 $ kV. The measured velocity in experiments without the imposed external flow is of the order of $ U^* = 40$ cm/s. The liquid used is Benzyl Neocaprate (BNC) with the following physical properties, namely, relative permittivity $ \epsilon_r = 3.8 $, electrical conductivity $ \sigma^*  = 10^{-9}$ S/m and ionic mobility $ K^* = 5\times10^{-9}\mathrm{m^2/(V\bcdot s)} $. These are also typical values used in some of our simulations. Taking $ L_y^* = 1 $mm as the typical distance between the wire and the plates, and $ L_x^*=10L_y^* = 10 $mm as the streamwise length scale, we have  
\begin{equation}
	\begin{split}
		\text{the transit drift time} \ \ \ &\tau_K^*=L_y^{*2}/(K^*\phi^*_0)\simeq 0.04 \mathrm{s},\\
		\text{the convective time}   \ \ \ &\tau_c^*=L_x^*/U^*\simeq 0.025 \mathrm{s},\\
		\text{the diffusion time}   \ \ \ &\tau_D^*=L_y^{*2}/D_\nu^*=L_y^{*2}e^*_0/(K^*k^*_BT^*)\simeq 10^4 \mathrm{s},\\
		\text{the ohmic time}  \ \ \ &\tau_\sigma^*=\varepsilon^*/\sigma^*\simeq 0.03 \mathrm{s},		
	\end{split}
	\label{eq.time}
\end{equation}
where, the Einstein equation $\frac{D^*_{\nu\pm}}{K^*_\pm}=\frac{k_B^*T^*}{e^*_0}$ \citep{melcher1981continuum} has been used to describe the relation between the diffusion coefficients and ionic mobilities, where $ k_B^* $ is the Boltzmann constant, $ T^* $ is the temperature in Kelvins ($ T^*\approx 295 $K in \cite{mccluskey}), $ e_0^* $ denotes the elementary electric charge.
We can see from equations (\ref{eq.time}) that the recombination time $\tau_\sigma^*$ is close to the transit times $ \tau_K^* $ and the convective time $ \tau_c^* $, meaning that dissociation and recombination of the species cannot be neglected. Particularly, they become more considerable when the charges are entrained by the velocity rolls into the bulk of the fluid, where the charges have a longer time to recombine. On the other hand, because $ \tau_D^* $ is much larger than other times, the diffusion effect will be less important in the bulk region, but its effect in the region close to the bluff body and boundaries should not be neglected.

In nonpolar dielectric liquids (relative permittivity $ \epsilon_r \leq 5 $) the ions injected from the electrodes are the same as those of the same polarity involved in the dissociation-recombination process \citep{denat1979ion}. Then, we will consider a 1-1 conduction model similar to the one described in \cite{atten2003electrohydrodynamically,vazquez2019depth}. There are two ionic species in the dielectric liquid, one positive and one negative, and their volumetric densities are $ N_+^* $ and $ N_-^* $, respectively, with the same ionic mobilities $ K_+^* = K_-^* $, and the same diffusion coefficients $ D_{\nu+}^* = D_{\nu-}^* $. The species are weakly dissociated, so the concentration of the neutral species $ c^* $ can be considered constant. In equilibrium
\begin{equation}
	k_D^*c^*=k^*_RN_+^{eq*}N_-^{eq*}=k^*_RN_{eq}^{*2},
\end{equation}
where $ k_D^* $ and $ k_R^* $ denote the dissociation and recombination rate constants, respectively, and $ k_R^* = (K_+^*+K_-^*)/\epsilon^* $, which is the upper limiting value derived by \cite{langevin1902recherches}. $ N_{eq}^{*} $ is the equilibrium charge concentration. 

Due to electroneutrality, $ N_+^{eq*}=N_-^{eq*}=N_{eq}^*=\sigma^*/(2e^*_0K^*) $. Since the electric field near the wire is strong, the Onsager-Wien effect has to be considered, which describes the enhancement effect of the electric field on dissociation \citep{onsager1934deviations}. The dissociation constant depends on the magnitude of the electric field as \citep{castellanos2007electrohydrodynamic}
\begin{equation}
	k_D^*(|\Eb^*|)=k_D^{0*}F(|\Eb^*|)=k_D^{0*}\frac{I_1(4\omega(|\Eb^*|))}{2\omega(|\Eb^*|)},  \ \ \ \ \text{with} \ \ \ \ \omega(|\Eb^*|)=(\frac{e_0^{*3}|\Eb^*|}{16\pi\varepsilon^*k_B^{*2}T^{*2}})^{1/2},
	\label{eq.onsager}
\end{equation}
where $ I_1 $ is the modified Bessel function of the first kind and of order 1 and $ \omega(|\Eb^*|) $ is the enhanced dissociation rate coefficient. We assume that the positive species are injected from the wire with a constant concentration $ N_+^*=Q_0^*/e_0^* $, where $ Q_0^* $ is the injected charge density.

We obtain the governing equations for the wire-plate EHD-Poiseuille flow as follows (similar to the equations in \cite{vazquez2019depth}), consisting of the Poisson equation for the electric potential $ \phi^* $, the definition equation of the electric field $ \Eb^* $, the transport equations for the species concentrations $ N_\pm^* $, the momentum equation and the continuity equation for the flow field
\begin{subequations}
	\begin{align}
		\nabla^*\bcdot(\varepsilon^*\nabla^*\phi^*)  =&  - e_0^*(N_+^*-N_-^*),
		\label{eq.wiredim1}\\
		\Eb^* =&  - \nabla^* \phi^* ,
		\label{eq.wiredim2}\\
		\frac{{\partial N_+^*}}{{\partial t^*}} +  \nabla^*  \bcdot (N_+^*\Ub^*+K_+^*N_+^*\Eb^*-D_{\nu+}^*\nabla^* N_+^*) =& \frac{e_0^*(K_+^*+K_-^*)(n_{eq}^{0*})^2}{\varepsilon^*}\left(F(\omega(|\Eb^*|))-\frac{N_+^*N_-^*}{(n_{eq}^{0*})^2}\right),
		\label{eq.wiredim3}\\
		\frac{{\partial N_-^*}}{{\partial t^*}} +  \nabla^*  \bcdot (N_-^*\Ub^*-K_-^*N_-^*\Eb^*-D_{\nu-}^*\nabla^* N_-^*) =& \frac{e_0^*(K_+^*+K_-^*)(n_{eq}^{0*})^2}{\varepsilon^*}\left(F(\omega(|\Eb^*|))-\frac{N_+^*N_-^*}{(n_{eq}^{0*})^2}\right),
		\label{eq.wiredim4}\\				
		\rho^*\frac{{\partial \Ub^*}}{{\partial t^*}} + \rho^*\Ub^* \bcdot \nabla^*\Ub^* =&  - \nabla^* P^*+ \mu^*{\nabla^{*2}}\Ub + e_0^*(N_+^*-N_-^*)\Eb^*,\\
		\nabla^*  \bcdot \Ub^* =& 0,
		\label{eq.wiredim6}
	\end{align}
	\label{eq.dimen}
\end{subequations}
where $ \varepsilon^* $ is the permittivity of the liquid, $ \rho^* $ the density, $ n_{eq}^{0*} $ the concentration of the ionic species at equilibrium, $ P^*$ the pressure and $ \mu^* $ the viscosity.

Regarding the boundary conditions for the velocity field, at the inlet, we impose a parabolic Poiseuille flow. The electric potential obeys a zero normal derivative at the inlet. The densities of ion species $ N_+^*$ and $ N_-^*$ at the inlet are fixed at a constant which has the same value as the initial condition ($n_{eq}^{0*} $ in this work) within the domain. Therefore, the boundary conditions at the inlet read
\begin{equation}
\mathrm{Inlet}:  \;\;\;\;\; \Ub^*=-\frac{1}{2\mu^*}\frac{\partial P^*}{\partial x^*}(L_y^{*2}-y^{*2})\mathbf{e}_x ,\; \mathbf{n}\bcdot\nabla^* \phi^* =0,\; N_+^* =n_{eq}^{0*},\; N_-^* =n_{eq}^{0*}.
\end{equation}
At the outlet, we apply an open boundary condition for all the variables, that is
\begin{equation}
\mathrm{Outlet}:  \;\;\;\mathbf{n}\bcdot\nabla^* \Ub^* =0 ,\;\;\;\; \mathbf{n}\bcdot\nabla^* \phi^* =0,\;\;\;\;\;\; \mathbf{n}\bcdot\nabla^*N_+^* =0,\;\;\;\;\;\; \mathbf{n}\bcdot\nabla^*N_-^* =0.
\end{equation}
At the wire surface, a no-slip boundary condition is used for the velocity. In addition, a constant electric potential $ \phi_0^* $ is applied to the wire. For the positive species $ N_+^*$, a constant volumetric density $ Q_0^*/e_0^* $ is injected from the wire and not affected by the nearby electric field, according to the autonomous and homogeneous hypothesis of the injection mechanism. The electrode is an open boundary for opposite polarity ions. This means that we ignore in our computations the very thin layer near the metallic electrodes where the electron transfer between the ionic species and the electrodes occurs. This is a common assumption in EHD problems \citep{castellanos1991coulomb,perez2014electrohydrodynamic,vazquez2019depth}. Therefore, the boundary condition for the negative ions is zero normal derivative of the ion concentration. We summarize the boundary conditions at the wire as follows
\begin{equation}
\mathrm{Wire}:   \Ub^* =0 ,\;\;\;\;\;\;\;\;\;\;\;\;\;  \phi^* =\phi^*_0,\;\;\;\;\;\;\;\;\;\;\;\;\;\; N_+^* =Q_0^*/e_0^*,\;\;\;\;\;\;\;\;\; \mathbf{n}\bcdot\nabla^*N_-^* =0.
\end{equation} 
At the plate electrodes, the boundary condition for velocity is also no-slip. The electric potential is zero, meaning that the plates are grounded. Similar to the process of negative ions on the wire, the normal derivative of the positive ion concentration at the plate electrodes is zero, whereas the concentration of the negative species is zero due to Coulomb repulsion. Thus, the boundary conditions at plates are
\begin{equation}
\mathrm{Plates}:   \;\;\;\Ub^* =0 ,\;\;\;\;\;\;\;\;\;\;\;\;\;\;  \phi^* =0,\;\;\;\;\;\;\;\;\;\;\;\;\;\;\;\; \mathbf{n}\bcdot\nabla^*N_+^* =0,\;\;\;\;\;\; N_-^* =0.
\end{equation}


\subsection{Non-dimensionalized governing equations}
In this section, we discuss the nondimensionalization of the equations. We nondimensionalize (\ref{eq.wiredim1})-(\ref{eq.wiredim6}) by the following scales. The length is nondimensionalized by $ R^* $ (the radius of the wire), the time $ t^* $ by $ R^{*2}/(K_+^*\phi_0^*) $, the electric potential $ \phi^* $ by $ \phi_0^* $, the electric density $ N^*_\pm $ by $ n_{eq}^{0*} $, the velocity $ \Ub^* $ by $ K^*_+\phi_0^*/R^* $, the pressure by $ \rho^*K^{*2}_+\phi_0^{*2}/R^{*2} $ and the electric filed $ \Eb^* $ by $ \phi_0^*/R^* $. In addition, due to the electroneutrality, the electrical conductivity $ \sigma^* $ satisfies the equation \citep{langevin1903recombinaison}
\begin{equation}
	\frac{\sigma^*}{K^*_++K^*_-}=e_0^*n_{eq}^{0*}.
\end{equation}
Therefore, the nondimensional governing equations can be obtained as follows
\begin{subequations}
	\begin{align}
		{\nabla ^2}\phi  &=  - C_0\lambda^2(N_+-N_-),
		\label{eq.wiredimless1}\\
		\Eb &=  - \nabla \phi, 
		\label{eq.wiredimless2}\\
		\frac{{\partial N_+}}{{\partial t}} +  \nabla  \bcdot (N_+\Ub+N_+\Eb-\alpha\nabla N_+) &= \lambda^2C_0(1+K_r)(F(\omega(|\Eb|))-N_+N_-),
		\label{eq.wiredimless3}\\
		\frac{{\partial N_-}}{{\partial t}} +  \nabla  \bcdot (N_-\Ub-K_rN_-\Eb-K_r\alpha\nabla N_-) &= \lambda^2C_0(1+K_r)(F(\omega(|\Eb|))-N_+N_-),
		\label{eq.wiredimless4}\\		
		\frac{{\partial \Ub}}{{\partial t}} + \Ub \bcdot \nabla\Ub =  - \nabla P&+ \frac{1}{Re^E}{\nabla ^2}\Ub + \lambda^2C_0M^2(N_+-N_-)\Eb,\\
		\nabla  \bcdot \Ub &= 0.
		\label{eq.wiredimless6}
	\end{align}
	\label{eq.nonlinear}
\end{subequations}
The Onsager function (equation \ref{eq.onsager}) becomes
\begin{equation}
	F(|\Eb|)=\frac{I_1(4\omega(|\Eb|))}{2\omega(|\Eb|)} \ \ \ \ \text{with}  \ \ \ \ \omega(|\Eb|)=O_s^{1/2}|\Eb|^{1/2}.
\end{equation}
The corresponding profile of the parabolic 2-D Poiseuille flow can be written as
\begin{equation}
	U_x(y)=\frac{3}{2}\frac{Re^W}{Re^E}(1-\frac{y^2}{L_y^2})=\frac{3}{2}U_{0}(1-\frac{y^2}{L_y^2}).
\end{equation}
The non-dimensional parameters are
\begin{equation}
	\begin{split}
		C_0=\frac{\sigma^*L_y^{*2}}{(K_+^*+K_-^*)\varepsilon^*\phi_0^*},\; K_r=\frac{K_-^*}{K_+^*},\;\alpha=\frac{D^*_{\nu\pm}}{K^*_{\pm}\phi_0^*}=\frac{k_B^*T^*/e_0^*}{\phi_0^*},\;Re^W=\frac{\rho^* U_0^*R^*}{\mu^*},\; \\ Re^E=\frac{\rho^* K_+^*\phi_0^*}{\mu^*},\;U_0=\frac{Re^W}{Re^E}=\frac{U_0^*R^*}{K_+^*\phi_0^*},\; M = \frac{\sqrt{\frac{\varepsilon^*}{ \rho^*}}}{K_+^*},\; C_I=\frac{Q_0^*R^{*2}}{\varepsilon^*\phi_0^*},\; \\O_s=\frac{e_0^{*3}\phi_0^*}{16\pi\varepsilon^*k_B^{*2}T^{*2}R^*},\;\lambda=\frac{R^*}{L_y^*},\; \Lambda_1=\frac{L_{x1}^*}{R^*},\;\Lambda_2=\frac{L_{x2}^*}{R^*}.
	\end{split} \label{eq.nonnumbers}
\end{equation}
In the above, $ C_0 $ is the conduction number, which is the ratio of the drift time $ L_y^{*2}/K^*\phi_0^* $ and recombination time $ \varepsilon^*/\sigma^* $. The parameter $ K_r $ is the ionic mobilities ratio of positive and negative ionic species. $ \alpha $ is the charge diffusion coefficient.  The hydrodynamic Reynolds number $ Re^W $ is the Reynolds number related to the velocity of the cross-flow. The electric Reynolds number $ Re^E $ is defined with the drift velocity (ionic transit velocity). The ratio between $ Re^W $ and $ Re^E $, i.e. $ U_0 $ measures the nondimensional mean velocity of the Poiseuille flow. The mobility number $ M $ is the ratio between hydrodynamic mobility and ionic mobility, and it depends only on the properties of the fluid. The number $ C_I $ is the nondimensional value of the injected charge at the wire. The parameter $ O_s $ is the Onsager constant, which describes the Onsager effect \citep{onsager1934deviations}. In addition, the blockage ratio $ \lambda $ measures the relative sizes of the radius of the wire to the distance between the wire and the plates. Finally, $\Lambda_1$ and $\Lambda_2 $ denote the ratio of the length of the left or right half of the plates to the radius of the wire.

\begin{figure}
	\centering
	\includegraphics[width = 0.9\textwidth]{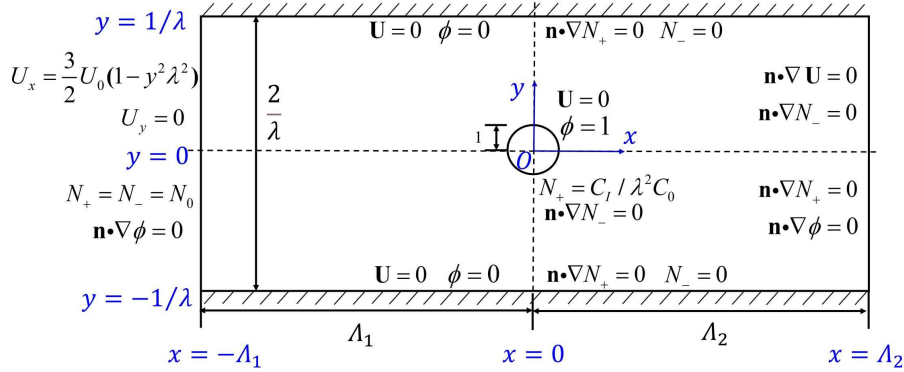}
	\caption{Geometry and boundary conditions of wire-plate EHD-Poiseuille flow.}\label{fig.wire}
\end{figure}
For the initial conditions, we start the nonlinear numerical simulation with no flow ($ \Ub=0 $) and $ N_+=N_-=1 $. In addition, figure \ref{fig.wire} depicts the non-dimensional domain with the boundary conditions, which are also listed as follows
\begin{equation}\label{bc2d14}
	\begin{split}
		&\mathrm{Inlet}:  \;\;\;\;\; \Ub=\frac{3}{2}U_{0}(1-\frac{y^2}{L_y^2})\mathbf{e}_x ,\; \mathbf{n}\bcdot\nabla \phi =0,\; N_+=N_0= 1,\; N_-=N_0 =1;\\
		&\mathrm{Outlet}:  \;\;\;\mathbf{n}\bcdot\nabla \Ub =0 ,\;\;\;\;\;\;\;\;\;\;\;\;\;\;\; \mathbf{n}\bcdot\nabla \phi =0,\; \mathbf{n}\bcdot\nabla N_+ =0,\;\; \mathbf{n}\bcdot\nabla N_- =0;\\
		&\mathrm{Plates}:   \;\;\;\Ub =0 ,\;\;\;\;\;\;\;\;\;\;\;\;\;\;\;\;\;\;\;\;\;\;\;\;  \phi =0,\;\;\;\;\;\;\;\;\; \mathbf{n}\bcdot\nabla N_+ =0,\;\; N_- =0;\\
		&\mathrm{Wire}:  \;\;\;\;\; \Ub =0 ,\;\;\;\;\;\;\;\;\;\;\;\;\;\;\;\;\;\;\;\;\;\;\;\; \phi =1,\;\;\;\;\;\;\;\;\; N_+ =\frac{C_I}{\lambda^2 C_0},\;\;\; \mathbf{n}\bcdot\nabla N_- =0.
	\end{split}
\end{equation}

Similar to the classical Newtonian cylindrical wake flow, our wire-plate EHD flow will also become time-dependent once the driving parameters exceed some critical values. In the field of hydrodynamic instability, the global stability analysis \citep{theofilis2011global} has been well developed to study the stability of a certain base flow. This base flow could be the time-averaged flow or the unstable steady base flow. The latter means a solution to the Navier-Stokes equations and it can be obtained via the selective frequency damping (SFD) method \citep{aakervik2006steady}. This method adds a forcing term to the right-hand side of the governing equations
\begin{equation}
	-\chi(\mathbf{G}-\hat{\mathbf{G}})
\end{equation}
where $ \mathbf{G}=(N_+,N_-,\phi,\Ub)^T $ and $ \chi $ is the control coefficient.$ \hat{\mathbf{G}} $ is the modification of $ \mathbf{G} $ with reduced temporal fluctuations, which reads $\frac{\partial \hat{\mathbf{G}}}{\partial t}=(\mathbf{G}-\hat{\mathbf{G}})/\Delta$, where $ \Delta $ denotes the filter width, which is the inverse of the cut-off angular frequency $ \omega_c $. The convergence of the SFD method is influenced by the filter width $ \Delta $, as well as the control coefficient $ \chi $. The guideline for choosing the parameters has been given in \cite{aakervik2006steady} as $ \chi\geq  \omega_r  $, $ \Delta=1/\omega_c\geq2/\omega_i $, where $ \omega_r $  and $ \omega_i $ denote the real and imaginary part of the leading eigenvalue, respectively.

In addition, the lift force $ F_L $ and drag force $ F_D $ on the wire surface will be computed. They can be used to further define the lift and drag coefficients as
\begin{equation}
	C_l=\frac{F_L}{RU_{0}^2}, \;\;\;\;	C_d=\frac{F_D}{RU_{0}^2}.
\end{equation}

\subsection{Linearisation}

In order to understand the perturbative dynamics in the wire-plate EHD-Poiseuille flow, we perform a linear stability analysis by linearizing the above nonlinear governing equations \citep{schmid2000stability}. From a mathematical point of view, the linearization step is to find the linear approximation of the function at a fixed point, that is, the first-order Taylor series of the function around that point. In this work, the base flow is time-independent. When the EHD flow is steady, we can use the final steady state of the nonlinear simulation as the base flow. When the EHD flow is oscillatory, the SFD method is applied to obtain the steady solution. In some cases of the wire-plate EHD flows with a strong cross-flow, time-averaged mean flow is also used as the base flow. In this work, we use the lower-case variables for the perturbation and the variables with bars for the base flow. The linearisation step is based on the Reynolds decomposition, i.e., the total flow state is decomposed into the base flow component and the perturbation, i.e., $ \mathbf{f}_{total} = \mathbf{\bar{F}} + \mathbf{f}=(\bar\phi,\bar\Eb,\bar{N}_+,\bar{N}_-,\bar\Ub,\bar{P})^T + (\varphi,\eb,n_+,n_-,\ub,p)^T $. The Reynolds decomposition is substituted into the nonlinear equations and the nonlinear terms of the perturbation is neglected to arrive at the linearised equations. After the linearization step, the linear equations for the wire-plate EHD-Poiseuille flow read
\begin{subequations}
	\begin{equation}
		{\nabla ^2}\varphi  =  - C_0\lambda^2(n_+-n_-),
		\label{eq.wirelin1}
	\end{equation}
	\begin{equation}
		\eb =  - \nabla \varphi, 
		\label{eq.wirelin2}
	\end{equation}
	\begin{equation}
		\begin{split}
			\frac{{\partial n_+}}{{\partial t}} +  \nabla  \bcdot ((\bar\Eb+\bar\Ub)n_++(\eb+\ub)\bar N_+)-\alpha\nabla^2 n_+ =&\\ \lambda^2C_0(1+K_r)(X(|\bar\Eb|)\frac{\bar E_x}{|\bar\Eb|}e_x+X(|\bar\Eb|)\frac{\bar E_y}{|\bar\Eb|}e_y-\bar N_+n_--\bar N_-n_+),&		
		\end{split}
		\label{eq.wirelin3}
	\end{equation}
	\begin{equation}
		\begin{split}
			\frac{{\partial n_-}}{{\partial t}} +  \nabla  \bcdot ((-\bar\Eb+\bar\Ub)n_-+(-\eb+\ub)\bar N_-)-\alpha\nabla^2 n_-  =&\\ \lambda^2C_0(1+K_r)(X(|\bar\Eb|)\frac{\bar E_x}{|\bar\Eb|}e_x+X(|\bar\Eb|)\frac{\bar E_y}{|\bar\Eb|}e_y-\bar N_+n_--\bar N_-n_+),\\	    	
		\end{split}
		\label{eq.wirelin4}	
	\end{equation}
	\begin{equation}
		\frac{{\partial \ub}}{{\partial t}} + (\ub \bcdot \nabla)\bar\Ub+(\bar\Ub \bcdot \nabla)\ub =  - \nabla p + \frac{1}{Re^E}{\nabla ^2}\ub + \lambda^2C_0M^2[(n_+-n_-)\bar\Eb+(\bar N_+-\bar N_-)\eb],
		\label{eq.wirelin5}
	\end{equation}
	\begin{equation}
		\nabla  \bcdot \ub = 0,
		\label{eq.wirelin6}
	\end{equation}
	\label{eq.linear}
\end{subequations}
where
\begin{equation} \label{eq.X}
	X(|\bar\Eb|) =F'(|\bar\Eb|)\omega'(|\bar\Eb|)=\frac{I_1'(4\omega(|\bar\Eb|))\bcdot4\omega(|\bar\Eb|)-I_1(4\omega(|\bar\Eb|))}{4(\omega(|\bar\Eb|))^2}\bcdot\frac{1}{|\bar\Eb|^{1/2}}.
\end{equation}
It is noted that the derivative of the modified Bessel function of the first kind and $ \gamma $-th-order satisfies the relationship $I'_\gamma(x)=\frac{1}{2}(I_{\gamma-1}(x)+I_{\gamma+1}(x))$, where the prime denotes the derivative with respect to the argument, and we have $ \gamma=1 $ in this work. The linear boundary conditions read
\begin{equation}
	\begin{split}
		&\mathrm{Inlet}:  \;\;\;\;\; \ub=0,\;\;\;\;\;\;\;\;\;\;\;\mathbf{n}\bcdot\nabla \varphi =0,\; n_+= 0,\;\;\;\;\;\;\;\;\; n_- =0;\\
		&\mathrm{Outlet}:  \;\;\;\mathbf{n}\bcdot\nabla \ub =0 ,\;\;\mathbf{n}\bcdot\nabla \varphi =0,\; \mathbf{n}\bcdot\nabla n_+ =0,\; \mathbf{n}\bcdot\nabla n_- =0;\\
		&\mathrm{Plates}:   \;\;\;\ub =0 ,\;\;\;\;\;\;\;\;\;\;\; \varphi =0,\;\;\;\;\;\;\;\;\; \mathbf{n}\bcdot\nabla n_+ =0,\; n_- =0;\\
		&\mathrm{Wire}:  \;\;\;\;\; \ub =0 ,\;\;\;\;\;\;\;\;\;\;\; \varphi =0,\;\;\;\;\;\;\;\;\; n_+ =0,\;\;\;\;\;\;\;\;\; \mathbf{n}\bcdot\nabla n_- =0.
	\end{split}
\end{equation}
For simplification, the above linearized equations are written in the following form
\begin{equation}
	\frac{\partial \mathbf{f}}{\partial t}=\Lh \mathbf{f}
	\label{eq.ml}
\end{equation}
where $ \Lh$ represents the linearized operator for the wire-plate EHD-Poiseuille flow. When a time-independent base flow is considered in the linear stability analysis, a wave-like assumption in time can be made to the solution
\begin{equation}
	\mathbf{f}(x,y,t)=\mathbf{\tilde f}(x,y)e^{ \omega t}
\end{equation}
Inserting this expression into the (\ref{eq.ml}) leads to an eigenvalue problem
\begin{equation}\label{eigenproblem}
\omega\tilde{\mathbf{f}}=\Lh\tilde{\mathbf{f}}
\end{equation}
where $\omega$ is the complex-valued eigenvalue corresponding to the eigenvector $\tilde{\mathbf{f}}$. The real part $ \omega_r $ denotes the temporal growth rate of perturbations. The sign of the leading mode determines the stability/instability of the linearised system. If $ \omega_r>0 $, the perturbation will grow exponentially at a large time; otherwise, the perturbation will decay. The imaginary part $ \omega_i $ represents the frequency of the eigenmode.

\subsection{Numerical methods}\label{numerical}
The direct numerical simulations in this paper are conducted using the high-order open-source computational flow solver Nek5000 based on the spectral element method \citep{Paul2008}. The code adopts the $ P_N-P_{N-2} $ formulation \citep{fischer1991parallel} for the spatial discretisation with the polynomial order $ N=7 $ in our simulations.  The semi-implicit scheme $ BDF_2/EXT_2 $ is adopted for time integration. The time step $ \delta t $ satisfies the Courant-Friedrichs-Lewy (CFL) condition with the target Courant number being 0.5 for the nonlinear simulation and 0.25 for the linear simulation.
Meanwhile, the global eigenvalue problem for the linear system is solved by the matrix-free time-stepping method called the Implicitly Restarted Arnoldi Method (IRAM) \citep{edwards1994krylov,lehoucq1996deflation,tuckerman2000bifurcation} in the Nek5000 solver.

\section{Result and Discussions} \label{results}

In this section, we present the results of DNS and linear global stability analysis for the 2-D wire-plate EHD-Poiseuille flow. As mentioned earlier, we will study three different flow settings, i.e., without a cross-flow, with a weak cross-flow and with a strong cross-flow. For the case without a cross-flow, the aim is to examine the flow structure and instability of the EHD flow in a wire-plate configuration. The case with a weak cross-flow is highly related to the flow dynamics in ESP. Finally, when a strong cross-flow is imposed, the overall flow resembles the classical cylindrical wake confined between two walls. This part will reveal how the electric field influences the wake behind the wire and its instability. We have fixed the ratio of the wire diameter to the channel height as 0.2. This value can be changed to study the confinement effect. Nevertheless, this is not considered in this work as our focus is on the three flow patterns to be discussed below.

The validation of our numerical simulations is shown in the appendices. In Appendix \ref{valinon}, we verify the nonlinear code by comparing our results with those in the experimental work \citep{mccluskey,verhelst2004visco} and the numerical work \citep{xiong2013numerical}. In Appendix \ref{valili}, we prove the robustness of the linear solver by comparing the leading eigenvalues of confined cylinder wake flow obtained by our code with those in \cite{li2022reinforcement}. Moreover, the grid independence tests are reported in Appendix \ref{valimesh}. 

\subsection{Wire-plate EHD flow without a cross-flow}
We first investigate the stability of wire-plate EHD flow without a cross-flow. Theoretically, there exists a critical value of the electric voltage beyond which the EHD flow transitions from the conduction regime to the injection regime. When the electric field is weaker than this threshold, only the conduction mechanism related to the dissociation-recombination process is at play. Once the electric field at the wire electrode excedes this critical value, the injection can occur, and it will play a leading role as the electric field becomes stronger. In the experiment of \cite{mccluskey}, this critical voltage was tested to be around 2kV, and the corresponding critical mean electric field between wire and plates was approximately $ 1.2\times10^6\mathrm{V/m} $. This critical value, expressed in our nondimensionalisation method, corresponds to the dimensionless critical $ Re^E =$ 0.23. That is, when $ Re^E\textless0.23 $, only the conduction mechanism is significant; otherwise, both conduction and injection of the species need to be considered, and the latter dominates, which we call the injection regime. Thus, the study of this EHD flow without a cross-flow is divided into two parts, namely the conduction regime ($ Re^E\textless0.23, C_I=0 $) and the injection regime ($ Re^E\textgreater0.23, C_I=0.2 $). Besides, this flow is featured by many parameters and we are not able to consider the effect of all of them, so the values of the other parameters are fixed as $ C_0=3, M=37, K_r=1, O_s=8.6, \lambda=0.2 $ and $ \alpha=0.001 $. 

It will be seen that the nonlinear flow in the conduction regime is steady whereas those in the injection regime may be oscillatory at large $Re^E$. We attribute the instability in the injection regime to a global instability mechanism. Thus, we conduct the global stability analysis of the flows in these two regimes.
\subsubsection{Conduction regime}
\begin{figure}
	\centering
	\subfigure{
	\begin{minipage}[h]{0.4\linewidth}
		\centering
		\includegraphics[height=2.5cm]{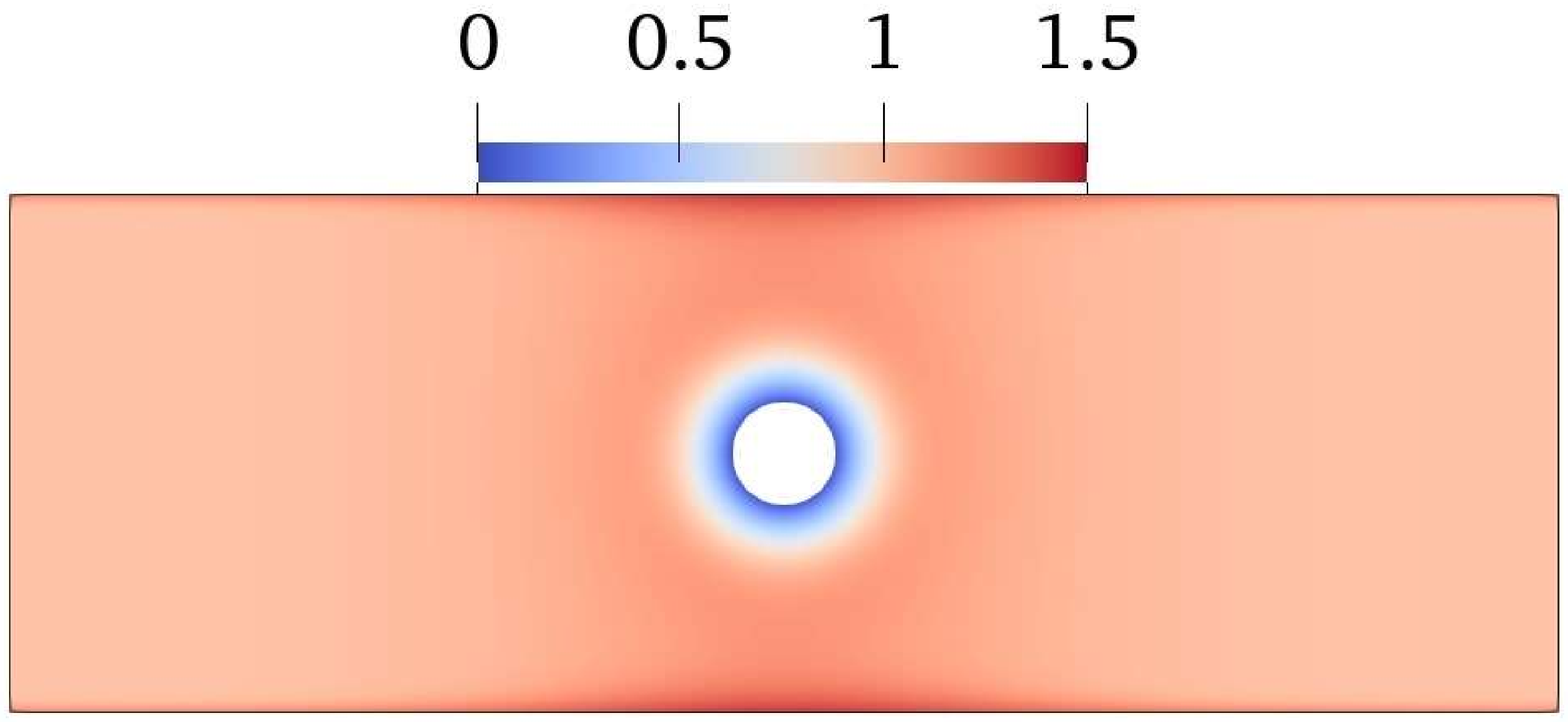}
		\put(-160,60){(a)}
		\label{fig.q101}
	\end{minipage}}
	\hspace{0pt}
	\subfigure{
	\begin{minipage}[h]{0.4\linewidth}
	\centering
	\includegraphics[height=2.5cm]{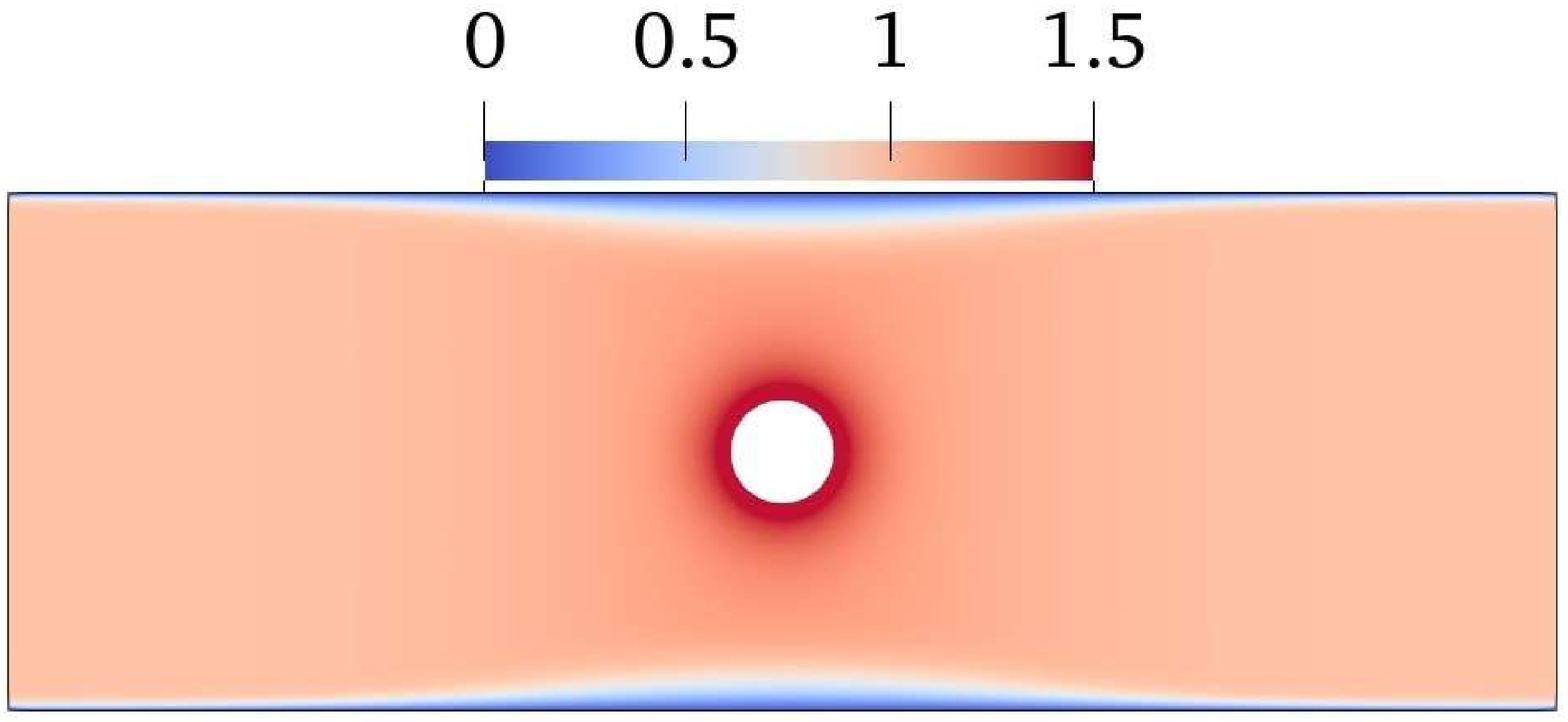}
	\put(-160,60){(b)}
	\label{fig.q201}
     \end{minipage}}
     \hspace{0pt}
     \subfigure{
     \begin{minipage}[h]{0.4\linewidth}
     	\centering
     	\includegraphics[height=2.5cm]{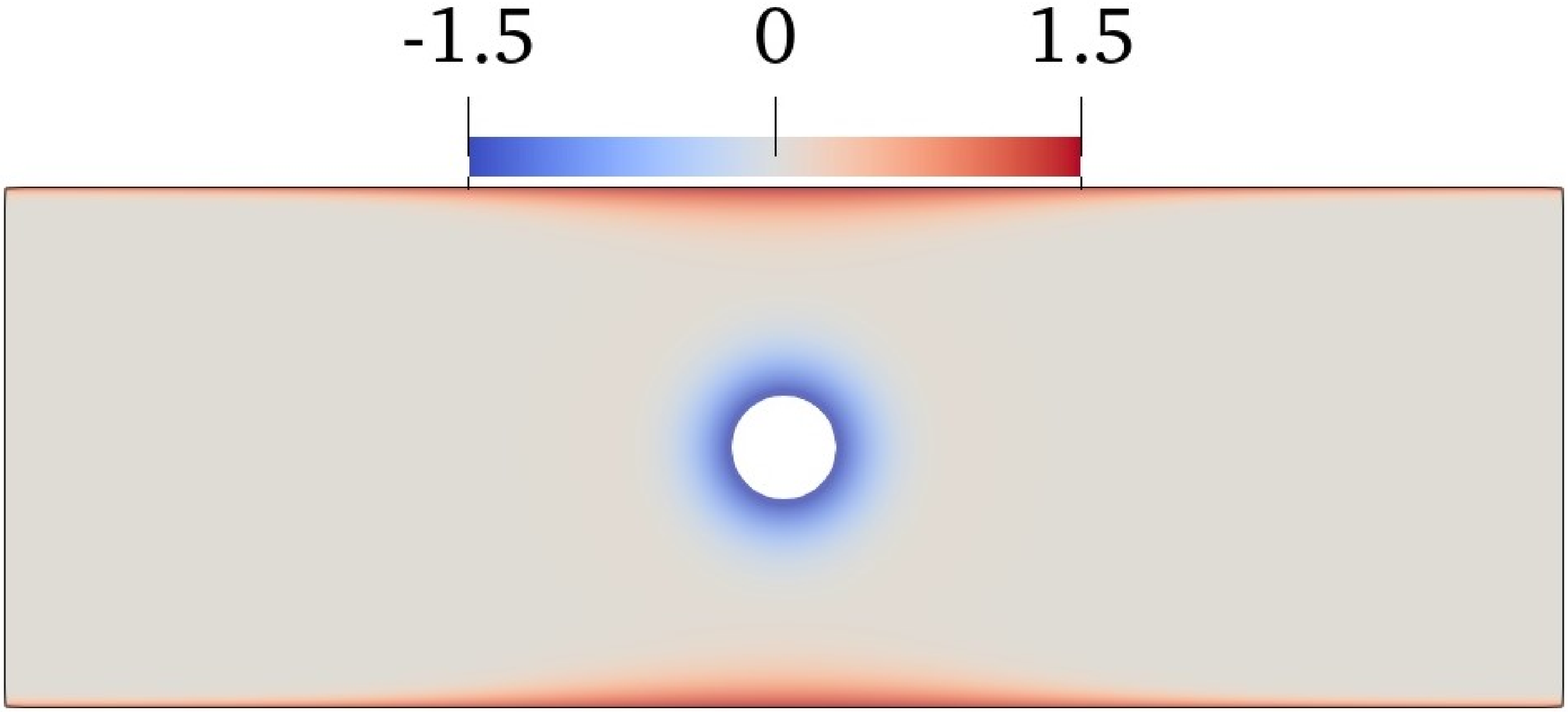}
     	\put(-160,60){(c)}
     	\label{fig.netq}
     \end{minipage}}
     \hspace{0pt}
     \subfigure{
	\begin{minipage}[h]{0.4\linewidth}
	\centering
	\includegraphics[height=2.58cm]{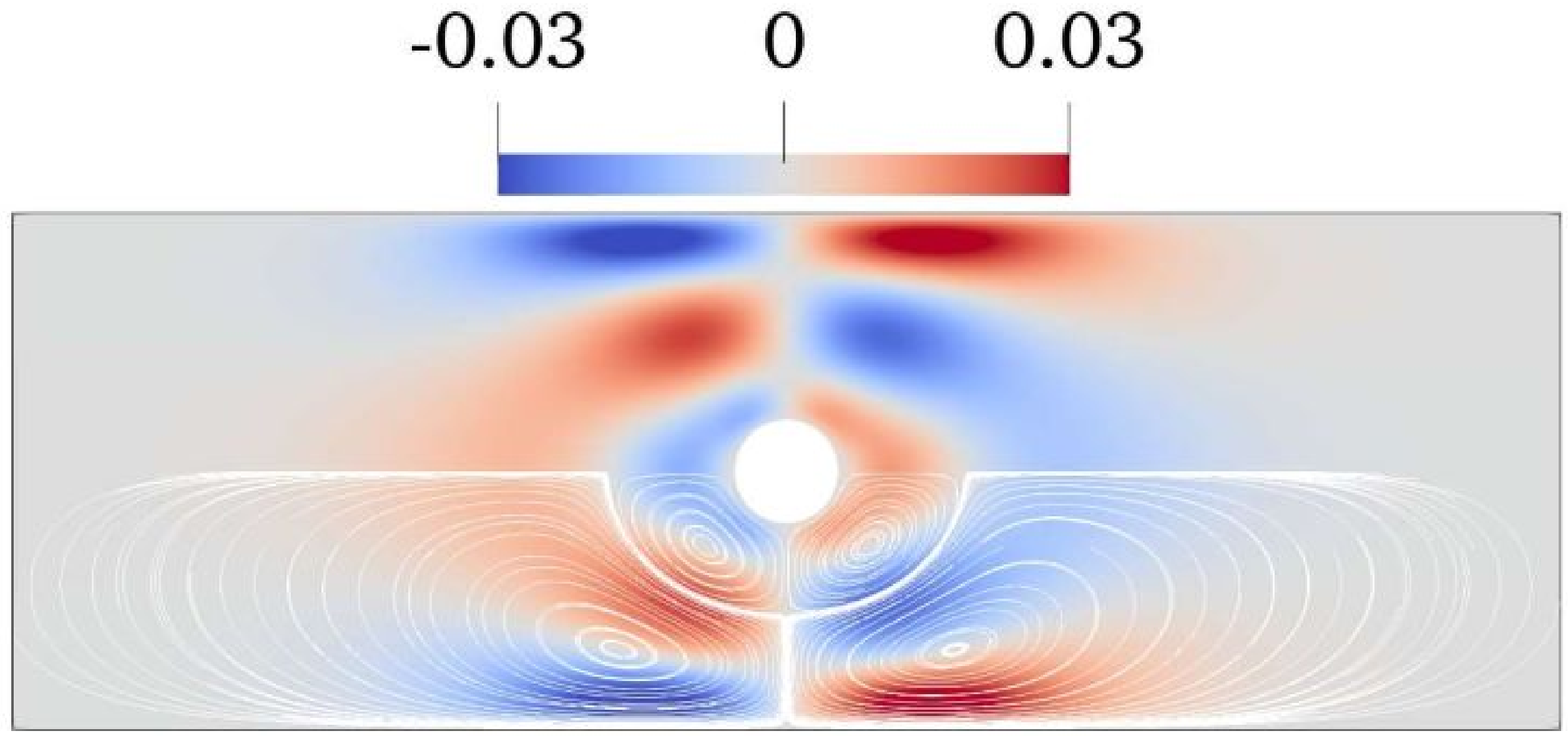}
	\put(-160,60){(d)}
	\label{fig.vx01}
     \end{minipage}}
    \hspace{0pt}
    \subfigure{
	\begin{minipage}[h]{0.4\linewidth}
		\includegraphics[height=2.5cm]{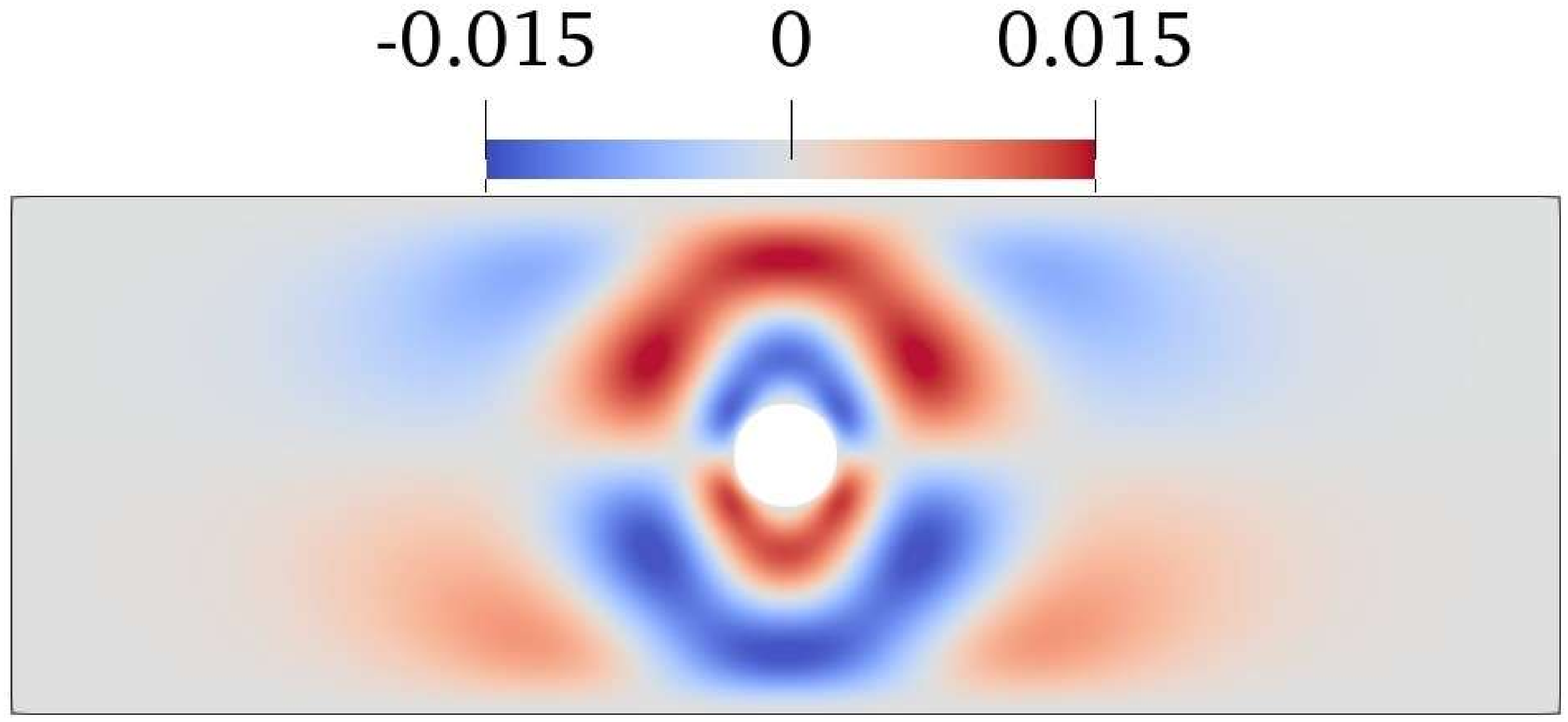}
		\put(-160,60){(e)}
		\label{fig.vy01}
	\end{minipage}}
	\hspace{0pt}
	\subfigure{
	\begin{minipage}[h]{0.4\linewidth}
		\centering
		\includegraphics[height=4.5cm]{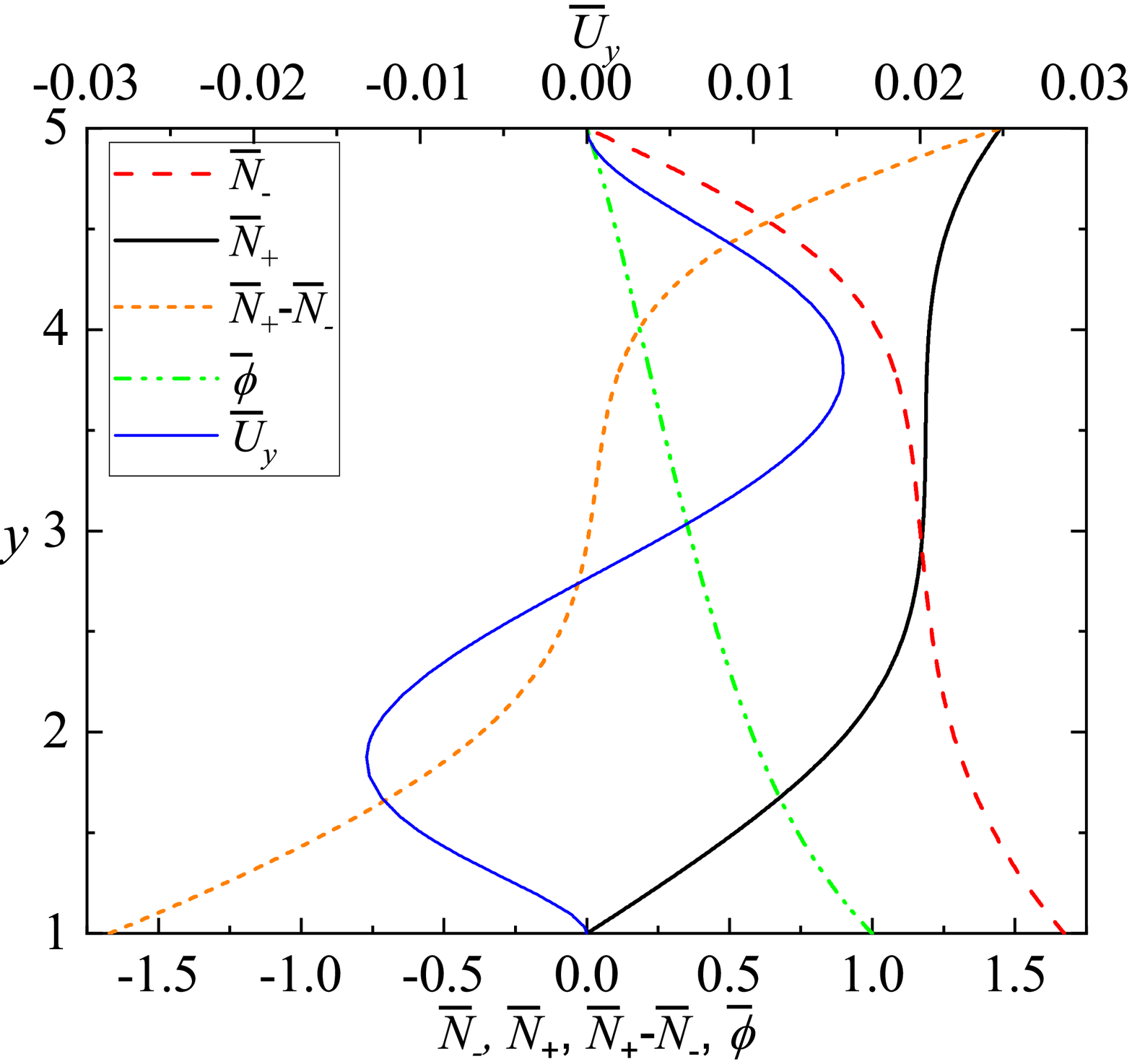}
		\put(-155,110){(f)}	
		\label{fig.x0}
	\end{minipage}}
	\caption{Steady state of the wire-plate EHD flow without cross-flow at $ Re^E=0.1, C_I=0 $ (thus, the ion generation mechanism is the dissociation process). Distributions of (a) positive species, (b) negative species, (c) net charges, (d) $ x $-velocity field and streamlines, (e) $ y $-velocity field and (f) positive and negative species $ \bar{N}_+, \bar{N}_- $, net charges $ \bar{N}_+-\bar{N}_- $, electric potential $\bar \phi$ and $ y $-velocity $\bar U_y$ along the line of $ x=0 $.}
	\label{fig.base01} 
\end{figure}

Figure \ref{fig.base01} depicts the final steady state of the distribution of positive and negative species, net charges ($ \bar{N}_+-\bar{N}_- $), $ x $-velocity and $ y $-velocity  at $ Re^E=0.1 $ (conduction regime). The solid lines in panel (d) are the streamlines. In the conduction regime, charges are generated everywhere in the domain. It can be found from panels (a) and (b) that the charges accumulate at the electrode of the opposite polarity.  The distribution of charges is controlled by electric drift ($ N_\pm\Eb $) and liquid convection ($ N_\pm\Ub $). The concentration of net charges is displayed in panel (c), which shows that negative charges accumulate near the wire and positive charges exist near the plates. From panels (d) and (e), we can see that there are two pairs of vortices in the lower half of the region, one large and one small near the wire, rotating in opposite directions. The whole flow field is symmetrical about the center line. The panel (f) presents the distribution of several variables at $ x=0 $, including the density of positive and negative species, net charges, electric potential and $ y $-velocity. It can be clearly seen that approaching the plate electrodes, the potential decreases from 1 to 0, and the positive (negative) species density increases (decreases), respectively, and the density of net charges changes from negative to positive. It is observed from the $ y $-velocity that the flow near the wire is directed toward the cylinder.

\begin{figure}
	\centering
		\subfigure{
	\begin{minipage}[h]{0.4\linewidth}
			\centering
			\includegraphics[height=4.5cm]{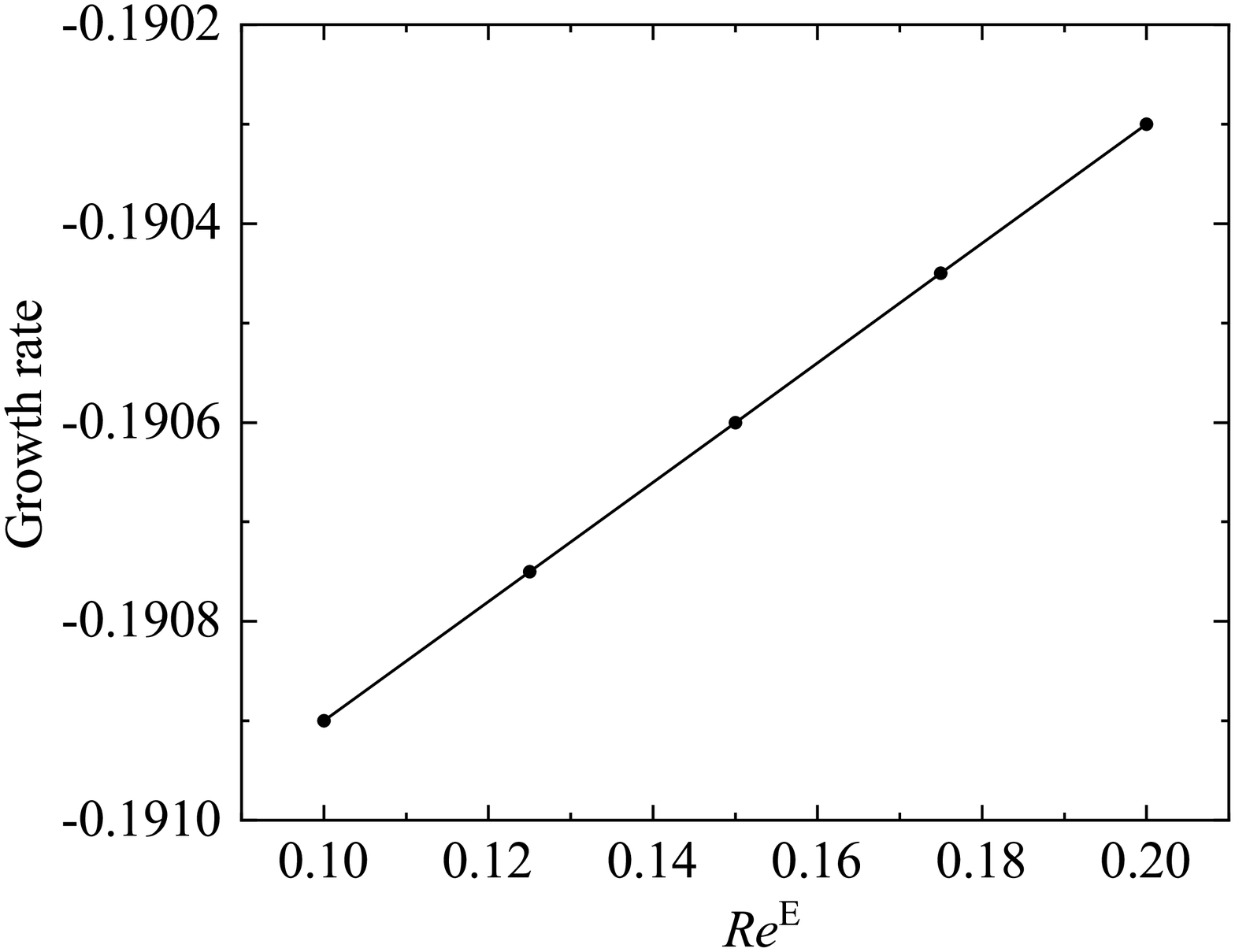}
			\put(-170,110){(a)}
			\label{fig.p1gr01}
	\end{minipage}}
	\hspace{200pt}
	\subfigure{
	\begin{minipage}[h]{0.4\linewidth}
		\centering
		\includegraphics[height=2.5cm]{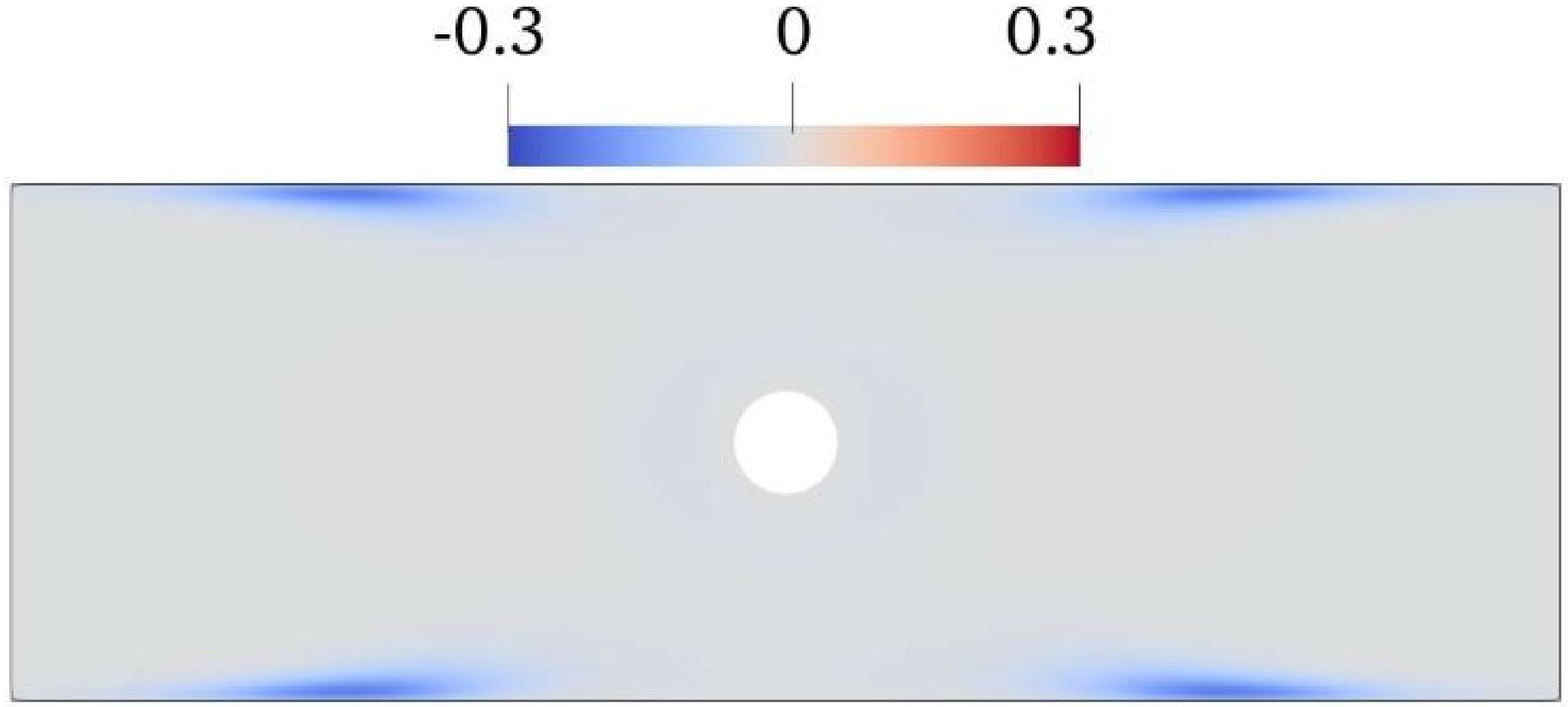}
		\put(-160,60){(b)}
		\label{fig.Aq101}
	\end{minipage}}
	\hspace{20pt}
	\subfigure{
	\begin{minipage}[h]{0.4\linewidth}
		\centering
		\includegraphics[height=2.5cm]{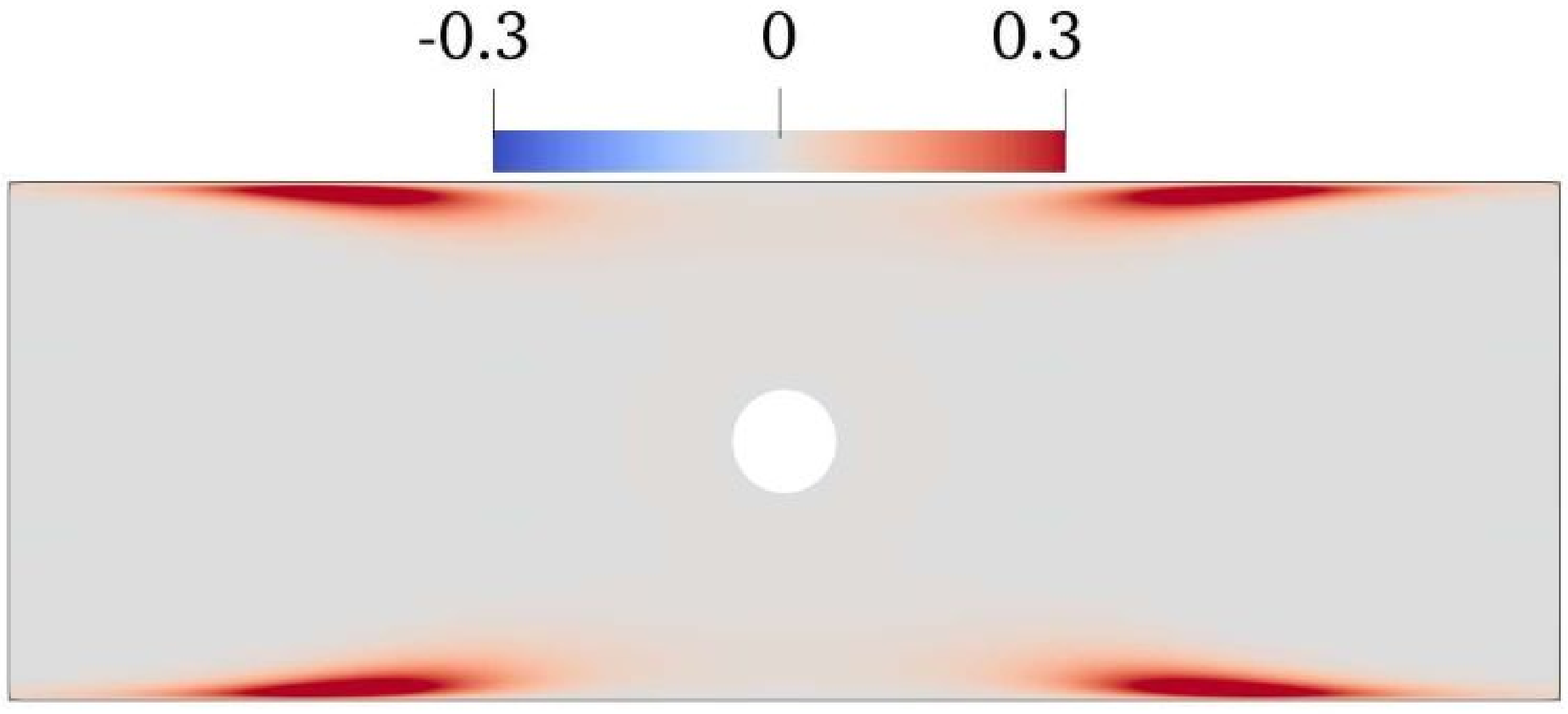}
		\put(-160,60){(c)}
		\label{fig.Aq201}
	\end{minipage}}
	\hspace{30pt}
	\subfigure{
	\begin{minipage}[h]{0.4\linewidth}
		\centering
		\includegraphics[height=2.45cm]{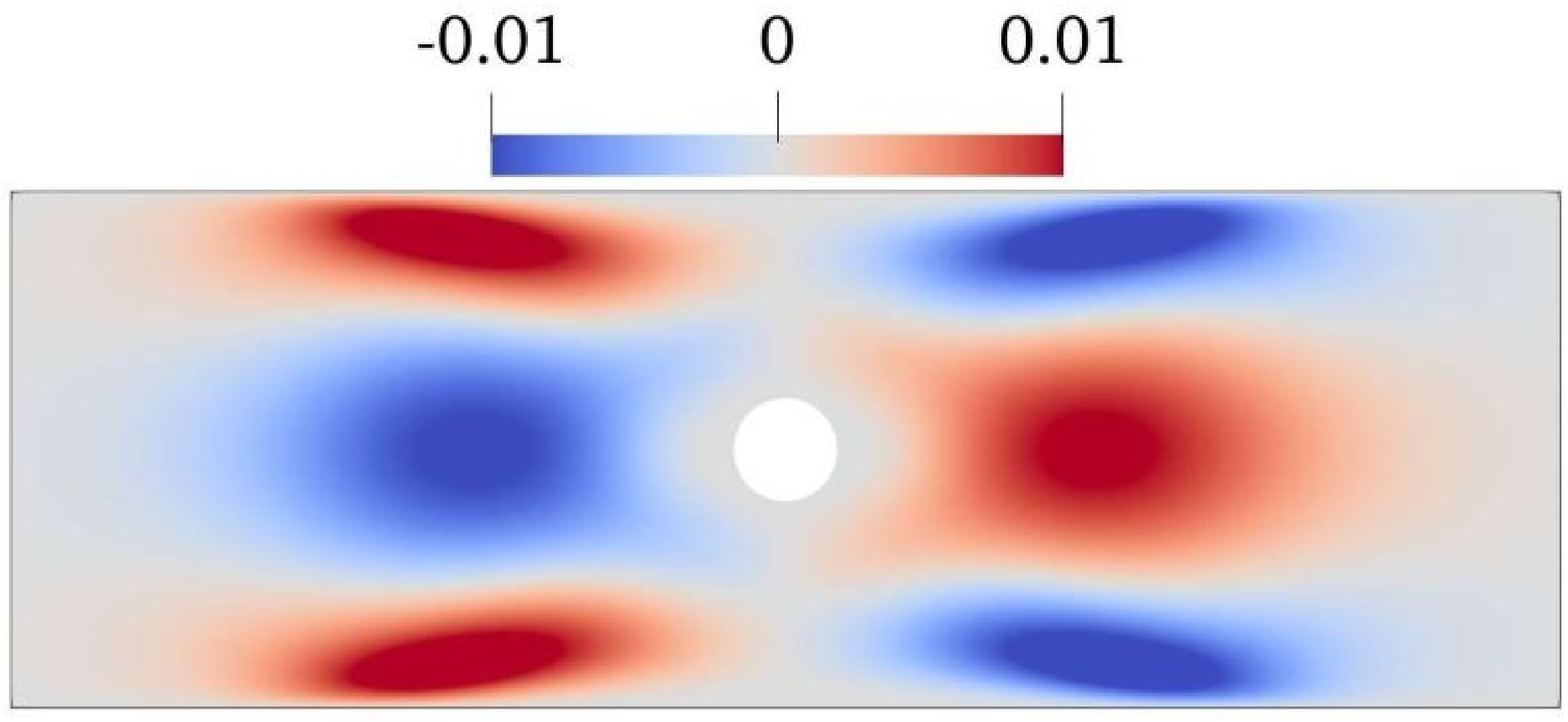}
		\put(-160,60){(d)}
		\label{fig.Avx01}
	\end{minipage}}
	\hspace{25pt}
	\subfigure{
	\begin{minipage}[h]{0.4\linewidth}
		\centering
		\includegraphics[height=2.55cm]{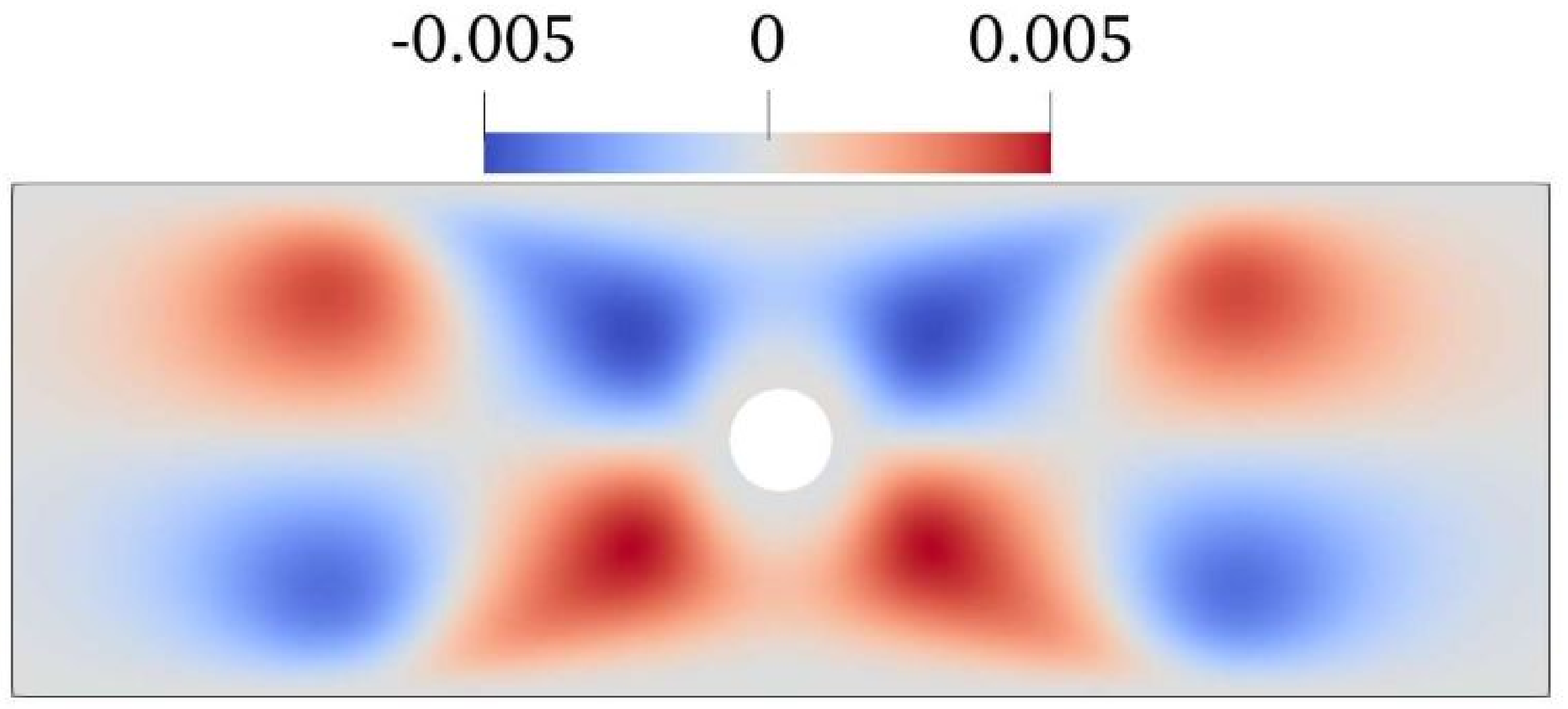}
		\put(-160,60){(e)}
		\label{fig.Avy01}
	\end{minipage}}
	\caption{(a)Growth rates of conduction regime of wire-plate EHD flow without cross-flow at different electric Reynolds numbers, the frequencies are all zero; and the corresponding leading eigenvectors at $ Re^E=0.1, C_I=0 $ for (b) positive charge density; (c) negative charge density; (d) $ x $-velocity; (e) $ y $-velocity.}
	\label{fig.eigenF01}
\end{figure}

Figure \ref{fig.p1gr01} shows the leading growth rates of the linearised wire-plate EHD flow in the conduction regime at different electric Reynolds numbers. The imaginary parts of the eigenvalues, representing the frequency of the base flow, are zero. We find that the growth rate increases with increasing $ Re^E $, indicating that the linear system becomes more unstable at larger electric intensity. Figure \ref{fig.eigenF01} (b-e) display the eigenvectors of the leading eigenmode for positive and negative ion species and velocity in $ x $ and $ y $ directions at $ Re^E=0.1 $. We can see that the perturbations of positive and negative ions are concentrated near the plate electrodes, and their values have opposite signs. Additionally, the eigenvectors of $x$-velocity and $y$-velocity are symmetric with respect to the vertical line $ x=0 $. These results may be instructive to gain more insight into the instability of the EHD conduction pumping in a wire-plate electrode configuration which uses the conduction mechanism to generate the ions. Particularly, the eigenvectors present the region where the perturbations accumulate and develop.

\subsubsection{Injection regime}
\begin{figure}
	\centering
	\subfigure{
	\begin{minipage}[h]{0.4\linewidth}\label{fig.Re09q1}
		\centering
		\includegraphics[height=2.5cm]{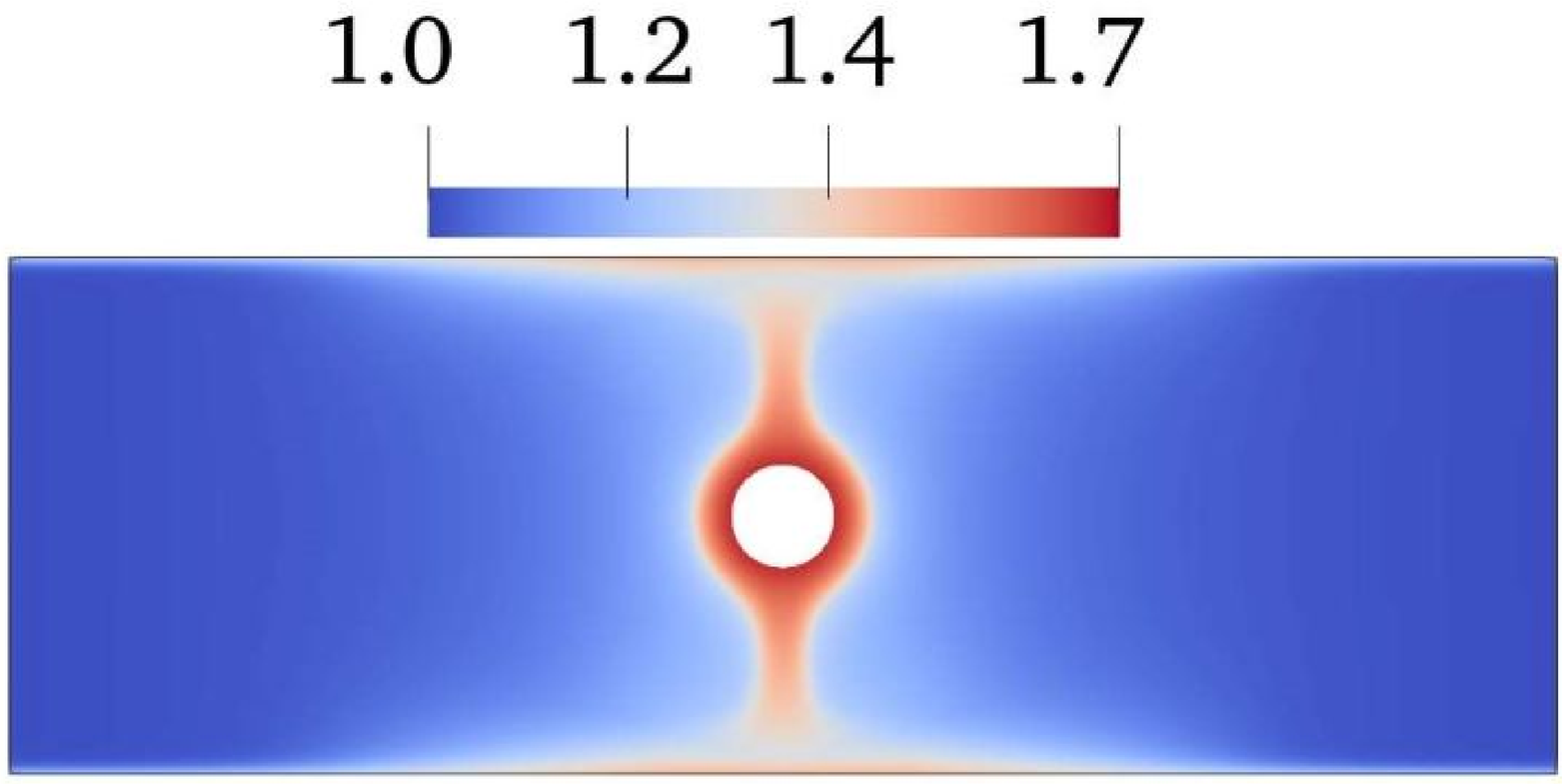}
		\put(-160,50){(a)}
	\end{minipage}}
    \hspace{15pt}
    \subfigure{
		\begin{minipage}[h]{0.4\linewidth}
	\centering
	\includegraphics[height=2.4cm]{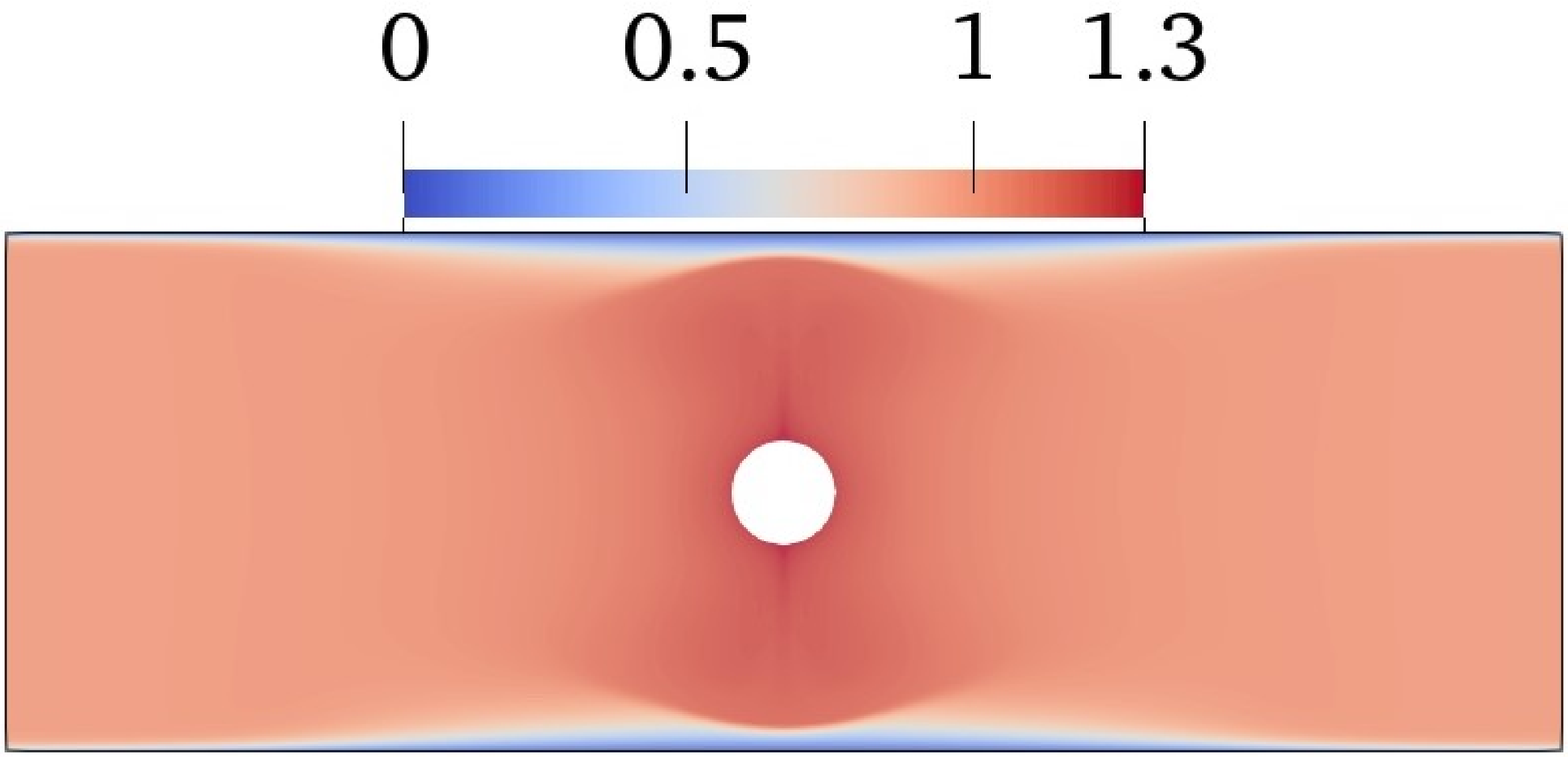}
	\put(-160,50){(b)}
	\label{fig.Re09q2}
    \end{minipage}}
    \hspace{15pt}
   \subfigure{
    \begin{minipage}[h]{0.4\linewidth}
	\centering
	\includegraphics[height=2.4cm]{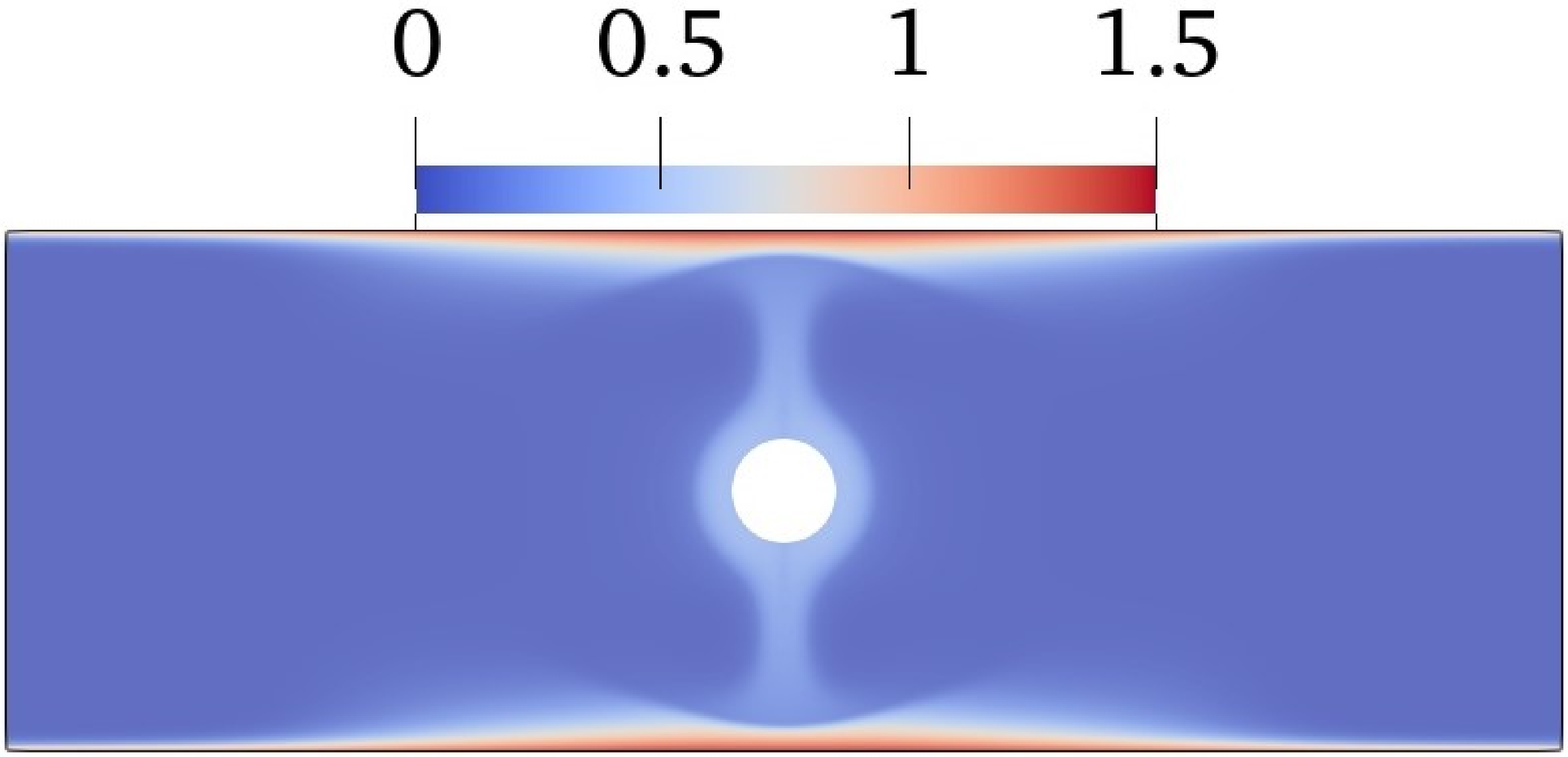}
	\put(-160,50){(c)}
	\label{fig.Re09netq}
    \end{minipage}}
	\hspace{15pt}
	\subfigure{
	\begin{minipage}[h]{0.4\linewidth}
		\centering
		\includegraphics[height=2.5cm]{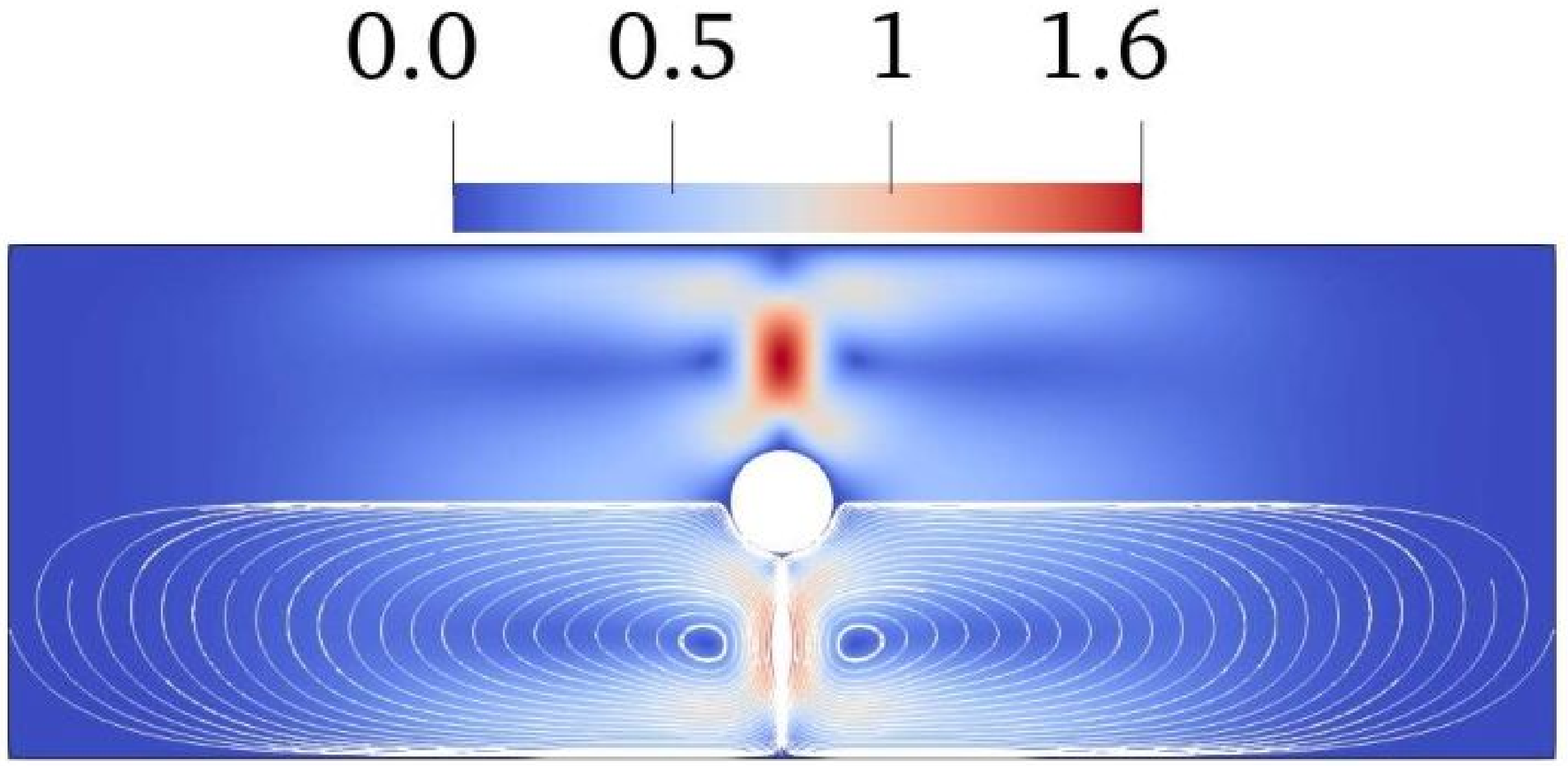}
		\put(-160,50){(d)}
		\label{fig.Re09v}
	\end{minipage}}
    \subfigure{
	\begin{minipage}[h]{0.4\linewidth}
	\centering
	\includegraphics[height=3.8cm]{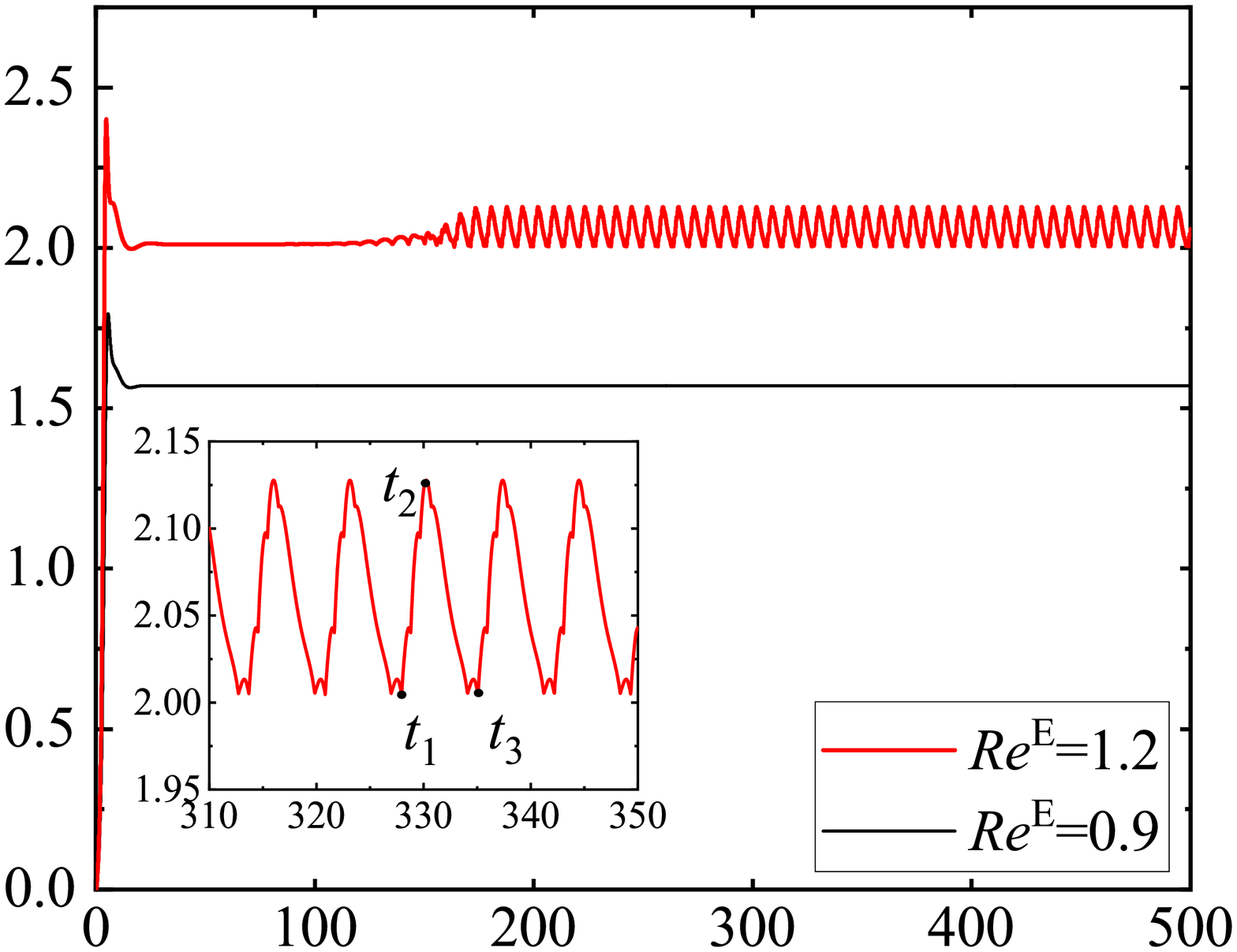}
	\put(-158,98){(e)}
	\put(-160,50){$ |U|_{max} $}
	\put(-70,-10){$ t $}
	\label{fig.p1Umax}
	\end{minipage}}	
    \hspace{15pt}
    \subfigure{
	\begin{minipage}[h]{0.4\linewidth}
		\centering
		\includegraphics[height=5.6cm]{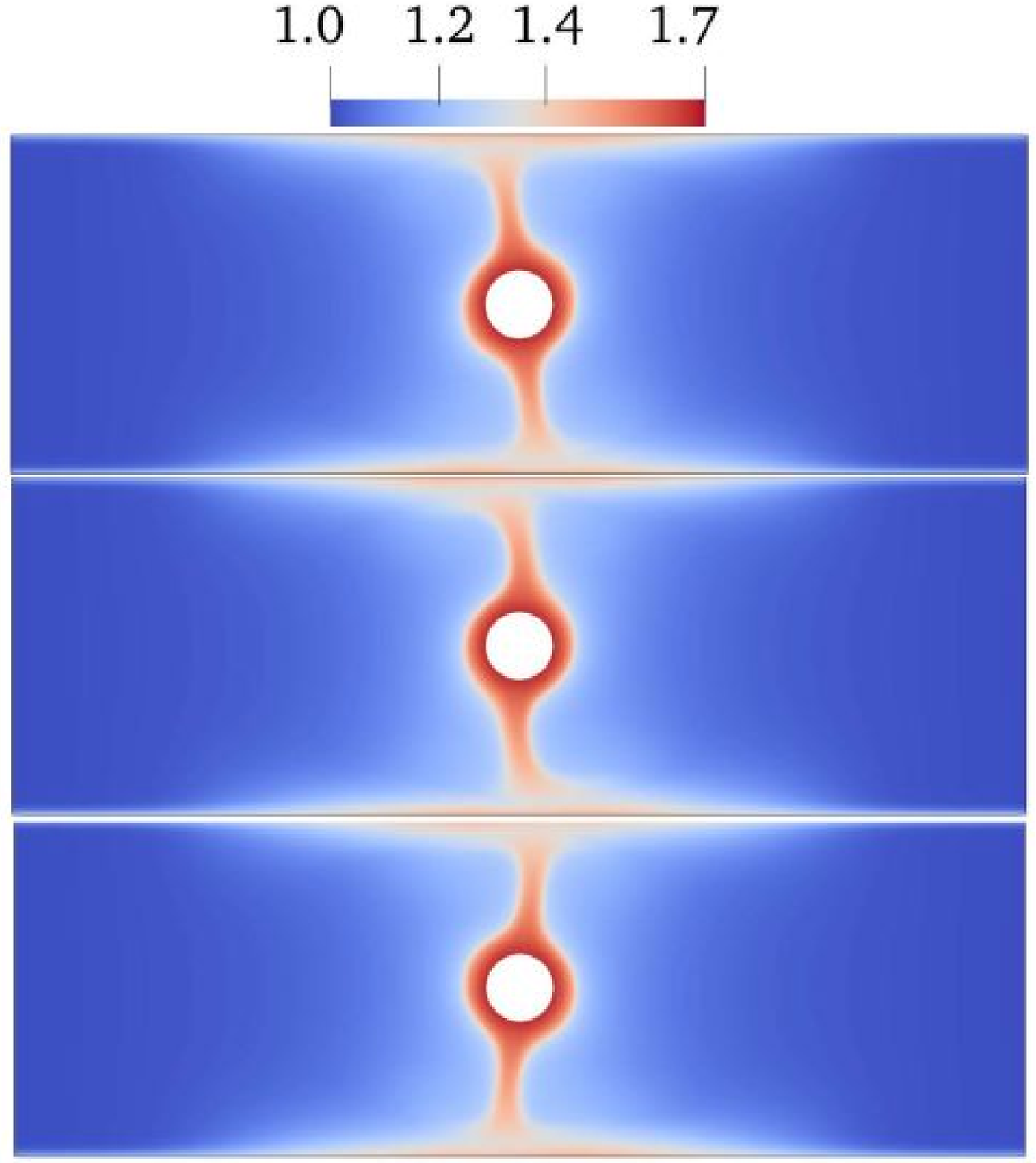}
		\put(-128,100){\textcolor{white}{$ t=t_1 $}}
        \put(-128,55){\textcolor{white}{$ t=t_2 $}}
        \put(-128,8){\textcolor{white}{$ t=t_3 $}}		
		\put(-160,130){(f)}
		\label{fig.t1t2t3}	
\end{minipage}}
	\caption{Nonlinear simulation of wire-plate EHD flow without a cross-flow at $ C_I=0.2 $ (thus, the dominant ion generation mechanism is injection. The same for the cases below with $C_I>0$.). Distribution of (a) positive species; (b) negative species; (c) net charges; and (d) velocity magnitude and streamlines at $ Re^E=0.9 $. (e) Time evolution of maximum velocity magnitude at different $ Re^E $. (f) Concentration of positive species at $ Re^E=1.2 $ at different times, from top to bottom, $ t=t_1, t_2, t_3 $, as shown in the inset of panel (e).}
	\label{fig.Re09}
\end{figure}

Then we explore the injection regime of wire-plate EHD flow without cross-flow. We plot the final steady state of nonlinear wire-plate EHD flow at $ C_I=0.2, Re^E=0.9 $ in figure \ref{fig.Re09} (a)-(d). We find from panel (a) that in the injection regime, the positive species hit the plate vertically under the action of the Coulomb force. The negative species generated by the dissociation process are also concentrated in the central region, as shown in panel (b). The distribution of the net charges is shown in panel (c), which resembles the pattern of positive species, leading to flow convection and the formation of two pairs of vortices (panel (d)). At larger $ Re^E $, the flow becomes oscillatory, as shown in figure \ref{fig.p1Umax}, which displays the time evolution of maximum velocity magnitude in the flows at $ Re^E=0.9 $ and $ Re^E=1.2 $, respectively. Three snapshots of the oscillatory flow at $ Re^E=1.2 $ are further presented in figure \ref{fig.t1t2t3}, from which we can see that the beam of positive species swings left and right.  

\begin{figure}
	\centering
	\includegraphics[height=4.5cm]{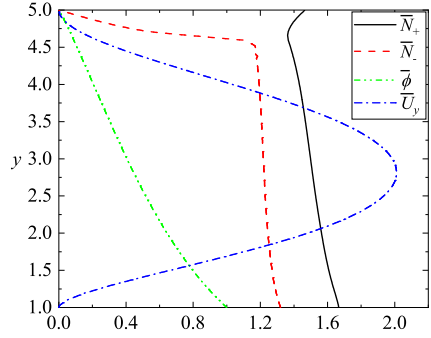}
	\label{fig.x01}
	\caption{Concentration of positive and negative species, electric potential and $ y $-velocity along the line of $ x=0 $ of SFD base flow of wire-plate EHD flow without cross-flow in the injection regime at $ Re^E=1.2, C_I=0.2 $. }
	\label{fig.base}
\end{figure}

As mentioned earlier, we adopt the global stability analysis to understand this oscillatory flow. In order to get a time-independent base flow for the linear stability analysis of the oscillatory flow at $ Re^E=1.2 $, the SFD method is applied. The base states in this case resemble those in figure \ref{fig.Re09} at $ Re^E=0.9 $ and thus will not be presented. We plot in figure \ref{fig.base} the concentration of positive and negative species, the electric potential as well as the $ y $-velocity in the upper part of the domain along the line of $ x=0 $. Compared to the conduction regime (figure \ref{fig.x0}), we find that the electric potential shows a similar trend. In addition, the densities of both kinds of ion species decrease with increasing $ y $ except that near the plate electrode, the positive species density increases due to the accumulation, and the negative ones continue to decrease to zero. Moreover, we find that the magnitude of the velocity in the injection mechanism is positive and two orders of magnitude greater than that in the conduction mechanism, which is also found in the experimental study of a needle-plate configuration \citep{sun2020experimental}.

\begin{figure}
	\centering
	\subfigure{
		\begin{minipage}[h]{0.4\linewidth}
			\centering
		\includegraphics[height=4.5cm]{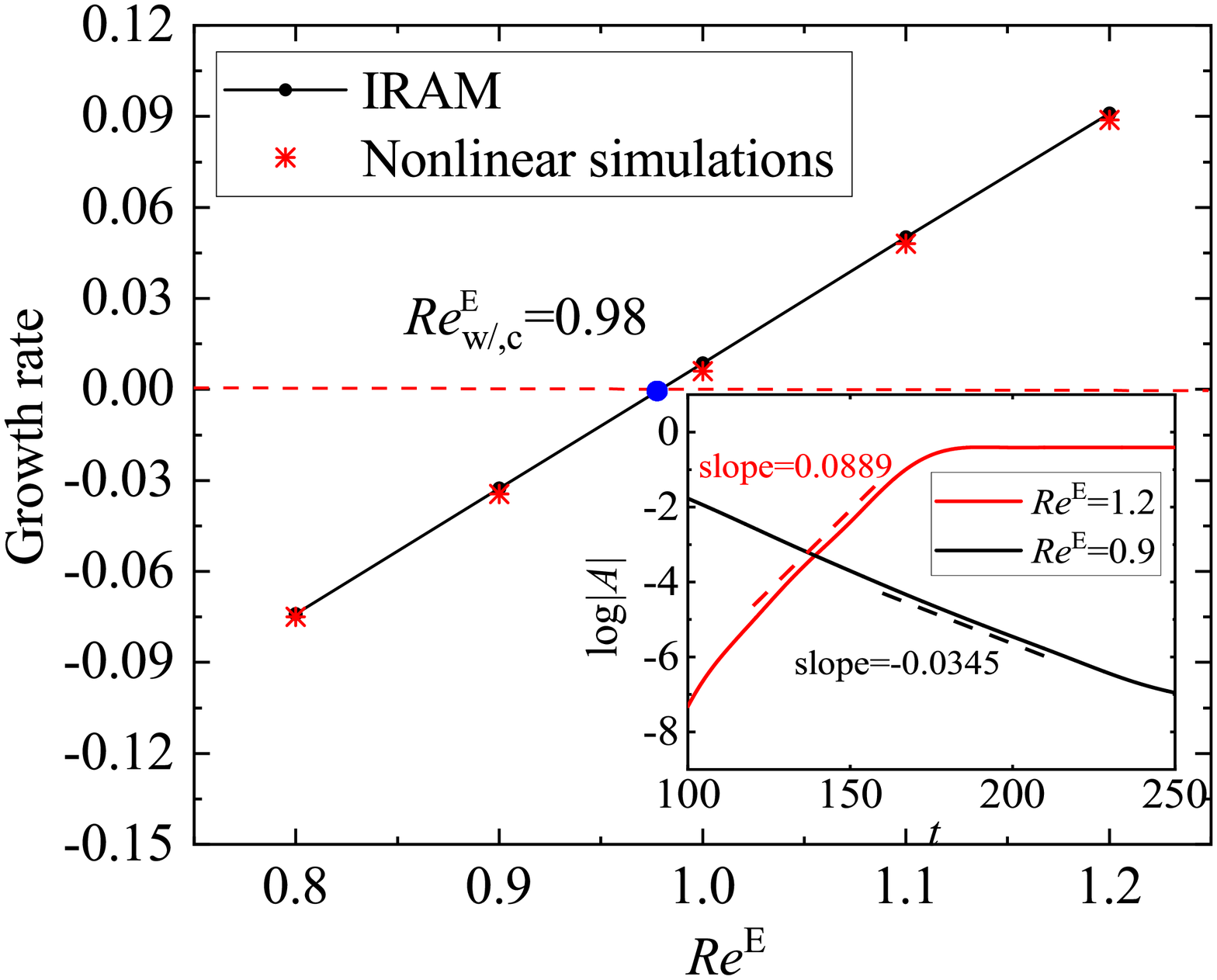}
        \put(-170,120){(a)}
        \label{fig.p1gr}
	\end{minipage}}
	\hspace{20pt}
	\subfigure{
	\begin{minipage}[h]{0.4\linewidth}
		\centering
		\includegraphics[height=4.5cm]{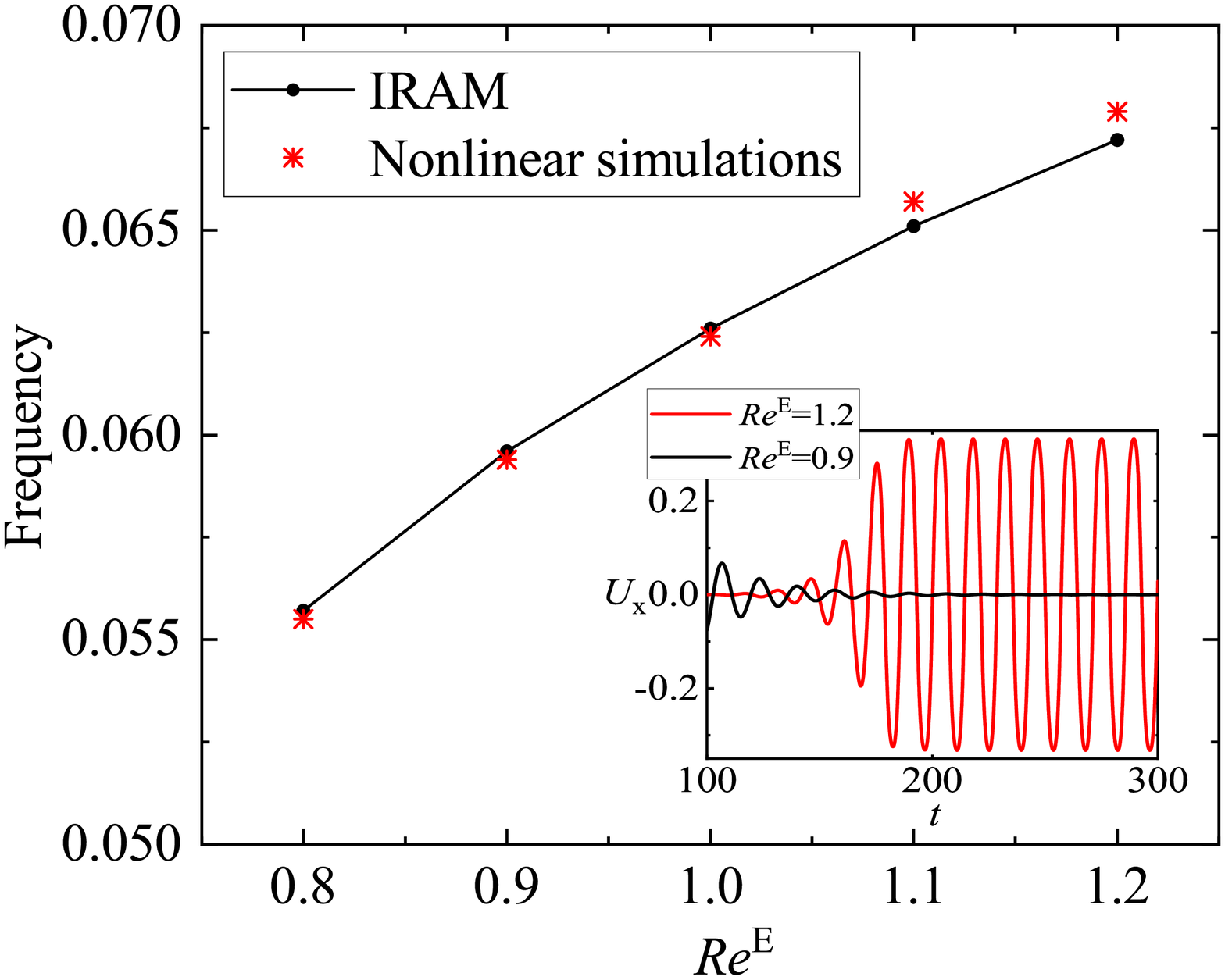}
        \put(-170,120){(b)}
        \label{fig.p1fr}
    \end{minipage}}
	\hspace{100pt}
	\subfigure{
	\begin{minipage}[h]{0.25\linewidth}
		\centering
		\includegraphics[height=3.5cm]{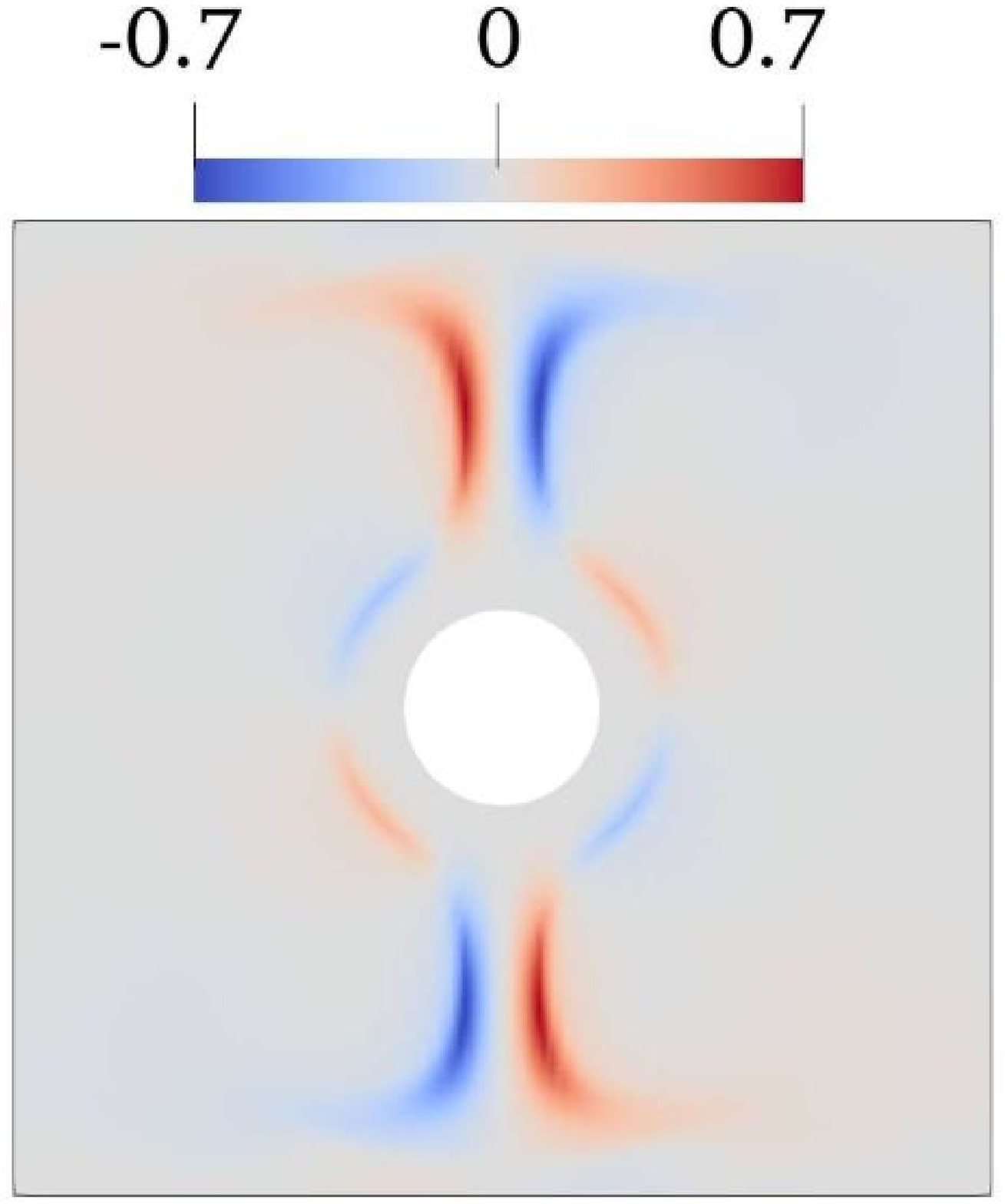}
		\put(-85,83){(c)}
		\label{fig.Aq1}
	\end{minipage}}
	\hspace{-10pt}
	\subfigure{
	\begin{minipage}[h]{0.25\linewidth}
		\centering
		\includegraphics[height=3.5cm]{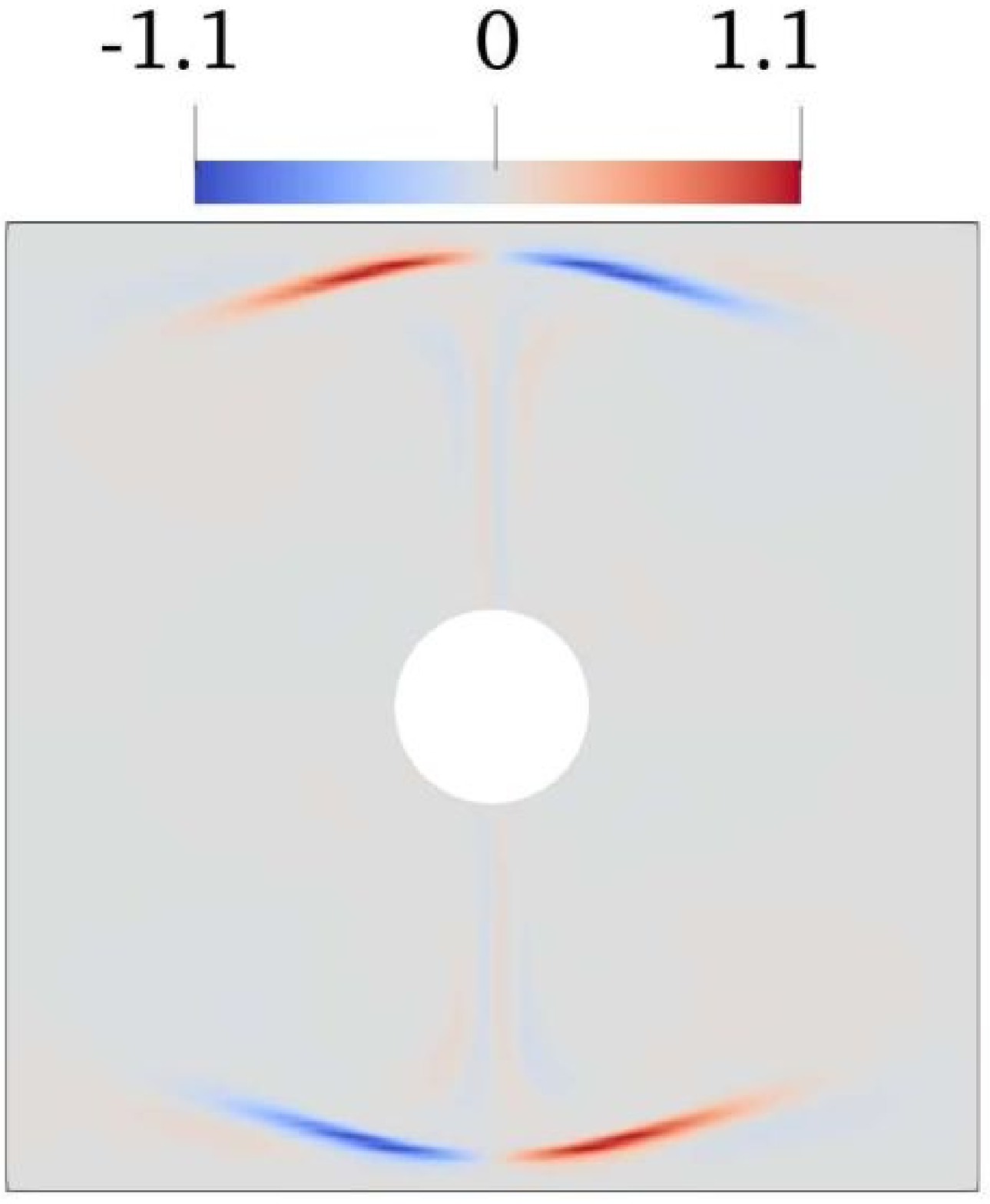}
		\put(-85,83){(d)}
		\label{fig.Aq2}
	\end{minipage}}
	\hspace{-10pt}
	\subfigure{
	\begin{minipage}[h]{0.25\linewidth}
		\centering
		\includegraphics[height=3.5cm]{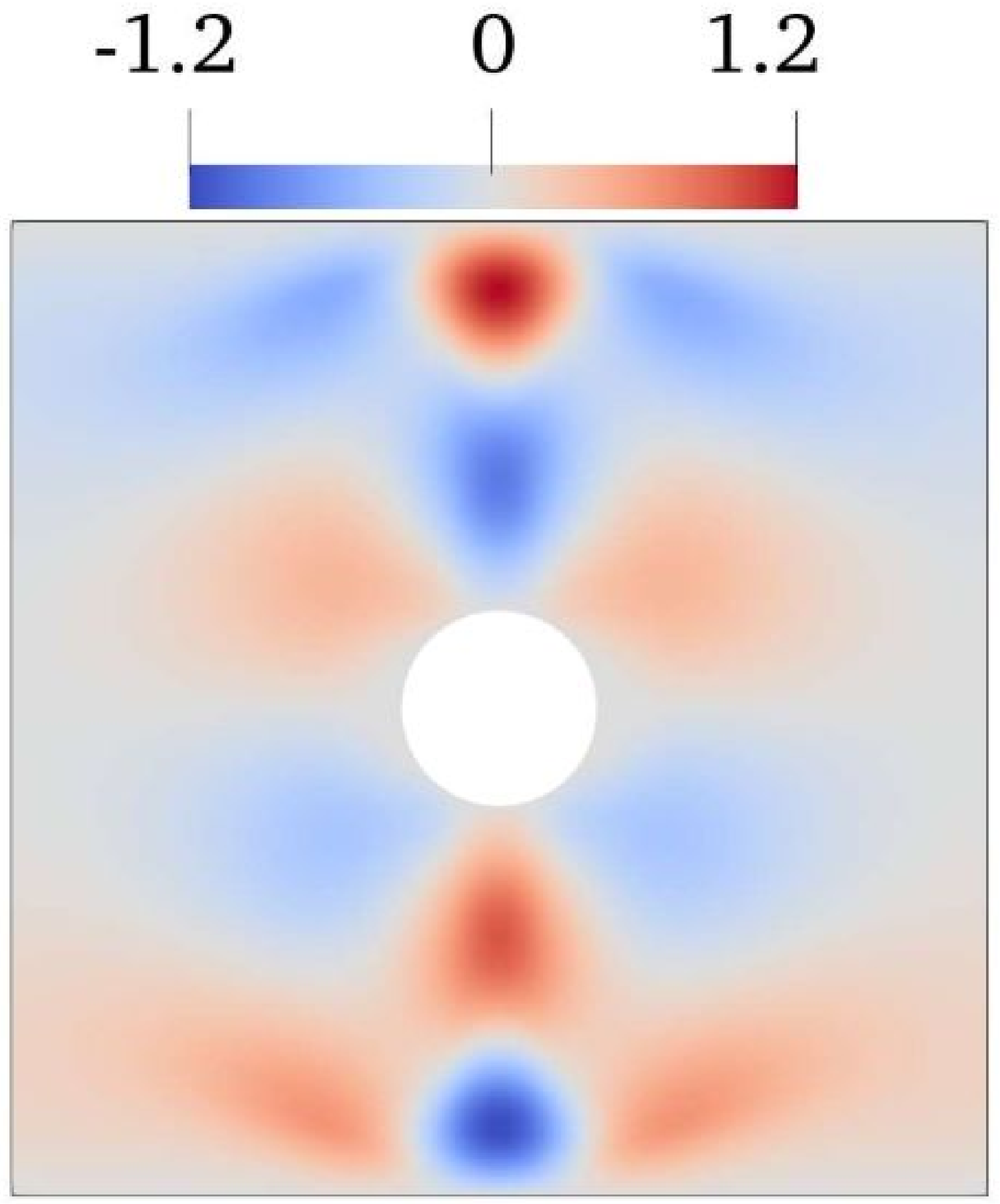}
		\put(-85,83){(e)}
		\label{fig.Avx}
	\end{minipage}}
	\hspace{-10pt}
	\subfigure{
	\begin{minipage}[h]{0.25\linewidth}
		\centering
		\includegraphics[height=3.5cm]{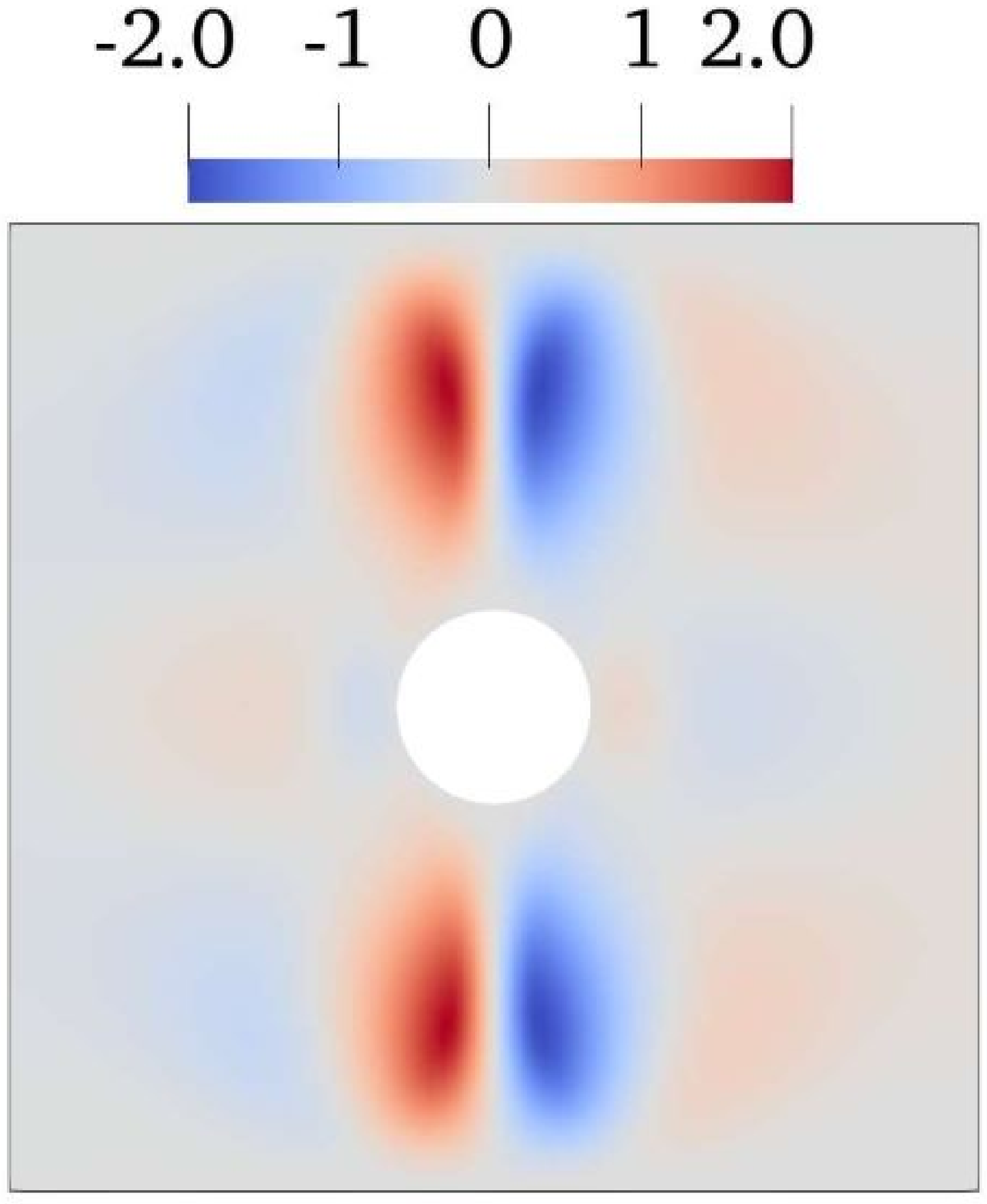}
		\put(-85,83){(f)}
		\label{fig.Avy}
	\end{minipage}}
	\caption{(a) Growth rates and (b) frequencies of wire-plate EHD flow without cross-flow in injection regime at different electric Reynolds numbers obtained by IRAM and nonlinear simulations. The inset of panel (a) is the log plot of the amplitude of $ U_x $ versus $ t $, and its slope in the linear phase gives the linear growth rate. The inset of panel (b) shows the $ x $-velocity evolution of point (0,0.25) in nonlinear simulation at $ Re^E=0.9, 1.2 $. The corresponding leading eigenvectors at $ Re^E=1.2 $ is shown: (c) positive charge density; (d) negative charge density; (e) $ x $-velocity; (f) $ y $-velocity.}
	\label{fig.eigenF}
\end{figure}

Figures \ref{fig.eigenF} (a-b) present the results of the global stability analysis of the flows using the IRAM, e.g., the growth rates and frequencies of the linearised flow in the injection regime at different electric Reynolds numbers. Panel (a) illustrates that the growth rate increases linearly with the increase of $ Re^E $, meaning that larger $ Re^E $ renders the flow more unstable. In addition, the critical electric Reynolds number is found to be $ Re^E_{w/,c}=0.98 $, above which the flow transitions from stable to unstable, and the oscillation occurs. Moreover, the frequency also increases with increasing $ Re^E $. Superposed on the IRAM results are the results of the nonlinear simulations of a slightly disturbed SFD base flow, represented by red stars. The perturbation will undergo a linear stage, as shown in the inset of panel (a), indicated by the dashed lines. The values of the slopes in the simulations are compared favourably to the growth-rate results of IRAM. In panel (b), the inset shows the evolution of the disturbance in $U_x$ at point (0,0.25) for $Re^E=0.9$ (stable) and $Re^E=1.2$ (unstable). The frequency information is extracted from the linear stage in these time series and also compared favourably to the frequency results of IRAM. Figures \ref{fig.eigenF} (c-f) show the eigenvectors of the leading eigenmodes at $ Re^E=1.2 $, and they display different patterns compared to those in conduction regime (figure \ref{fig.eigenF01} (b-e)). Additionally, it can be seen that there is a change of symmetry in the perturbation distributions for the positive and negative charge density as well as the velocity fields from figure \ref{fig.eigenF01} to figure \ref{fig.eigenF}. The former are symmetric with respect to the vertical central plane, whereas the latter are antisymmetric.


\begin{figure}
	\centering
	\subfigure{
	\begin{minipage}[h]{0.4\linewidth}
		\centering
		\includegraphics[height=2.5cm]{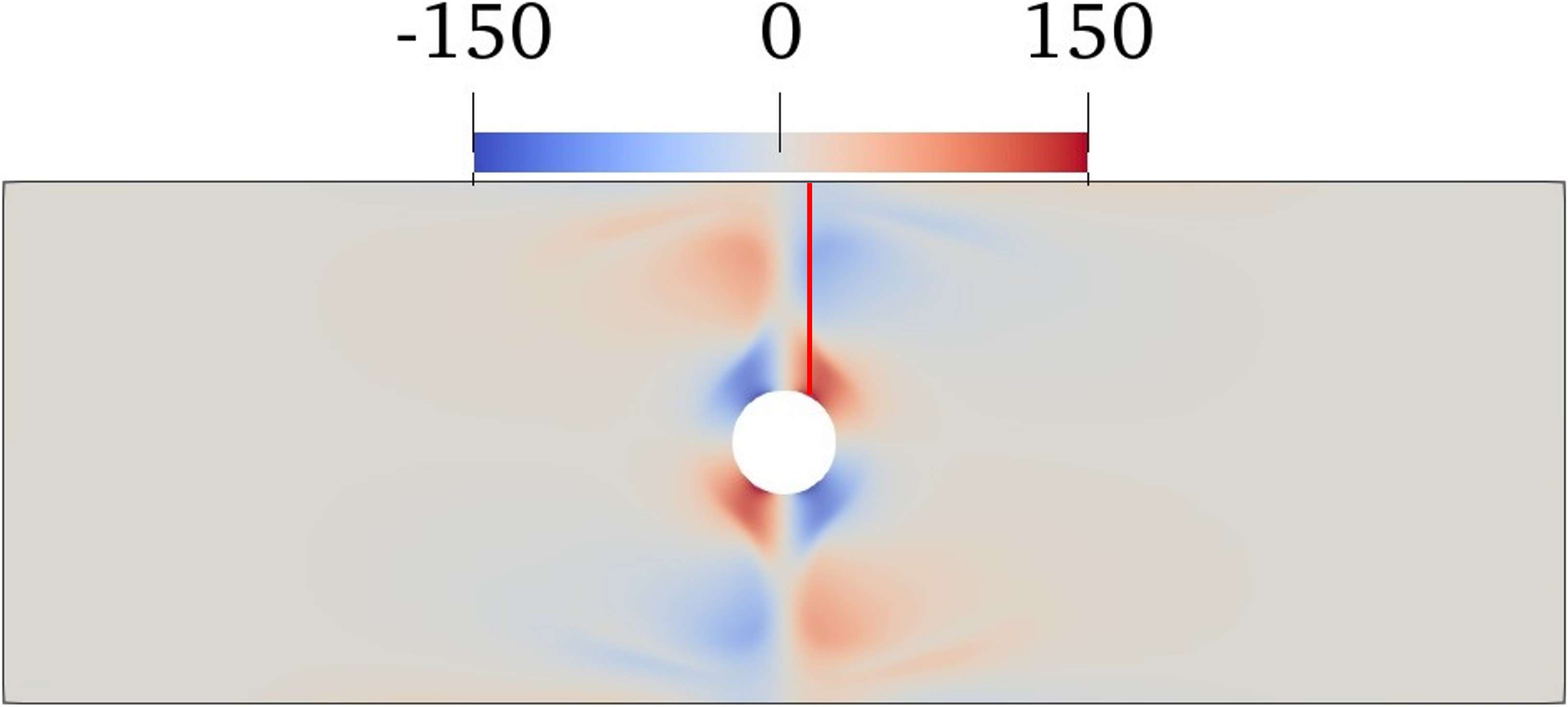}
		\put(-160,60){(a)}
		\label{fig.pressure}
	\end{minipage}}
	\hspace{20pt}
	\subfigure{
	\begin{minipage}[h]{0.4\linewidth}
		\centering
		\includegraphics[height=4cm]{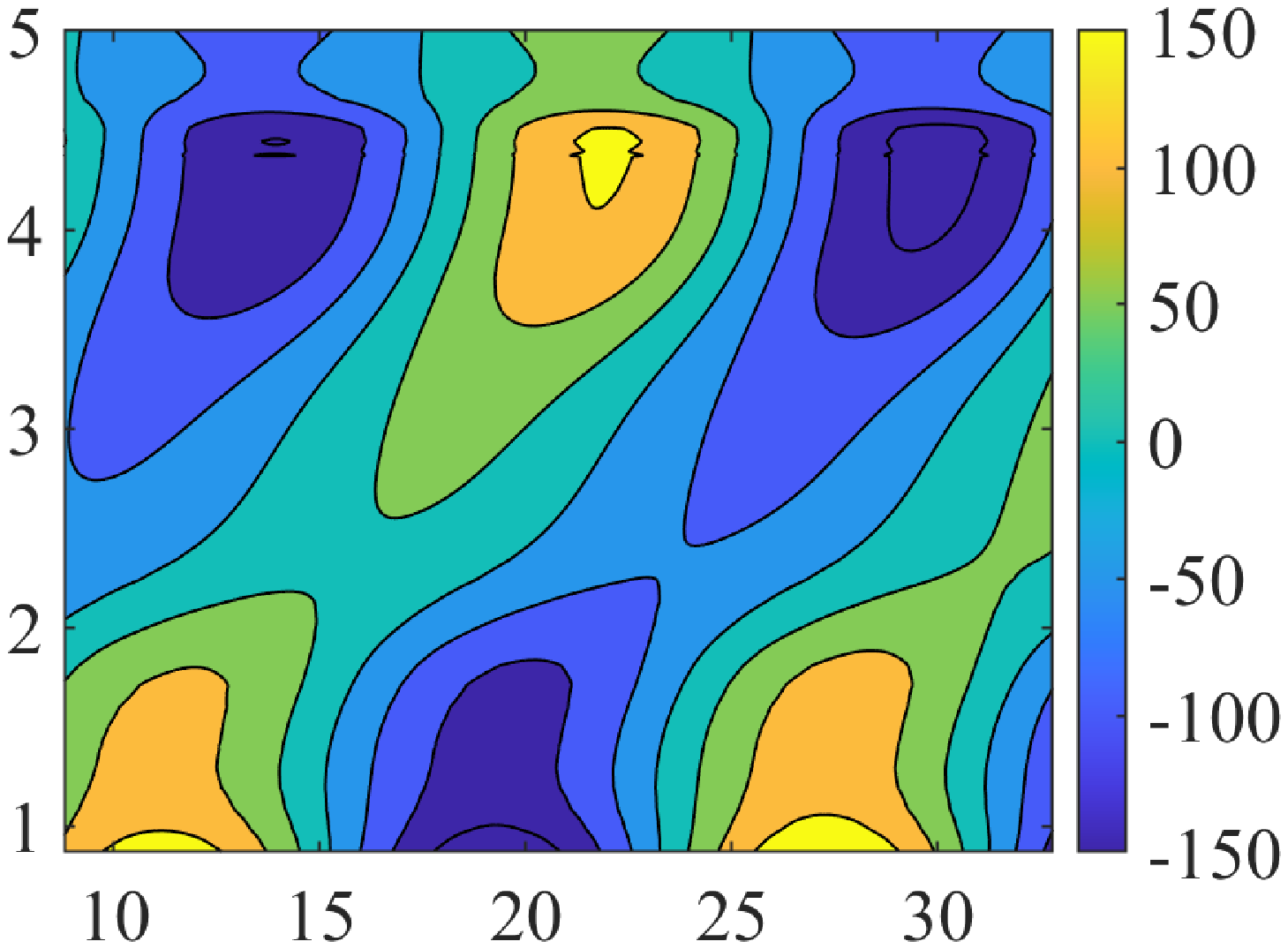}
		\put(-165,110){(b)}
		\put(-80,-10){$ t $}		
		\put(-160,60){$ y $}				
		\label{fig.pret}
	\end{minipage}}
	\caption{(a) The leading eigenvectors of wire-plate EHD flow without a cross-flow at $ Re^E=1, C_I=0.2 $ for pressure. (b) Pressure perturbation at $ x = 0.5 $ of the upper plate (red line in panel (a)) as a function of $ y $ and time.}
	\label{fig.prem} 
\end{figure}

A discussion on the instability mechanism of wire-plate EHD flow in the injection regime is in order. Figure \ref{fig.pressure} shows the linear eigenvector of the pressure of wire-plate EHD flow without a cross-flow at $ Re^E=1 $, which is slightly larger than the critical value, as shown in figure \ref{fig.p1gr}. We plot in figure \ref{fig.pret} the time-series of the pressure perturbation at $ x = 0.5 $ of the upper plate (red line in panel (a)). It illustrates that the maximum (minimum) pressure near the wire propagates upward to the plate electrode, indicating that there is a pressure feedback mechanism. It seems that the high pressure near the plate caused by the impinging flow leads to an increase in pressure near the wire, which creates a feedback loop. This pressure feedback mechanism may be a major cause for system instability, which is also observed in thermal plumes \citep{lesshafft2015linear} and cavity flows \citep{aakervik2007optimal}.

\subsection{Wire-plate EHD flow with a weak cross-flow}

\begin{figure}
	\centering
	\subfigure{
	\begin{minipage}[h]{0.4\linewidth}
	\centering
	\includegraphics[height=4.2cm]{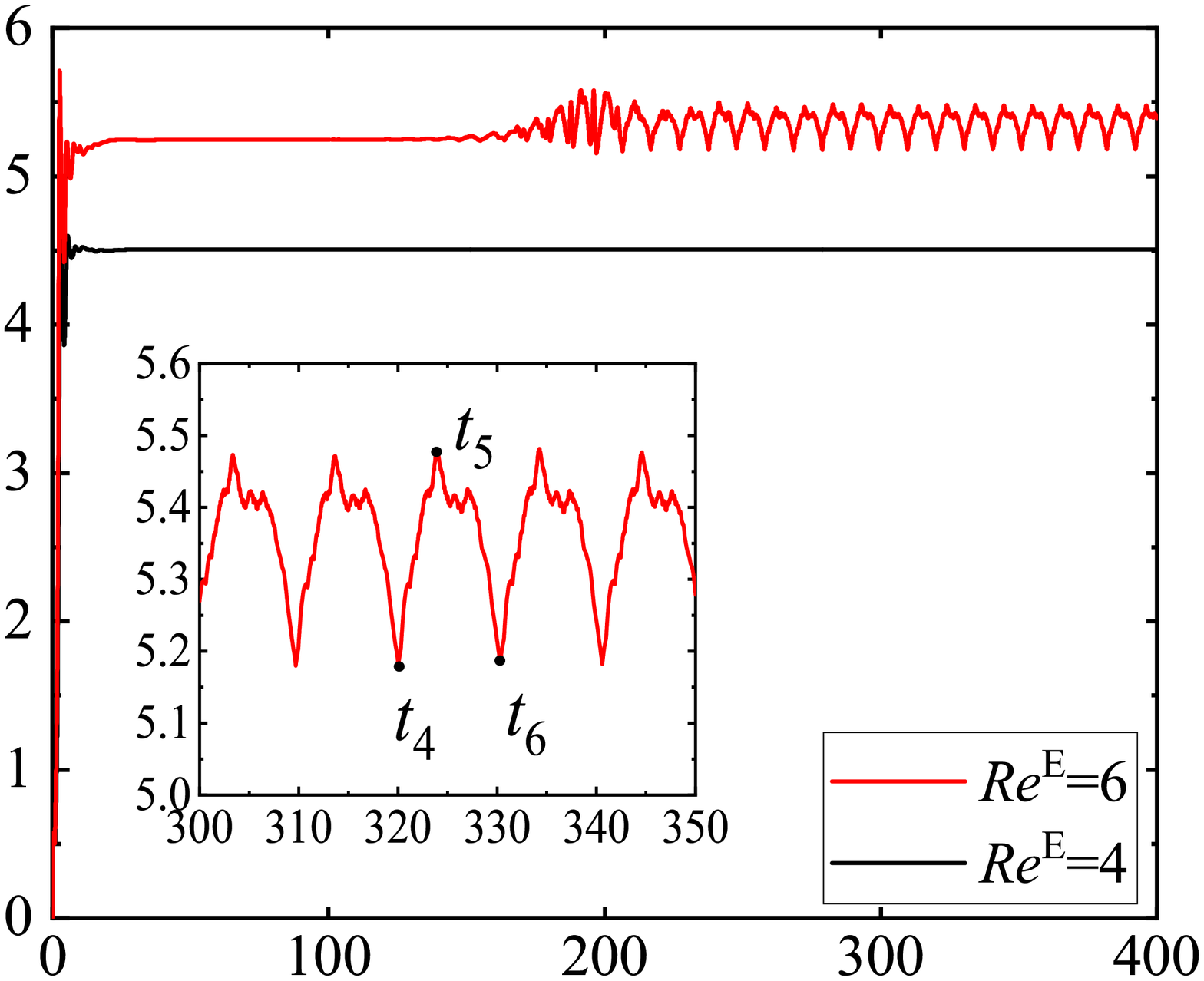}
	\put(-160,110){(a)}			
	\put(-170,65){$ |\Ub|_{max} $}
	\put(-70,-10){$ t $}
	\put(-90,-25){$ Re^E=4 $}
	\label{fig.p2Umax}
\end{minipage}}
	\hspace{150pt}
	\subfigure{
	\begin{minipage}[h]{0.4\linewidth}
			\centering
			\includegraphics[height=2.2cm]{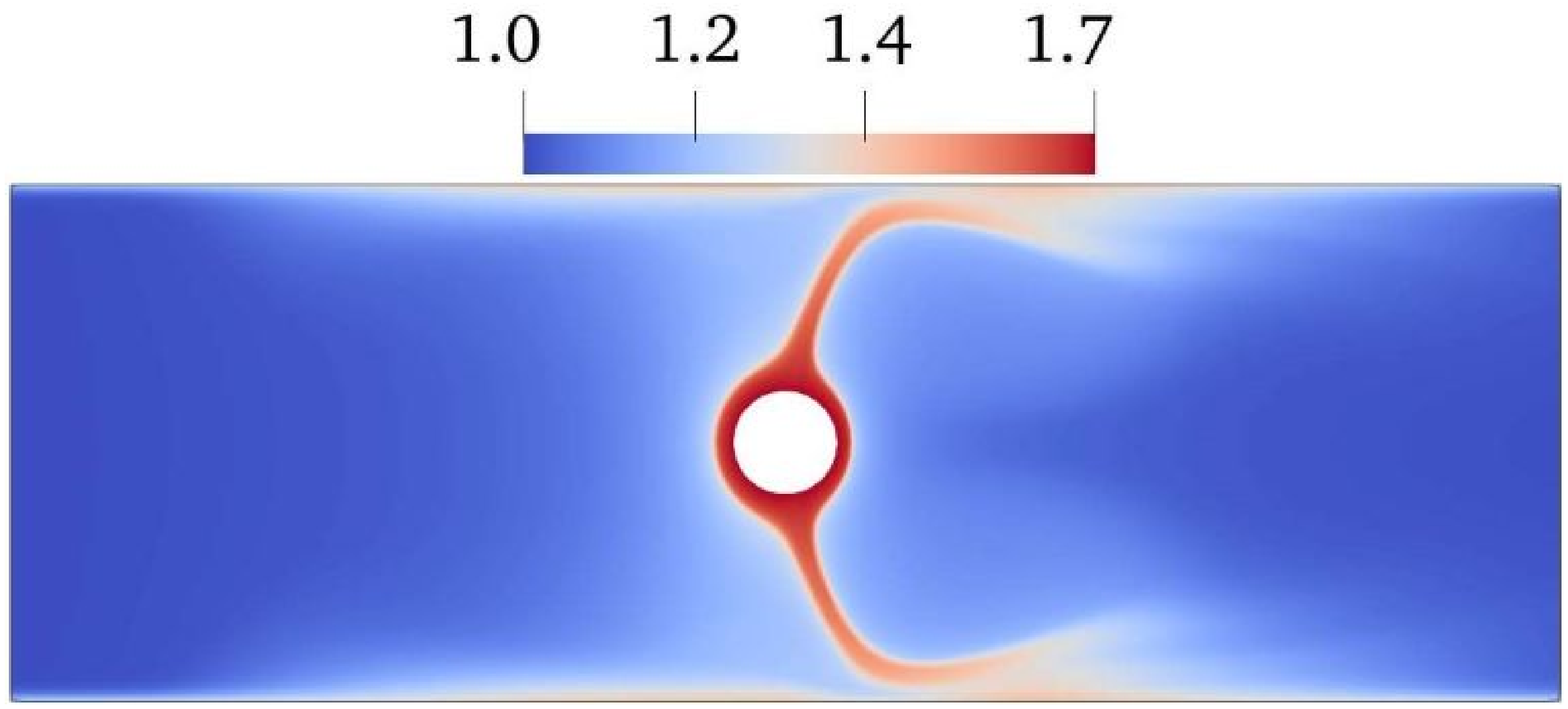}
			\put(-152,50){(b)}
			\label{fig.q1U03}
	\end{minipage}}
	\hspace{15pt}
	\subfigure{
		\begin{minipage}[h]{0.4\linewidth}
			\centering
			\includegraphics[height=2.2cm]{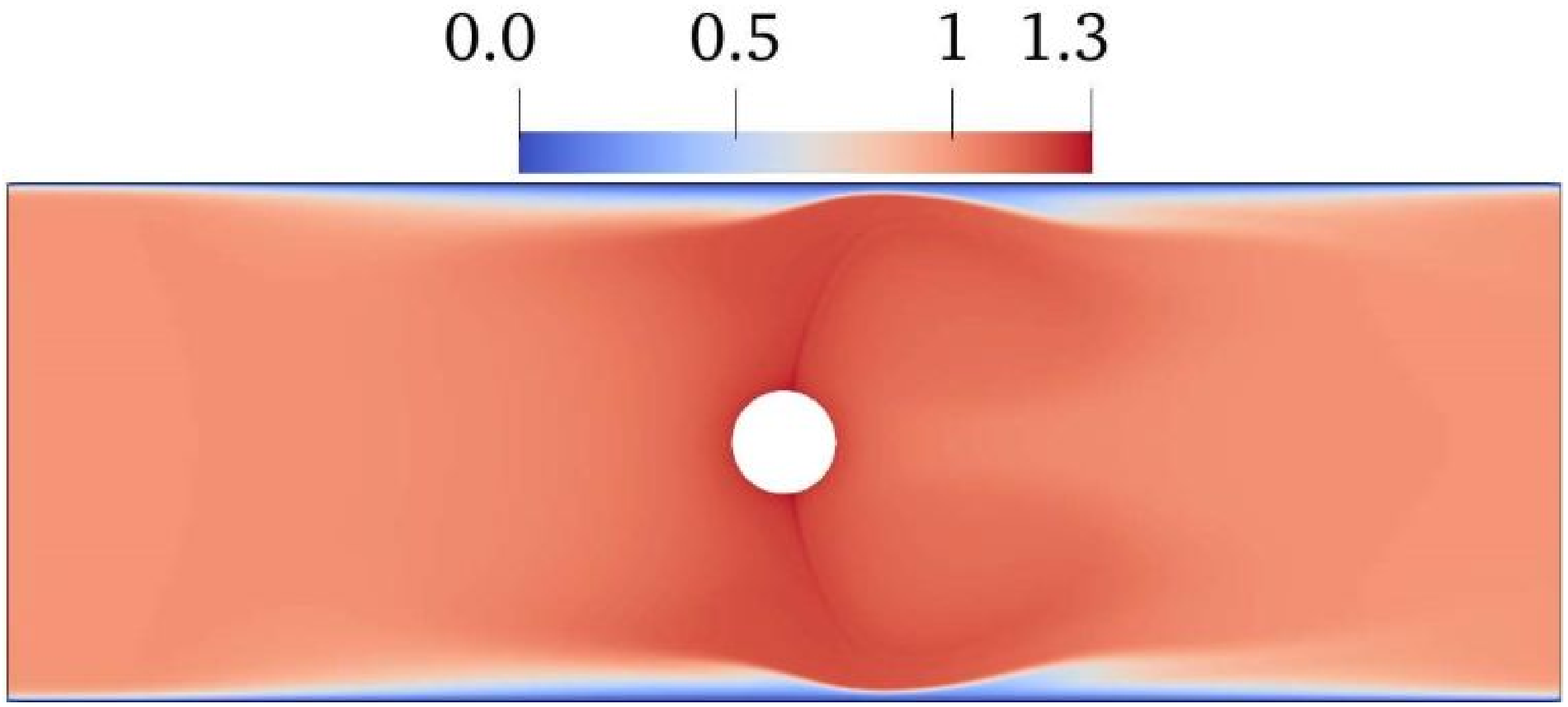}
			\put(-152,50){(c)}
			\label{fig.q2U03}
	\end{minipage}}
	\hspace{150pt}
	\subfigure{
		\begin{minipage}[h]{0.4\linewidth}
			\centering
			\includegraphics[height=2.2cm]{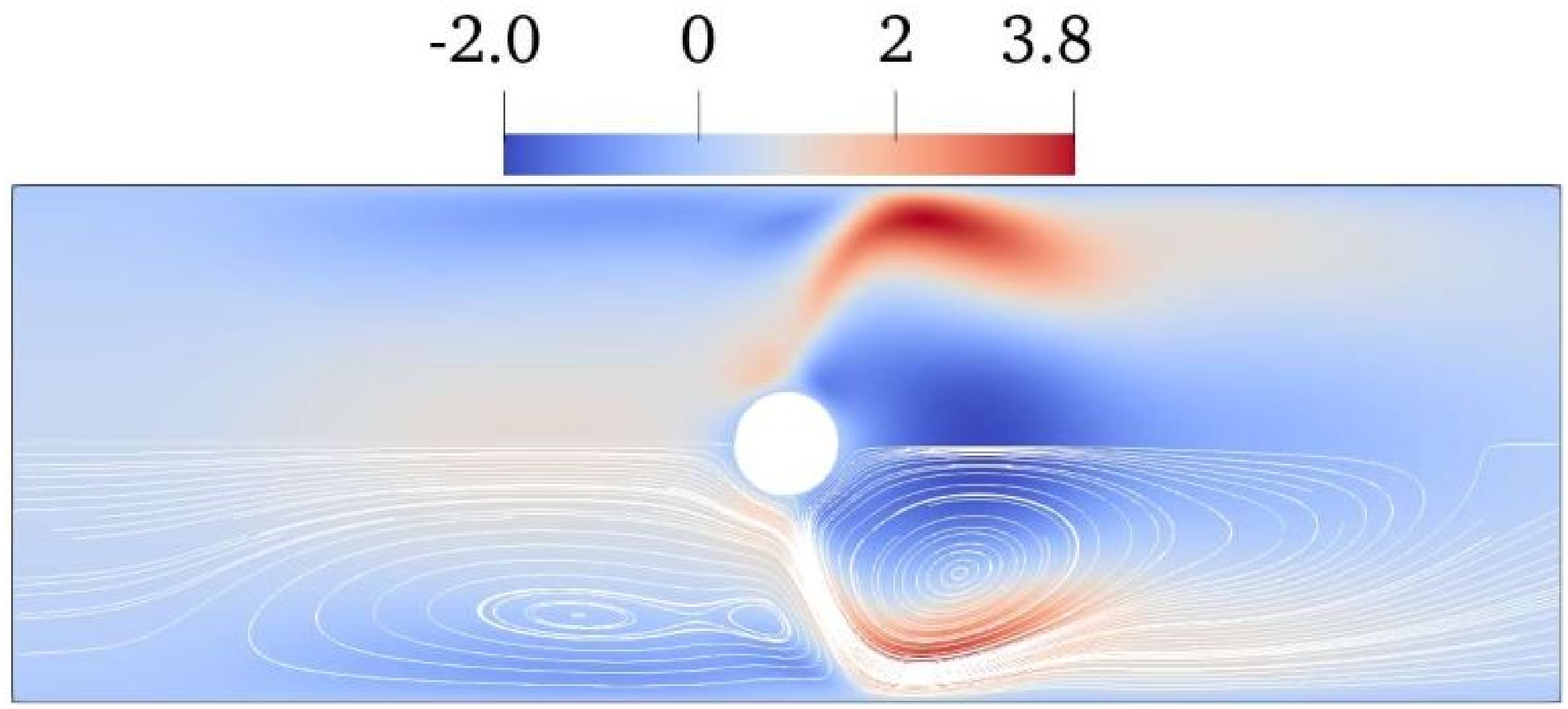}
			\put(-152,50){(d)}
			\label{fig.vxU03}
	\end{minipage}}
	\hspace{15pt}
	\subfigure{
		\begin{minipage}[h]{0.4\linewidth}
			\centering
			\includegraphics[height=2.2cm]{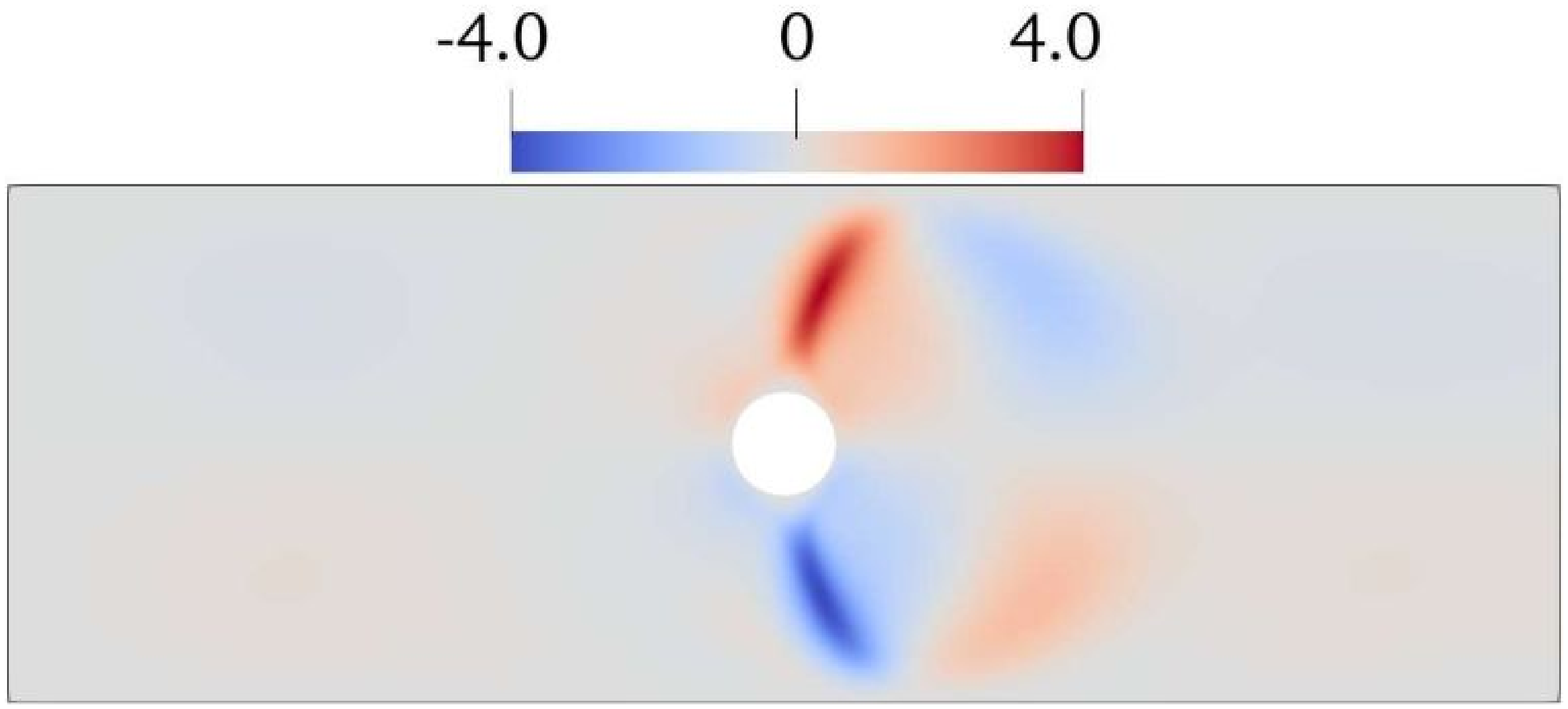}
			\put(-152,50){(e)}
			\label{fig.vyU03}
	\end{minipage}}
	\hspace{150pt}
	\subfigure{
	\begin{minipage}[h]{0.4\linewidth}
		\centering
		\includegraphics[height=5.5cm]{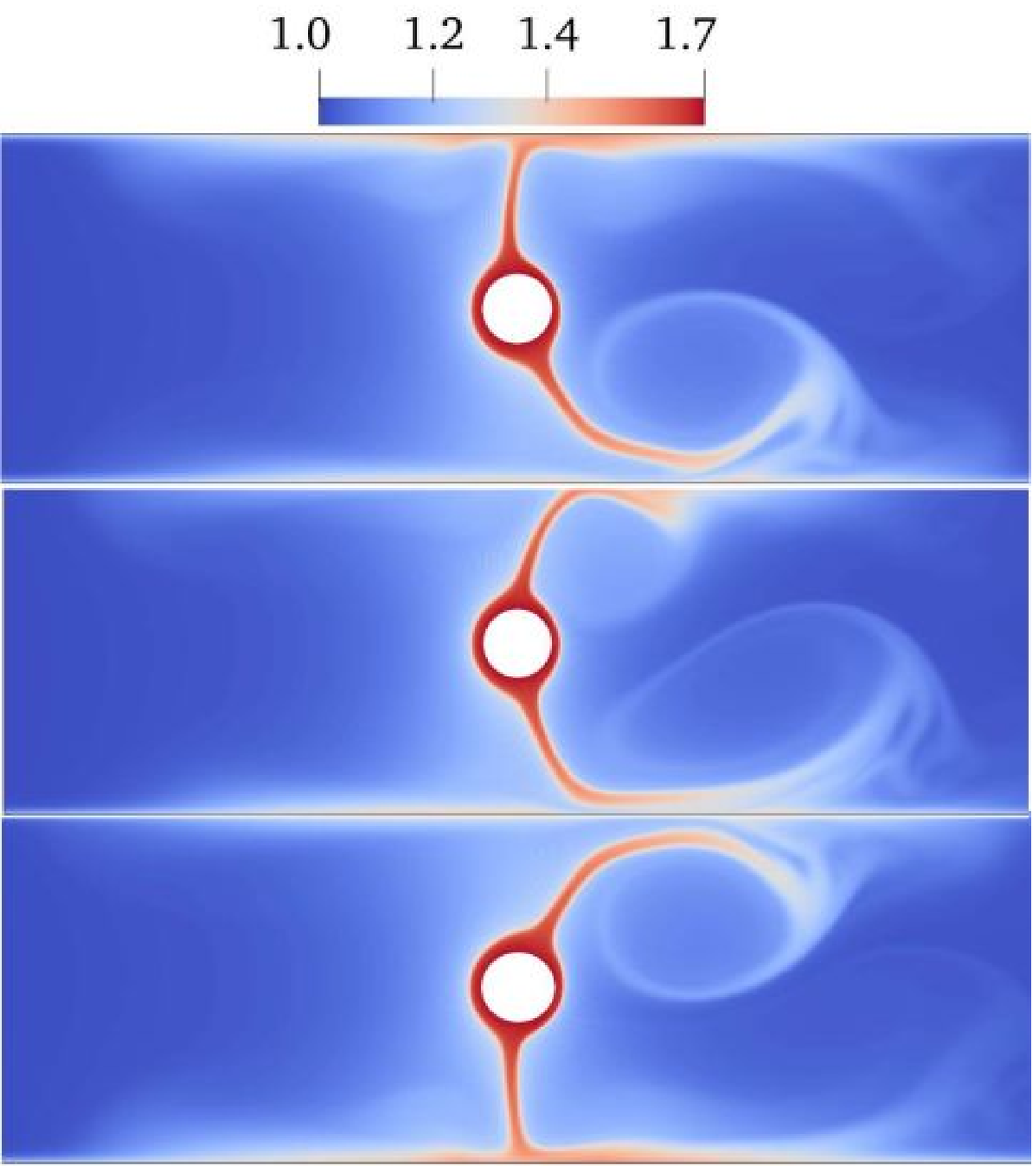}
		\put(-150,140){(f)}
		\put(0,170){$ Re^E=6 $}
		\put(-128,100){\textcolor{white}{$ t=t_4 $}}
		\put(-128,55){\textcolor{white}{$ t=t_5 $}}
		\put(-128,8){\textcolor{white}{$ t=t_6 $}}				
		\put(-115,160){Positive charge density}		
		\label{fig.t263}
	\end{minipage}}
    \subfigure{
	\hspace{15pt}
	\begin{minipage}[h]{0.4\linewidth}
		\centering
		\includegraphics[height=5.5cm]{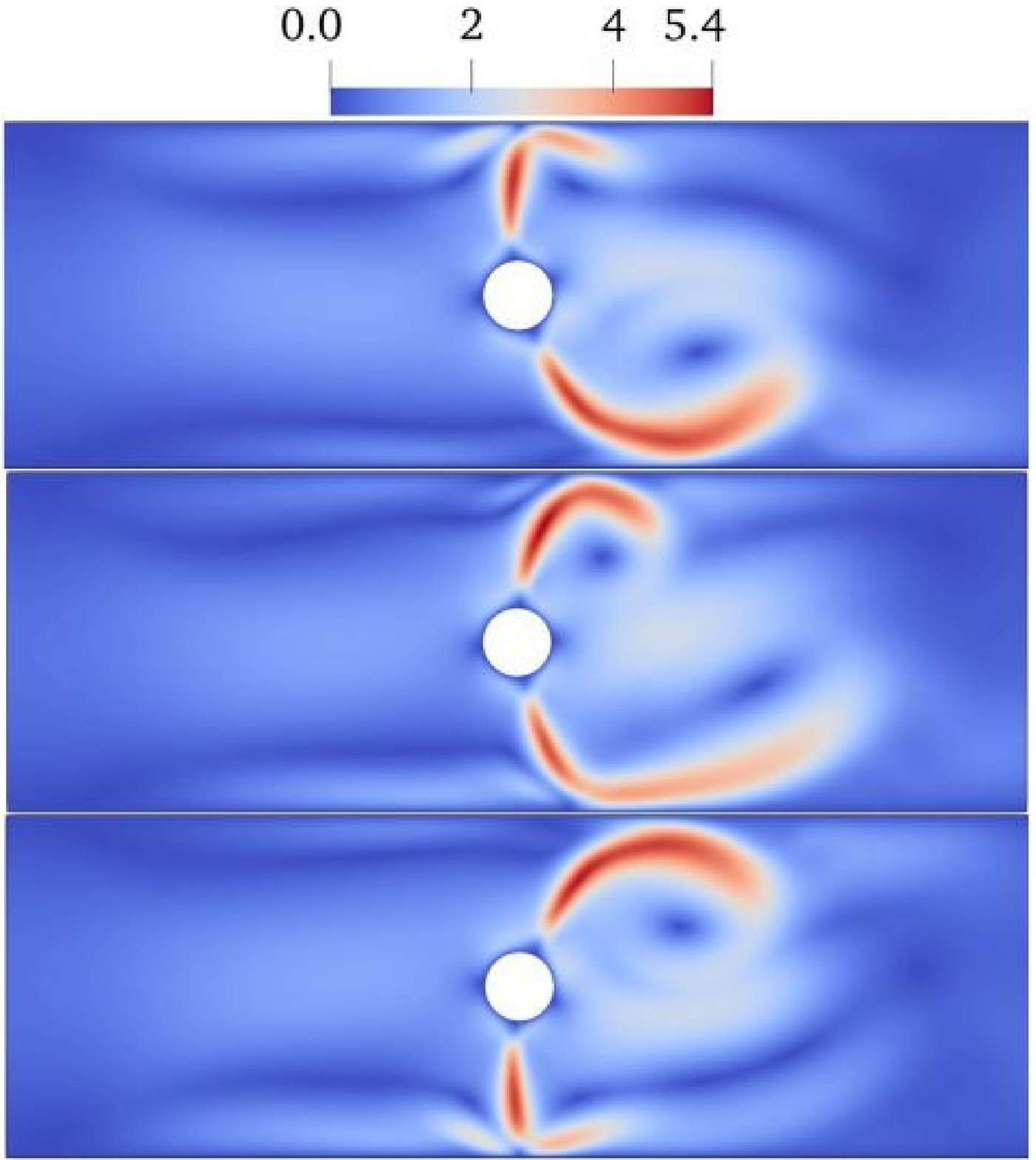}
		\put(-150,140){(g)}
	    \put(-10,180){}
		\put(-128,100){\textcolor{white}{$ t=t_4 $}}
        \put(-128,55){\textcolor{white}{$ t=t_5 $}}
        \put(-128,8){\textcolor{white}{$ t=t_6 $}}		
		\put(-110,160){Velocity magnitude}			
		\label{fig.t267}
	\end{minipage}}
	\caption{(a) Evolution of the maximum velocity norm of nonlinear wire-plate EHD-Poiseuille flow at $ U_0=0.3 $ versus $ Re^E $. The instants $ t_4, t_5, t_6 $ in the zoom-in view denote three sampling times which will be depicted in panels (f) and (g). Distribution of (b) positive species; (c)negative species; (d) $ x $-velocity field and streamlines; (e) $ y $-velocity field of the final steady state of wire-plate EHD flow with a weak cross-flow at $ U_0=0.3, Re^E=4 $ . Distribution of (f) positive species density and (g) velocity magnitude of wire-plate EHD flow with a weak cross-flow at a periodic state at $ U_0=0.3, Re^E=6 $. In each panel, from top to bottom, the time stamps are $ t=t_4 $, $ t=t_5 $, $ t=t_6 $, respectively ($ t_4-t_6 $ can be found in panel (a)). }
	\label{fig.paC}
\end{figure}

When a weak flow is imposed, the positive charges will be convected downstream and hit the plate electrodes obliquely. Such a pattern is common in ESP, in which the charged particles attached by ions will advect downstream and settle on the plates due to the EHD secondary flow. Via the study of this case, we hope to improve our understanding of the dynamics of the large-scale flow structures in ESP. Firstly, the streamlines of the flow field and the distribution of charge density may be helpful to unravel the interaction between electric field and flow field. Additionally, it has been pointed out that reducing the generation of turbulence helps increase the deposition efficiency of ESP \citep{leonard1983experimental}. Our global linear stability analysis of this flow is conducive to a deeper understanding of the flow instability in ESP.

\subsubsection{The EHD flow and the pressure field on the wire}
We first investigate the time evolution of the wire-plate EHD flow with a weak cross-flow at the parameters $ C_0=3, M=37, K_r=1,  C_I=0.2, O_s=8.6, U_0=0.3, \lambda=0.2 $ and $ \alpha=0.001 $. In figure \ref{fig.p2Umax}, we show the time evolution of the maximum velocity norm of nonlinear wire-plate EHD-Poiseuille flow at $ Re^E=4 $ and $ Re^E=6 $, respectively. We observe that at $ Re^E=4 $, the velocity field remains a steady state. The distribution of charges and the velocity field of the steady state are presented in figures \ref{fig.paC} (b-e). It can be seen from panels (b) and (c) that the beams of positive and negative species both tilt downstream and hit the plate electrodes. We find from panel (d) that the vortices also tilt downstream. At larger $ Re^E=6 $, a periodic oscillation emerges (figure \ref{fig.p2Umax}). Figures \ref{fig.paC} (f-g) exhibit the positive species concentration and velocity magnitude field of time-periodic state at different times at $ Re^E=6 $. It can be seen that the charge beams swing strongly. Such a flow condition is unfavourable for the ESP collecting the charged dusts. This is because the charged beams and the corresponding flow structures will bounce back after hitting the walls, bringing the charged dusts back to the bulk region. This seems to suggest the importance of choosing a proper voltage applied on the wire when operating the ESP. It is advisable to apply a voltage that is lower than the oscillatory threshold, so that on one hand the EHD secondary flow is intense enough to carry the dust to the plate electrodes, and on the other hand, it avoids the instability/chaotic flows caused by EHD effect, which may reduce the collection efficiency.

\begin{figure}
	\centering
	\subfigure{
	\begin{minipage}[h]{0.4\linewidth}
		\centering
		\includegraphics[height=3cm]{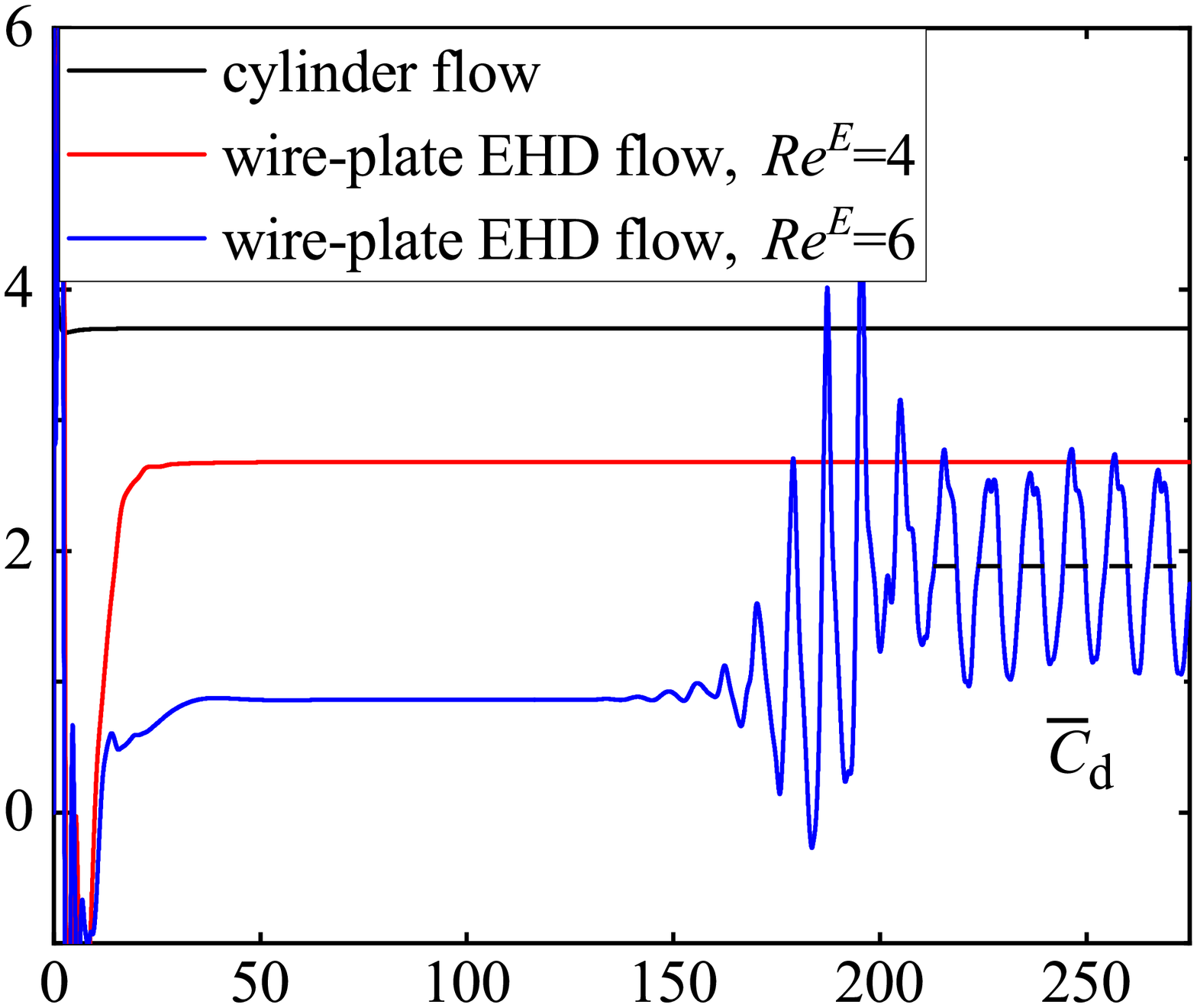}
		\put(-110,75){(a)}
		\put(-50,-5){$ t $}	
		\put(-110,45){$ C_d $}			
		\label{fig.weakcdw}
	\end{minipage}}	
	\hspace{-50pt}
	\subfigure{
	\begin{minipage}[h]{0.4\linewidth}
		\centering
		\includegraphics[height=3cm]{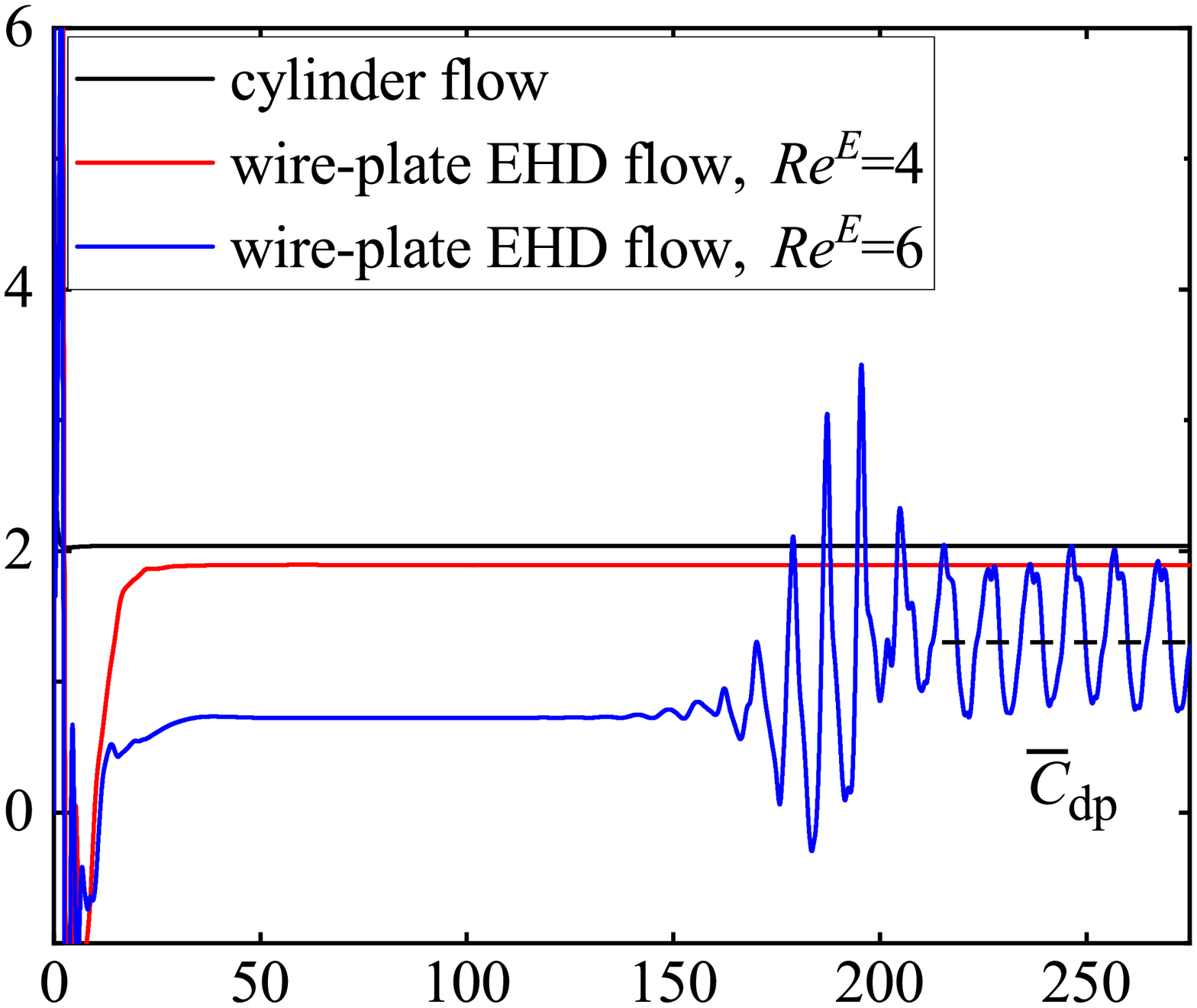}
		\put(-110,75){(b)}
        \put(-50,-5){$ t $}	
        \put(-110,45){$ C_{dp} $}							
		\label{fig.weakcdp}
	\end{minipage}}
	\hspace{-50pt}
	\subfigure{
	\begin{minipage}[h]{0.4\linewidth}
		\centering
		\includegraphics[height=3cm]{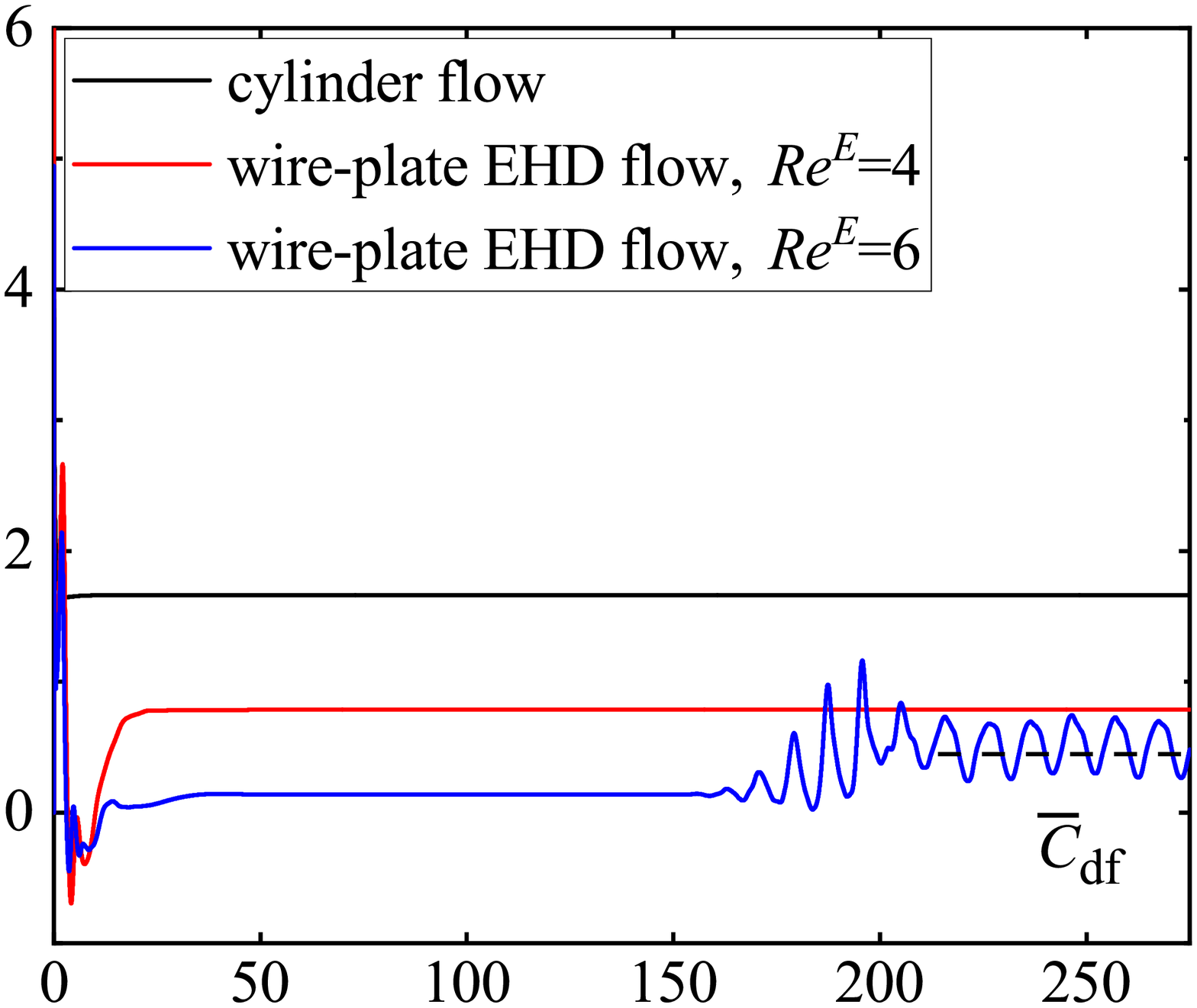}
		\put(-110,75){(c)}
        \put(-50,-5){$ t $}	
        \put(-110,45){$ C_{df} $}					
		\label{fig.weakcdf}
	\end{minipage}}
	\caption{Comparision for time evolution of drag coefficient (a) $ C_d $ and its decomposition of (b) pressure drag coefficient $ C_{dp} $ and (c) friction drag coefficient $ C_{df} $ between cylinder wake flow and wire-plate EHD-Poiseuille flow at $ Re^E=4 $ and $ Re^E=6 $ at $ U_0=0.3 $. The dash lines denote the time-averaged value of $ C_d $, $ C_{dp} $ and $ C_{df} $ in the oscillation state.}
	\label{fig.weakcd}
	\end{figure}

Now we would like to discuss the effect of EHD flow on the drag on the wire when the cross-flow is weak. It is noted that the drag on the bluff body is a net force in the streamwise direction due to the pressure and shear forces. Therefore, the drag coefficient can be divided into two parts, i.e., the distributions of pressure and shear forces, respectively. We define the pressure drag coefficient $ C_{dp} $ and the friction drag coefficient $ C_{df} $ following \cite{achenbach1968distribution}
\begin{equation}
	C_d=\frac{F_d}{RU_{0}^2}=\frac{F_{dp}}{RU_{0}^2}+\frac{F_{df}}{RU_{0}^2}=C_{dp}+C_{df},
\end{equation}
where $ F_{dp} $ is related to the pressure drag on the wire and $ F_{df} $ the shear stress on the wire. Figure \ref{fig.weakcd} shows the $ C_d $, $ C_{dp} $ and $ C_{df} $ at $ U_0=0.3 $ in the case of no EHD effect (cylinder flow), with EHD effect at $ Re^E=4 $ and $ Re^E=6 $, respectively. It can be seen from panel (a) that the presence of the EHD flow reduces the drag on the wire. When $ Re^E $ increases from 4 to 6, although the system starts to oscillate, the average drag coefficient is less than that at $ Re^E=4 $. This indicates that the EHD flow will cause a less pressure drop and a larger flow rate in ESP. In the study of turbulent flow in wire-plate ESP, \cite{soldati1998turbulence} also found that EHD flow reduces drag since the mean velocity increases at a constant pressure drop in the presence of the EHD flow. Additionally, from panels (b) and (c), we can observe that the reduction drag is mainly caused by the decrease of $C_{df} $, indicating that the EHD flow reduces the shear forces on the wire surface. A further note is on the negative dip of the $C_d, C_{dp}, C_{df}$ values when the flow is initiated, which may be related to the specific initial condition we considered in this weak cross-flow case. We observe that there are two plumes inclined at around 45 degrees with respect to the aft direction, which may give rise to the negative dip of the drag coefficients.  The results of this part illustrate that a proper EHD effect in ESP can drift particles to the plate electrodes and also increase the flow rate due to the reduced pressure drop. Animations have been provided for the weak cross-flow case as supplementary material.

\subsubsection{Global stability analysis of the EHD flow}
\begin{figure}
	\flushleft
	\subfigure{
	\begin{minipage}[h]{0.4\linewidth}
			\centering
			\includegraphics[height=4cm]{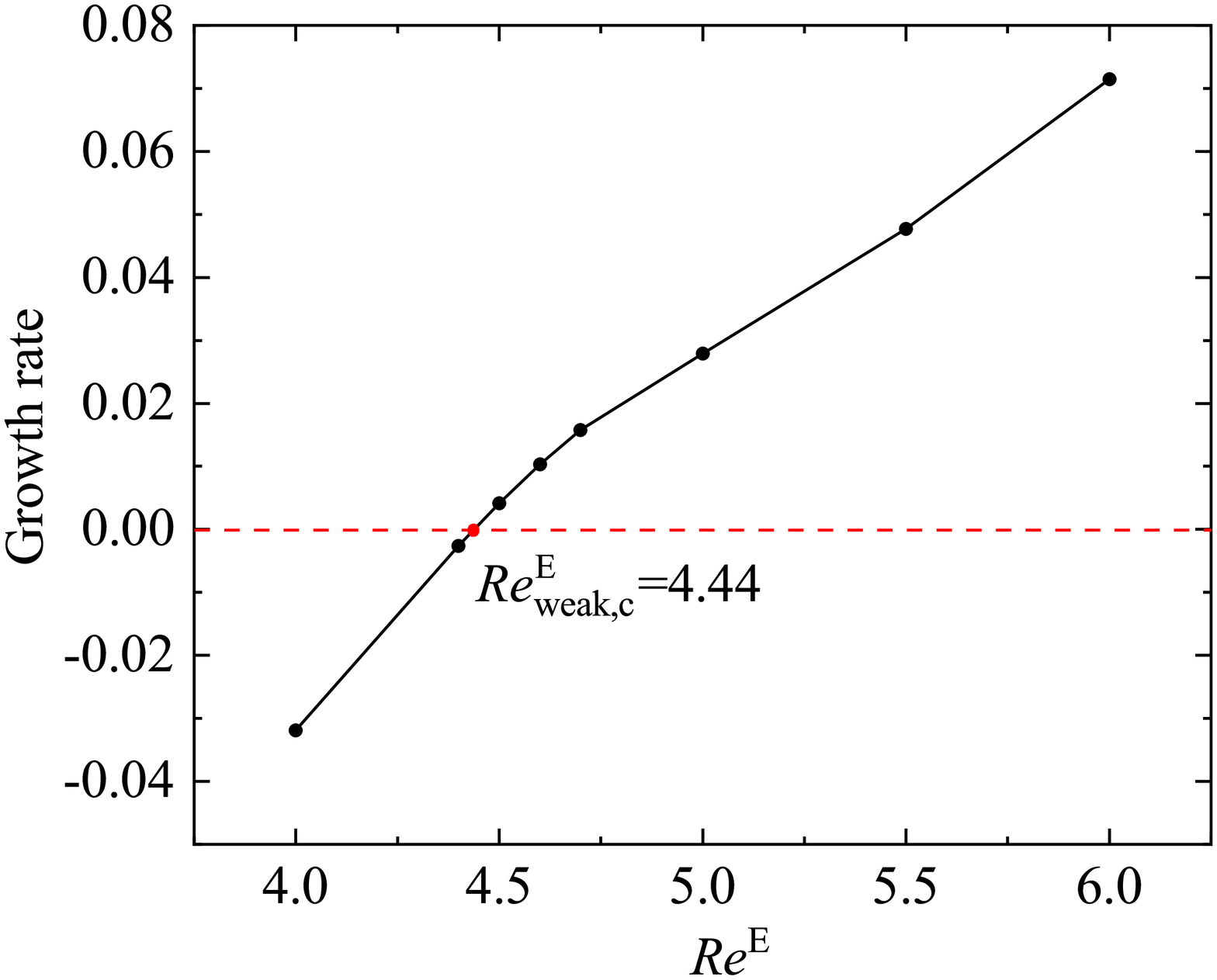}
			\put(-150,105){(a)}
			\label{fig.p2gr2}
	\end{minipage}}
	\hspace{30pt}
	\subfigure{
		\begin{minipage}[h]{0.4\linewidth}
			\centering
			\includegraphics[height=4cm]{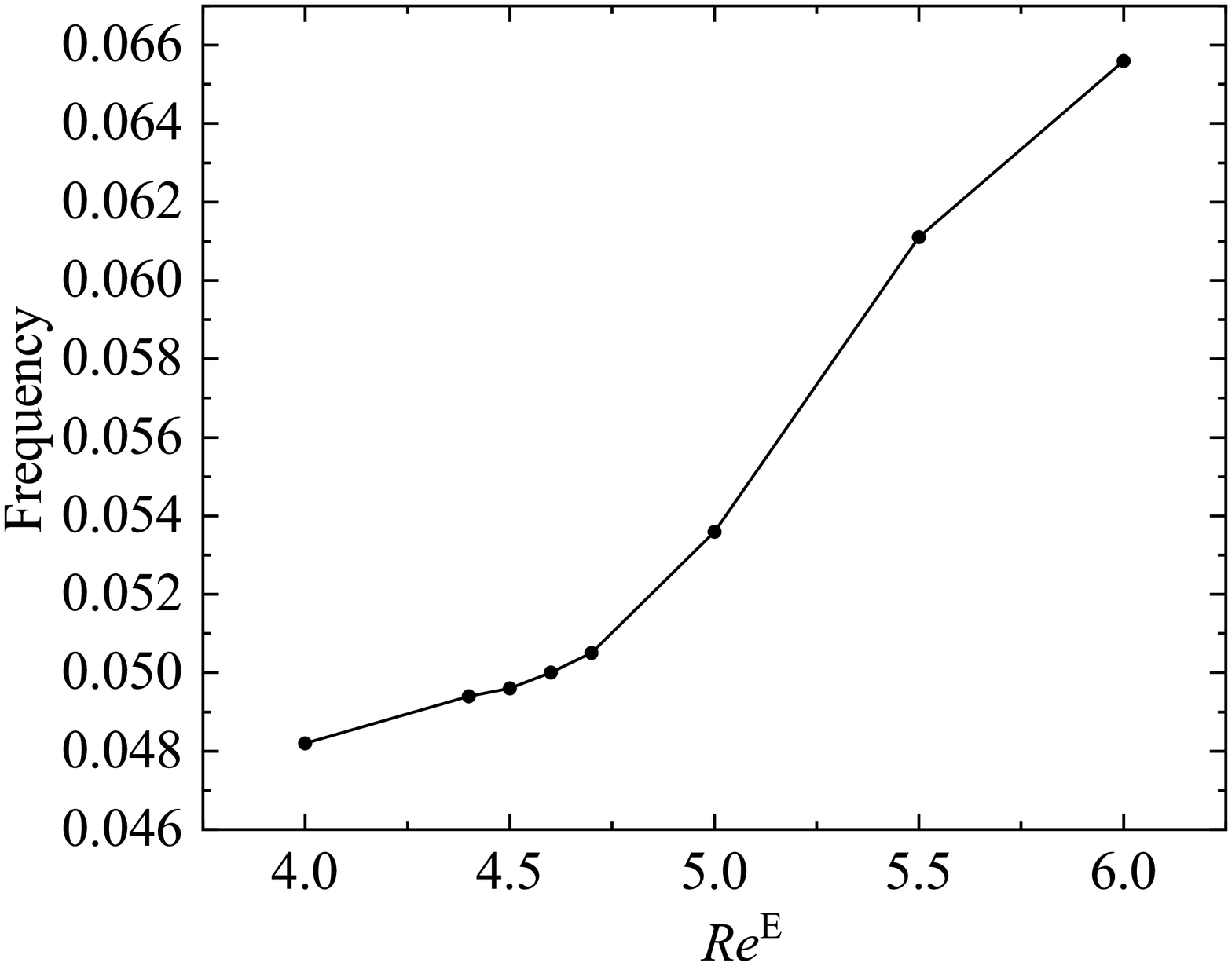}
			\put(-155,105){(b)}
			\label{fig.p2fr2}
	\end{minipage}}
	\subfigure{
	\begin{minipage}[h]{0.4\linewidth}
			\centering
			\includegraphics[height=4cm]{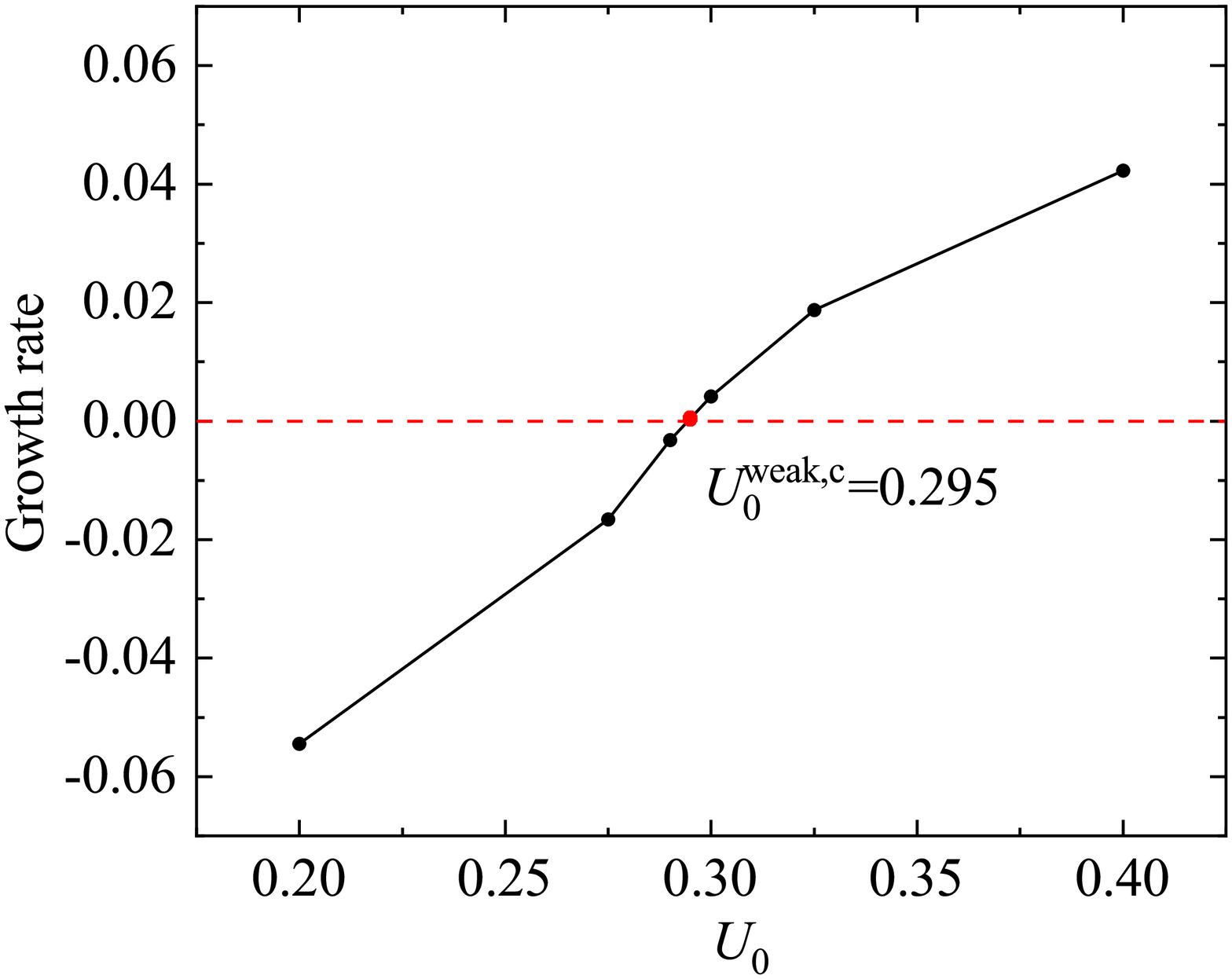}
			\put(-155,105){(c)}
			\label{fig.p2gr}
	\end{minipage}}	
	\hspace{30pt}
	\subfigure{
		\begin{minipage}[h]{0.4\linewidth}
			\centering
			\includegraphics[height=4cm]{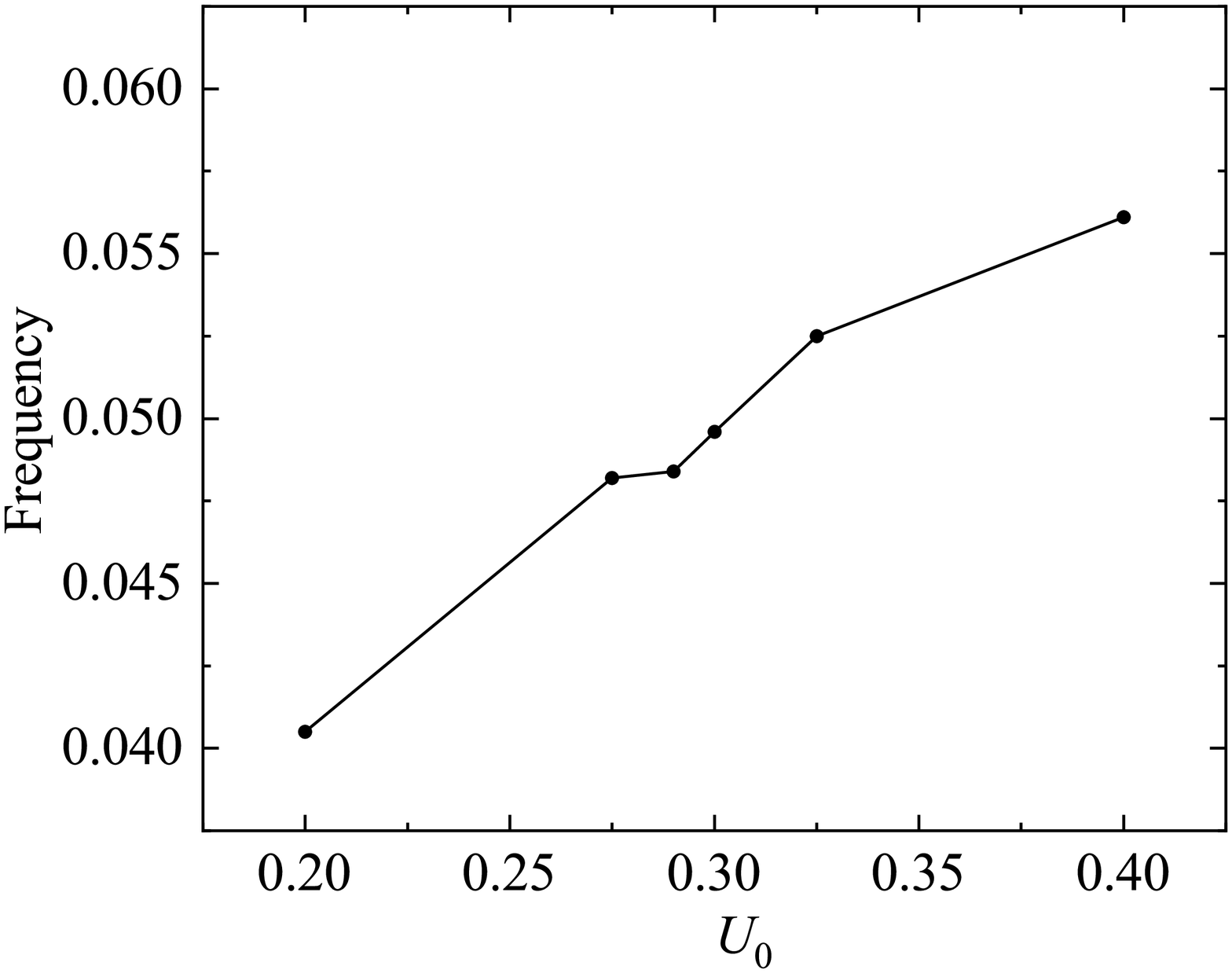}
			\put(-155,105){(d)}
			\label{fig.p2fr}
	\end{minipage}}
	\subfigure{
	\begin{minipage}[h]{0.4\linewidth}
		\centering
		\includegraphics[height=3cm]{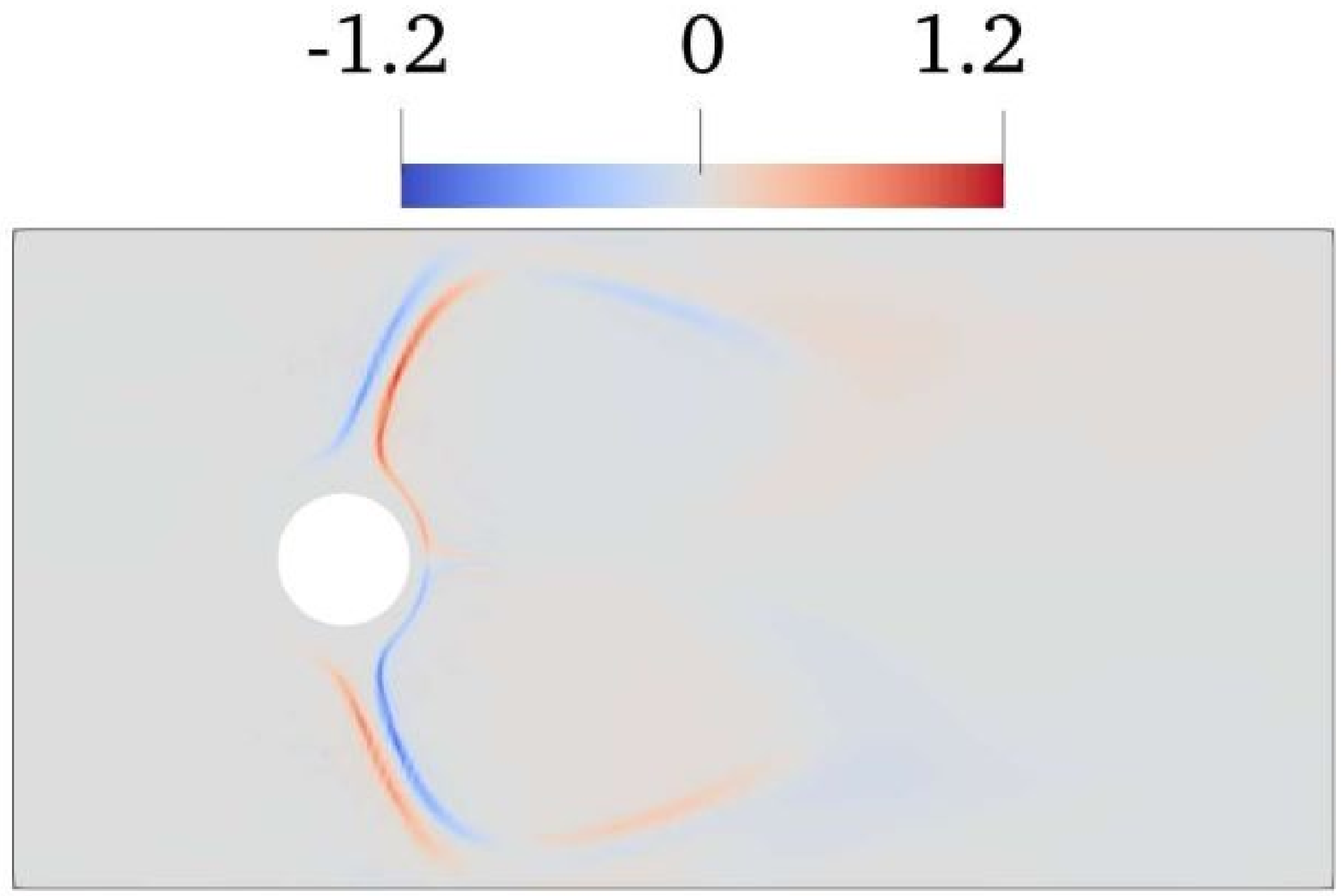}
		\put(-145,75){(e)}
		\label{fig.Aq1U03}
	\end{minipage}}
	\hspace{40pt}
	\subfigure{
	\begin{minipage}[h]{0.4\linewidth}
		\centering
		\includegraphics[height=3cm]{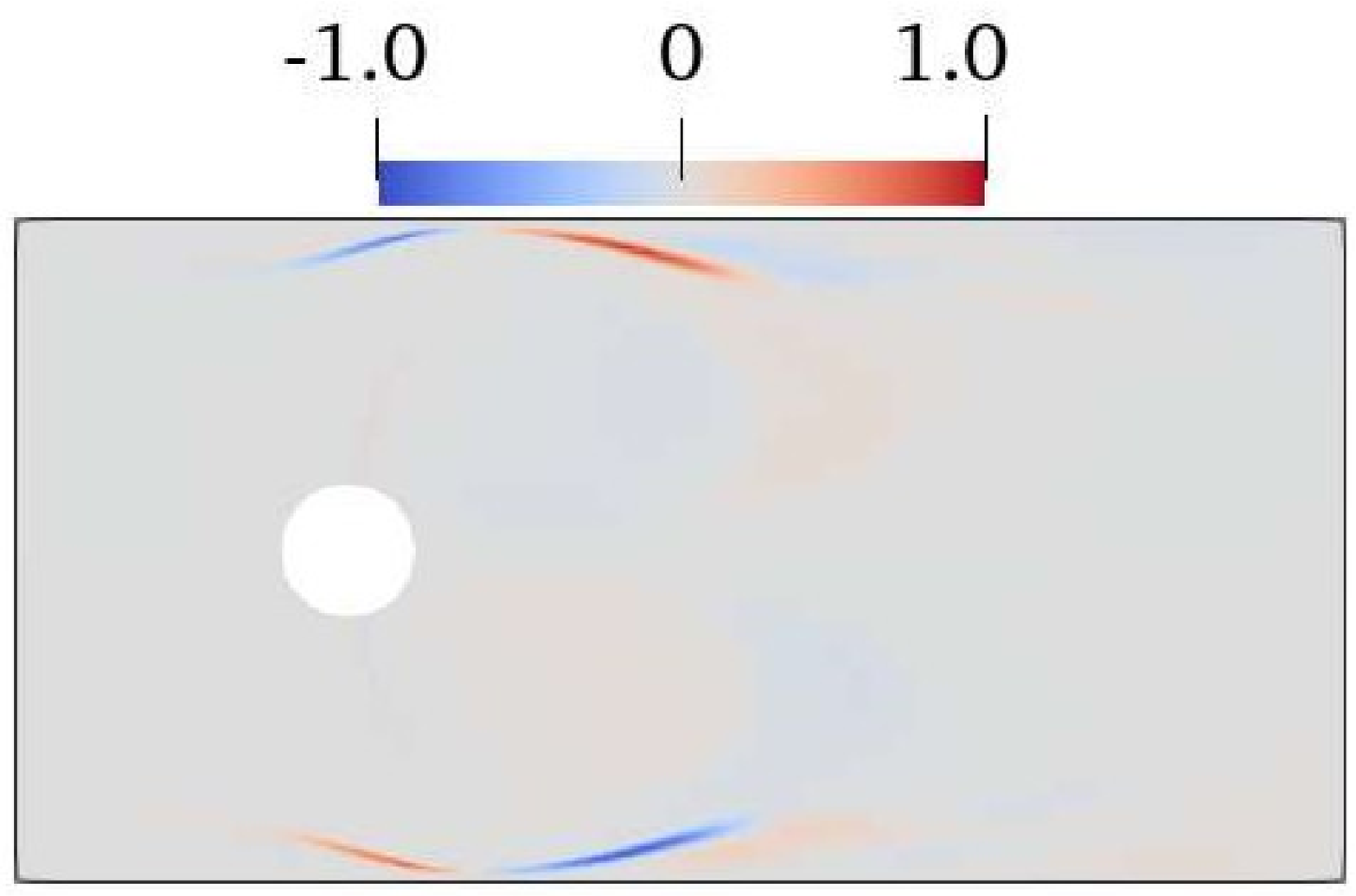}
		\put(-145,75){(f)}
		\label{fig.Aq2U03}
	\end{minipage}}
	\hspace{20pt}
	\subfigure{
	\begin{minipage}[h]{0.4\linewidth}
		\centering
		\includegraphics[height=3cm]{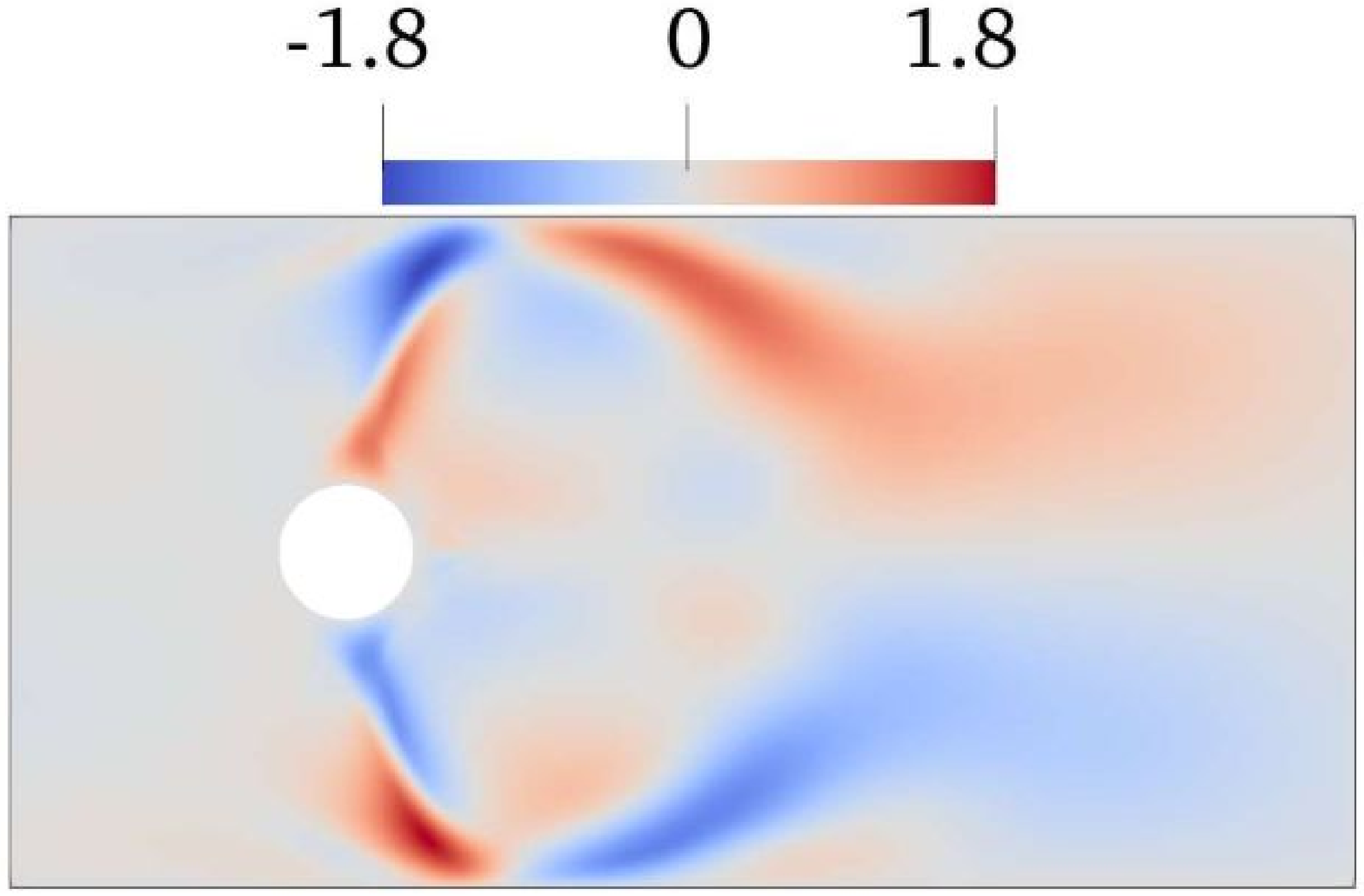}
		\put(-145,75){(g)}
		\label{fig.AvxU03}
	\end{minipage}}
	\hspace{40pt}
	\subfigure{
	\begin{minipage}[h]{0.4\linewidth}
		\centering
		\includegraphics[height=3cm]{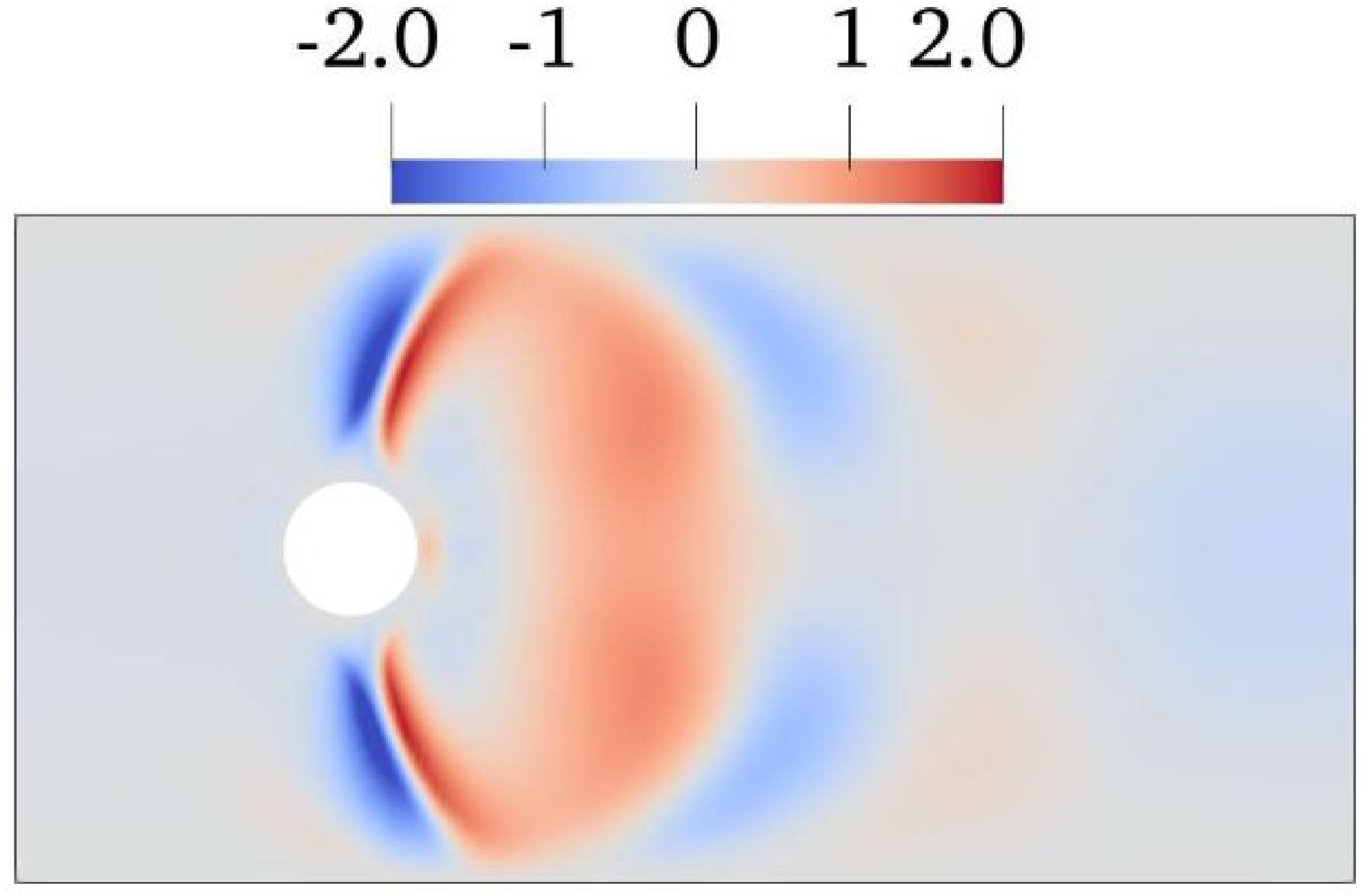}
		\put(-145,75){(h)}
		\label{fig.AvyU03}
	\end{minipage}}
	\caption{Growth rates and frequencies of wire-plate EHD flow with a weak cross-flow at (a)(b) different $ Re^E $ at $ U_0=0.3 $; and (c)(d) different $ U_0 $ at $ Re^E=4.5 $; and the corresponding leading eigenvectors at $ U_0=0.3, Re^E=4.5 $ for (e) positive charge density; (f) negative charge density; (g) $ x $-velocity; (h) $ y $-velocity.}
	\label{fig.eigenFU03}
\end{figure}

Next, we present the results on the linear stability analysis. Again, when the final state is a steady flow, we use that final state as the base flow. When the flow is unsteady, the SFD method is adopted to obtain the base flow. Firstly we examine the stability of the linearised flow at different electric field intensities. The parameters are $ C_0=3, M=37, K_r=1,  C_I=0.2, O_s=8.6, U_0=0.3, \lambda=0.2 $ and $ \alpha=0.001 $. Figures \ref{fig.eigenFU03} (a-b) show the growth rates and frequencies of wire-plate EHD flow with cross-flow for different electric Reynolds number $ Re^E $. We can see that with the increase of $ Re^E $, the growth rate increases, meaning that the flow becomes more unstable. Additionally, the critical electric Reynolds number is examined to be $ Re_{weak,c}^E=4.44 $, indicating that passing this critical value, the flow transitions from steady to time-periodic oscillation. This is consistent with the results in figure \ref{fig.p2Umax}. Moreover, the frequency increases with increasing $ Re^E $. 

We then investigate the effect of cross-flow intensity on the stability of the linear system. The parameters are $ C_0=3, M=37, K_r=1,  C_I=0.2, O_s=8.6, \lambda=0.2 $ and $ \alpha=0.001 $. Figures \ref{fig.eigenFU03} (c-d) display the growth rates and frequencies at different $ U_0 $ and $ Re^E=4.5 $ in the wire-plate EHD-Poiseuille flow. It can be observed that the growth rate of the perturbations increases with increasing $ U_0 $, and the critical $ U_0 $ is found to be 0.295. It indicates that the increase of cross-flow intensity leads to a more unstable flow. It may imply that the speed of the carrying flow in the ESP should not be too large; otherwise, it will lead to flow instability, rendering the flow oscillatory and reducing the working efficiency of ESP. This to some extent seems to be consistent with the conclusion in \cite{leonard1983experimental}. We can observe from the panel (d) that the frequency also increases with the increase of $ U_0 $ which may be due to the enhancement of inertial flow increasing the phase speed. The global eigenvectors of the most unstable mode for the wire-plate EHD-Poiseuille flow with a weak flow at $ U_0=0.3$ and $Re^E=4.5 $ are shown in figures \ref{fig.eigenFU03} (e-h). We can see that they are similar to those in the injection regime of wire-plate EHD flow without a cross-flow (figure \ref{fig.eigenF}), but tilted downstream, indicating that the EHD flow plays a more important role when the cross-flow is weak.

\subsection{Wire-plate EHD flow with a strong cross-flow}

When the cross-flow is strong, the Poiseuille flow plays a leading role in the system. Via this flow configuration, we can examine the effect of the electric field on the wake flow. 
In this section, we will mainly vary the cross-flow velocity $ U_0 $ and the remaining parameters are fixed as $ C_0=3, M=37, K_r=1,  C_I=0.2, O_s=8.6, \lambda=0.2, Re^E=2.4 $ and $ \alpha=0.001 $.  

\subsubsection{Oscillatory wake flow}
\begin{figure}
	\centering
	\begin{minipage}[h]{0.4\linewidth}
		\centering
		\includegraphics[height=2cm]{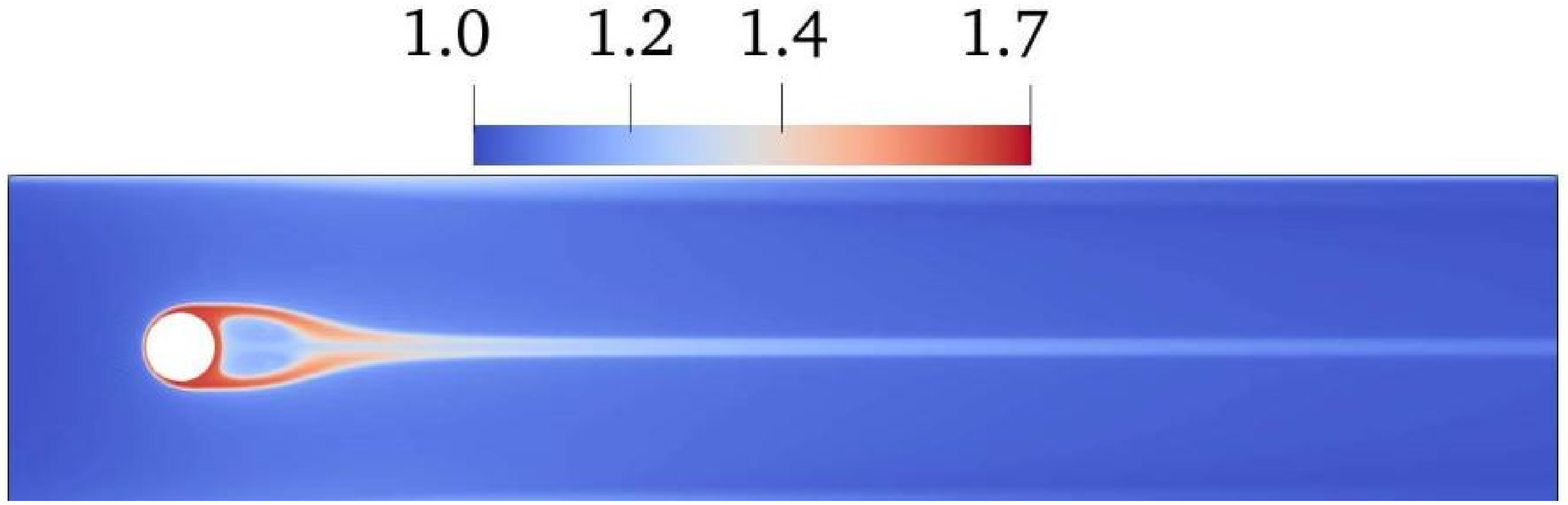}
		\put(-180,45){(a)}
		\label{fig.U6q1}
	\end{minipage}
	\hspace{20pt}
	\begin{minipage}[h]{0.4\linewidth}
		\includegraphics[height=2cm]{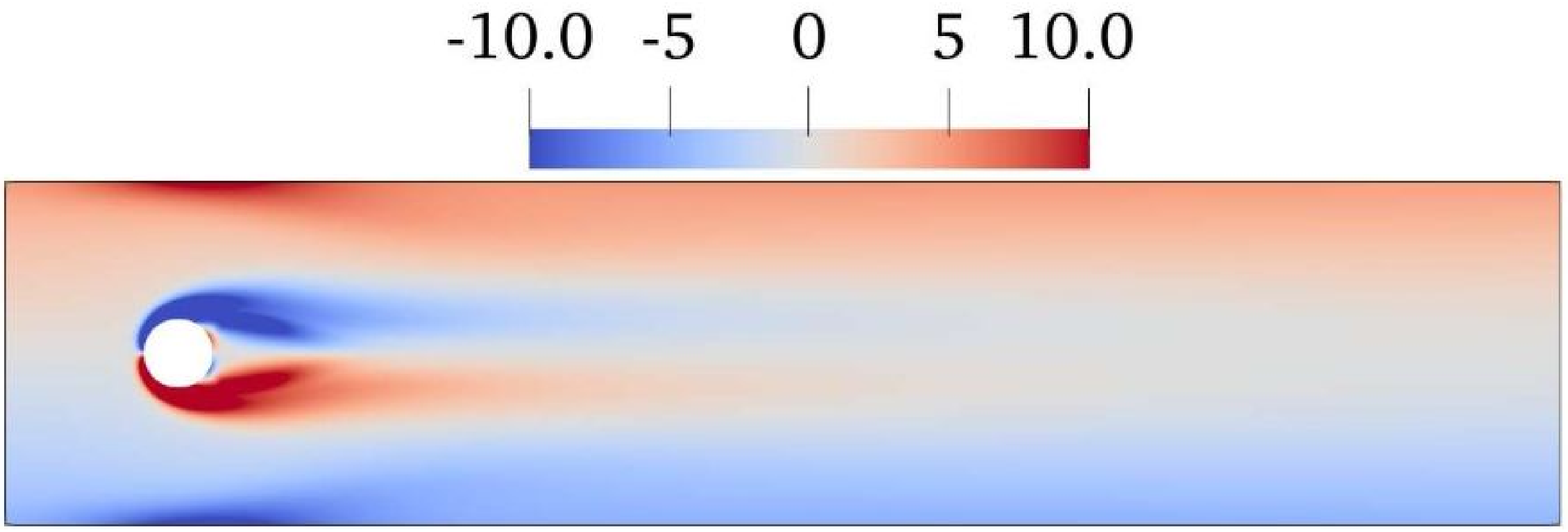}
		\put(-180,45){(b)}
		\label{fig.U6vor}
	\end{minipage}
	\caption{The final steady state of nonlinear simulation for wire-plate EHD-Poiseuille flow at $ U_0=7 $. (a)The concentration of positive species; (b) vorticity.  }
	\label{fig.base3}
\end{figure}

\begin{figure}
	\centering
	\subfigure{
	\begin{minipage}[h]{0.4\linewidth}
		\centering
		\includegraphics[height=4cm]{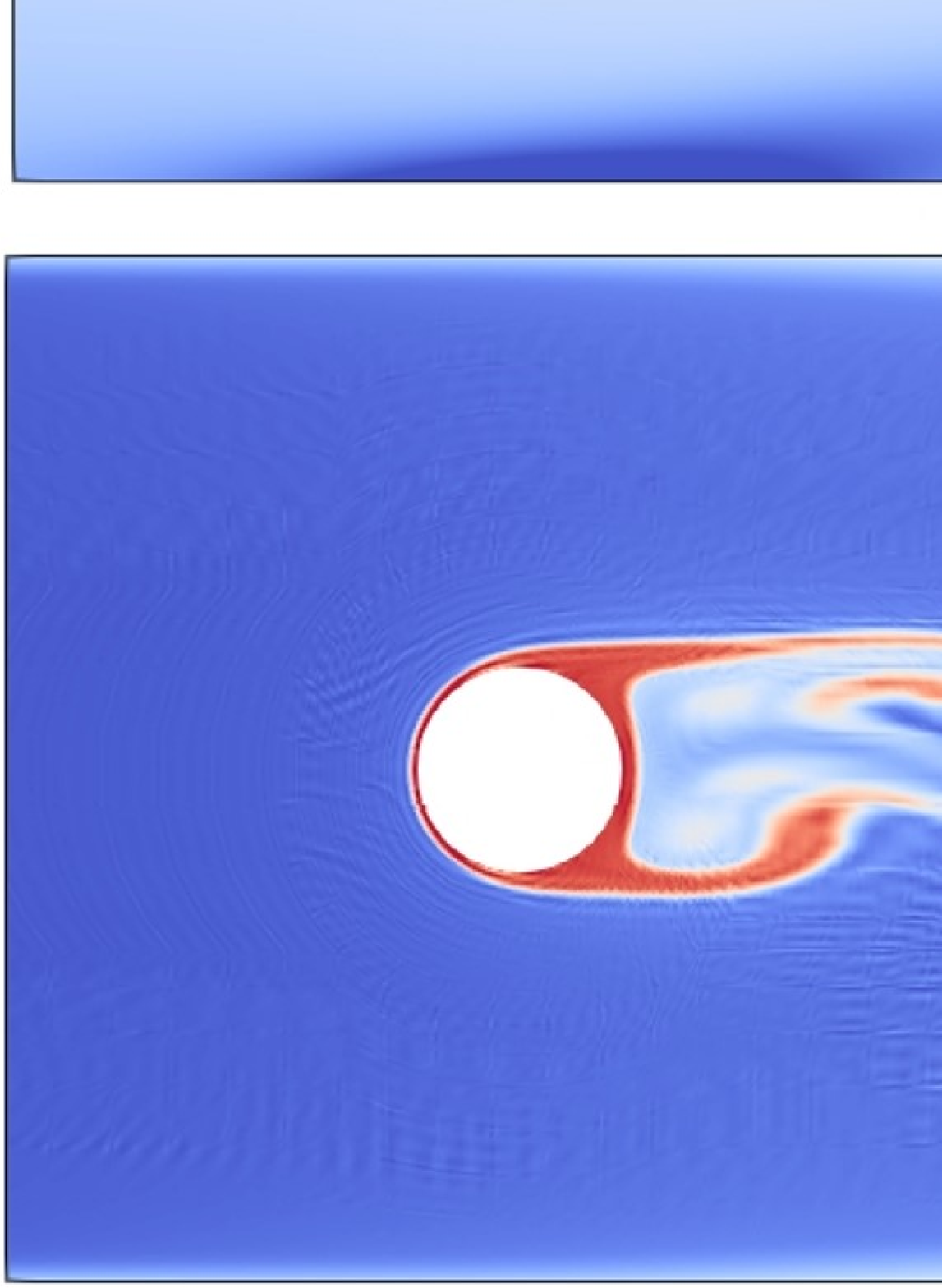}
		\put(-175,110){(a)}
		\label{fig.U12non}
	\end{minipage}}
	\hspace{20pt}
	\subfigure{
	\begin{minipage}[h]{0.4\linewidth}
		\centering
		\includegraphics[height=4cm]{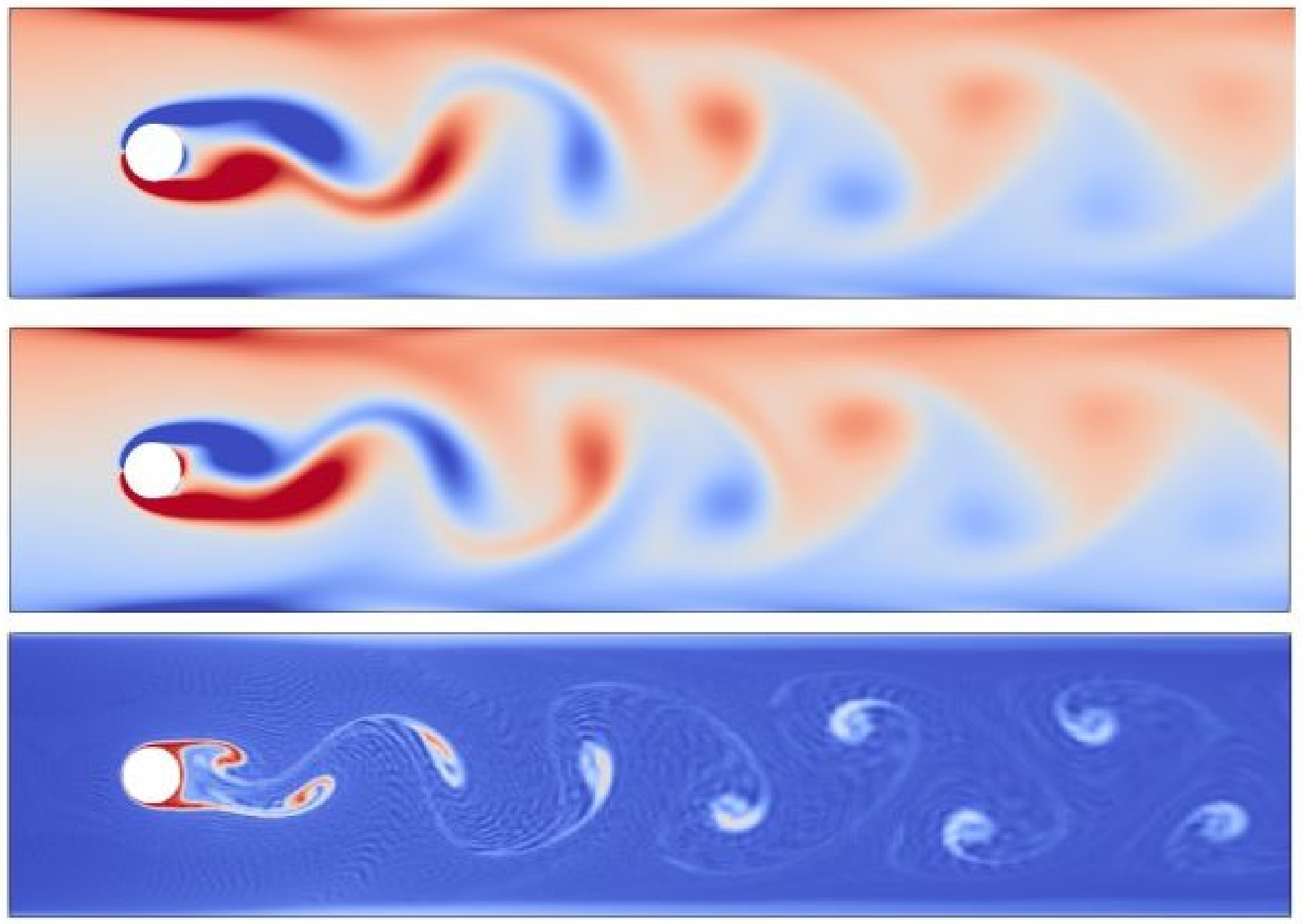}
		\put(-175,110){(b)}
		\label{fig.U15non}
	\end{minipage}}
	\hspace{120pt}
	\subfigure{
	\begin{minipage}[h]{0.4\linewidth}
		\centering
		\includegraphics[height=4cm]{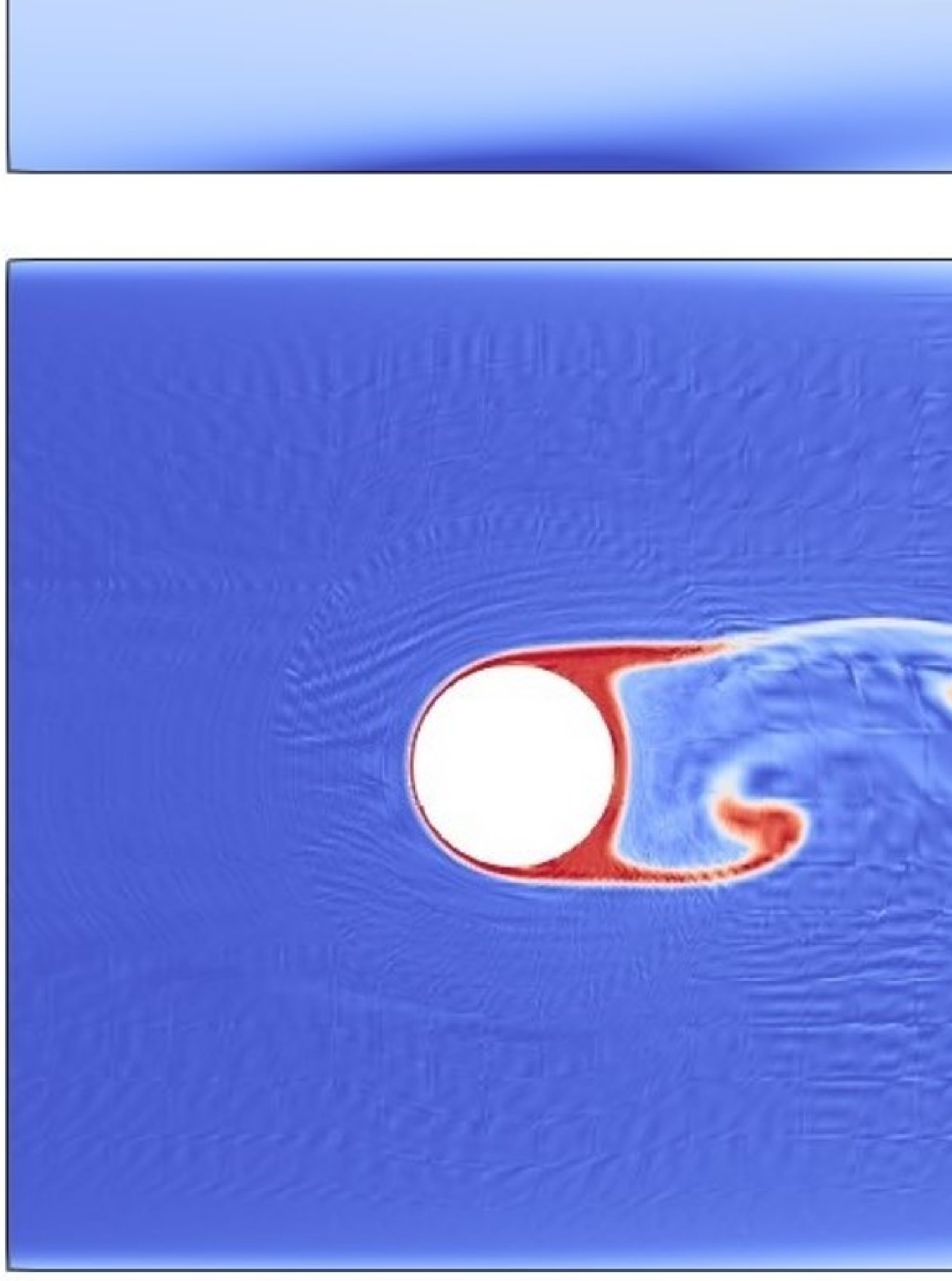}
		\put(-175,110){(c)}
		\label{fig.U18non}
	\end{minipage}}
	\hspace{20pt}
	\subfigure{
	\begin{minipage}[h]{0.4\linewidth}
		\centering
		\includegraphics[height=4.05cm]{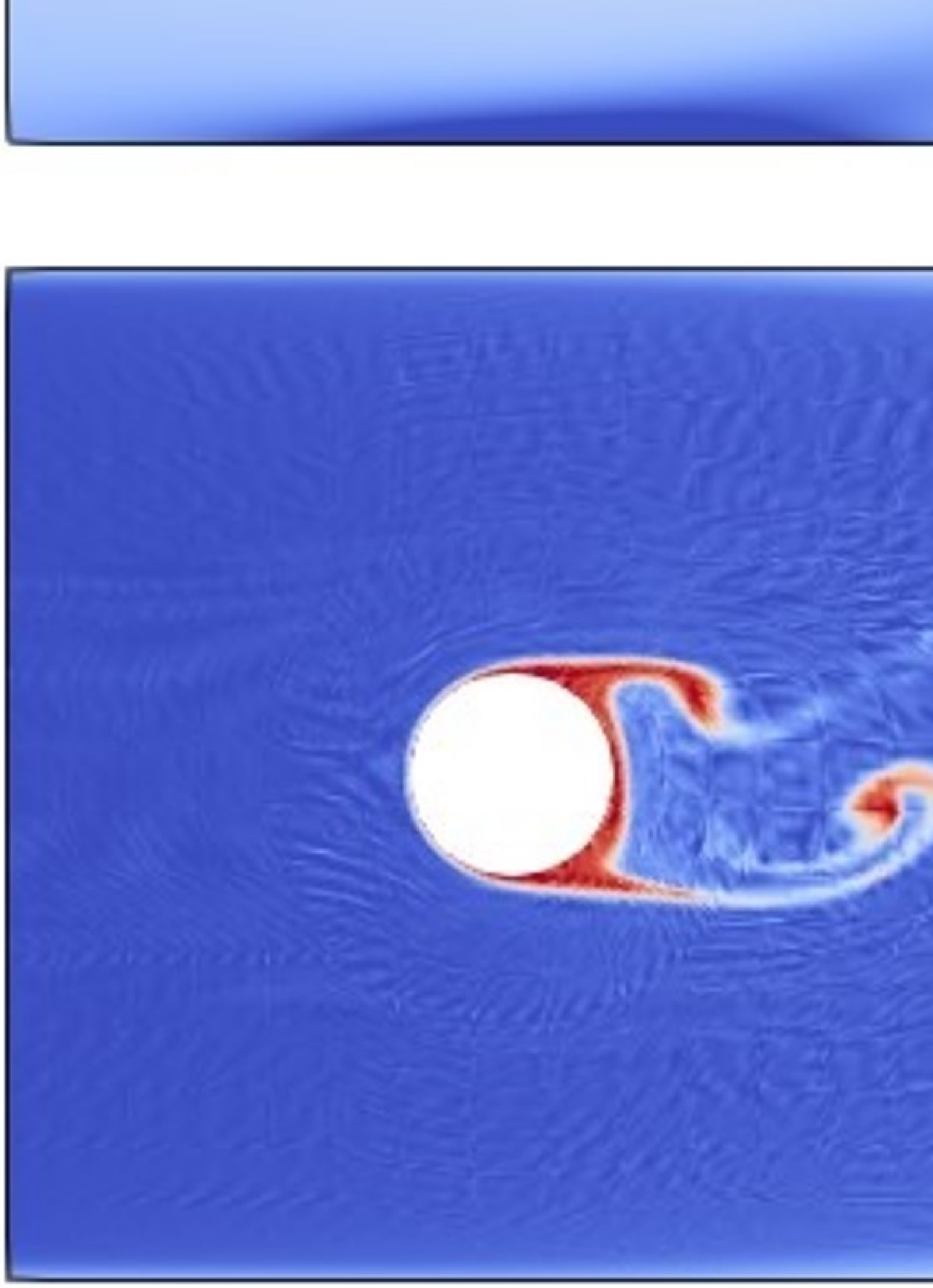}
		\put(-173,110){(d)}
		\label{fig.U21non}
	\end{minipage}}
	\caption{Nonlinear simulation for cylinder wake flow and wire-plate EHD-Poiseuille flow at (a) $ U_0=12 $ at $ t=75 $; (b) $ U_0=15 $ at $ t=40 $; (c)  $ U_0=18 $ at $ t=30 $; (d)  $ U_0=21 $ at $ t=20 $. In each panel, from top to bottom, they are vorticity of Newtonian cylinder wake flow; vorticity of wire-plate EHD-Poiseuille flow; positive charge density of wire-plate EHD-Poiseuille flow.}
	\label{fig.p3non}
\end{figure}

As shown in figure \ref{fig.base3}, at a steady state, the charge forms a long line behind the wire and flows to the outlet (with $ U_0=7 $). Additionally, the recirculation zone can be clearly distinguished from the distribution of the positive species behind the cylinder. When the cross-flow is stronger, vortex shedding will occur. Figure \ref{fig.p3non} illustrates the vortex shedding in the wake at different $ U_0 $ of the Newtonian cylinder wake and wire-plate EHD-Poiseuille flow, respectively. We find that the velocity fields are similar in the cases with and without the EHD effect. In addition, from the distribution of positive species (bottom figure in each panel), we observe that, interestingly, the charges in the wire-plate EHD-Poiseuille flow also undergo oscillatory motions and accumulate at the centers of the vortices. 

Similar to the cylinder wake flow, this unstable motion is caused by the Kelvin-Helmholtz instability due to the high shear layer that exists near the cylinder surface. This instability leads to a rolling up and a separation of fluid from the surface of the cylinder causing the formation of vortex shedding \citep{nair1996onset,nishioka1978mechanism}. To understand the phenomenon that the ions are trapped in the centers of the vortices and convected with them, one may resort to the similar dynamics of the (passive) particles in the sheared flow. It has been studied in the work of particle dispersion in a jet that the influence of large-scale structures on the particle motion mainly depends on the ratio of the particle aerodynamic response time to the flow characteristic time \citep{chung1988simulation}. If the ratio is much smaller than 1, the particles can quickly respond to the large-scale vortex, and will closely follow the flow streamlines.  The dynamics of the ions in our results may be understood similarly. In the case with a strong cross-flow, the effect of Coulomb force ($ N_\pm\mathbf{E} $) on the movement of ions is weak compared to the convection ($ N_\pm\mathbf{U} $). The (massless) ions are thus weakly influenced by the electric field, but strongly affected by the fluid motion. They act like the passive scalar in a strongly convected flow. Thus, the ions are entrained by the flow and move into the vortex centers.


\subsubsection{EHD effect on the wake flow: lift and drag coefficients}
\begin{figure}
	\flushleft
	\begin{minipage}[h]{0.4\linewidth}
		\centering
		\includegraphics[width=14cm]{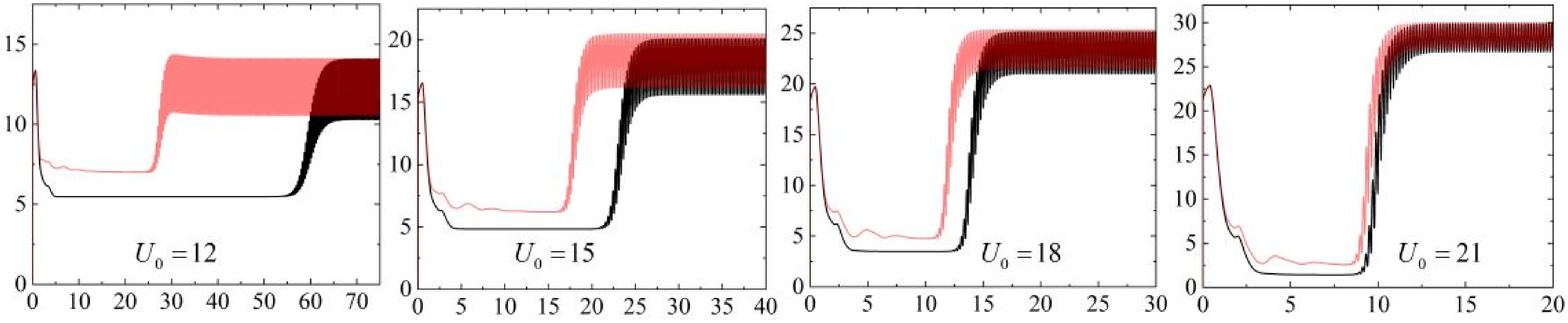}
		\put(-410,70){(a)}
		\put(-410,35){$ U_x $}
		\label{fig.p3vx}
	\end{minipage}
	\hspace{80pt}
	\begin{minipage}[h]{0.4\linewidth}
		\centering
		\includegraphics[width=14cm]{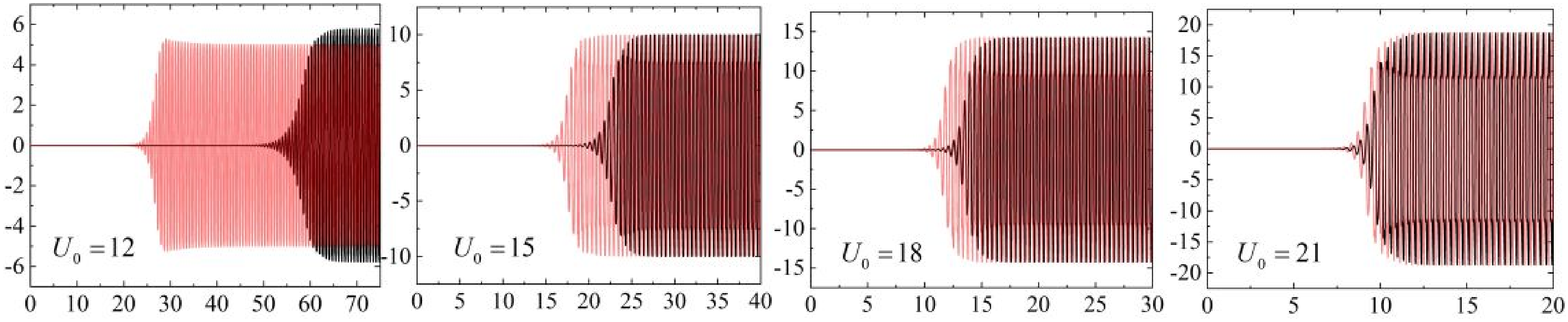}
		\put(-410,70){(b)}
		\put(-410,35){$ U_y $}		
		\label{fig.p3vy}
	\end{minipage}
	\hspace{80pt}
	\begin{minipage}[h]{0.4\linewidth}
		\centering
		\includegraphics[width=14cm]{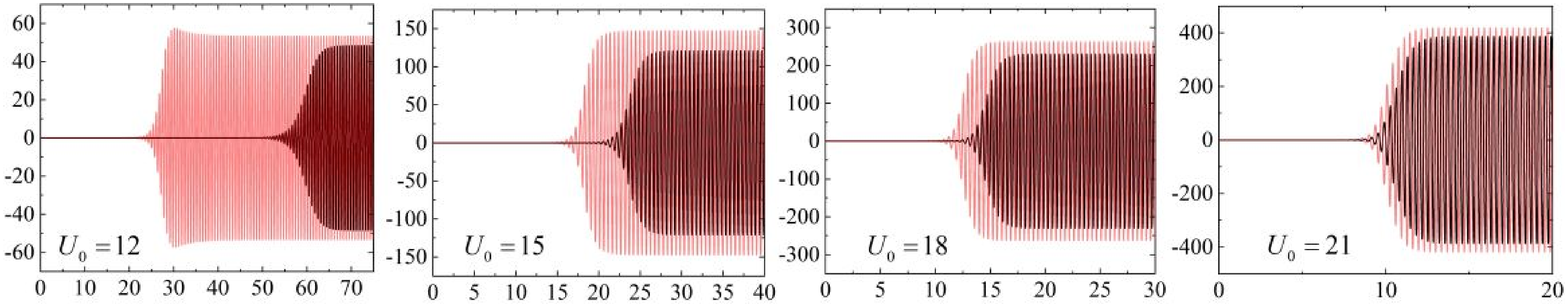}
		\put(-410,70){(c)}
		\put(-410,35){$ C_l $}
		\label{fig.p3cd}
	\end{minipage}
	\hspace{80pt}
	\begin{minipage}[h]{0.4\linewidth}
		\centering
		\includegraphics[width=14cm]{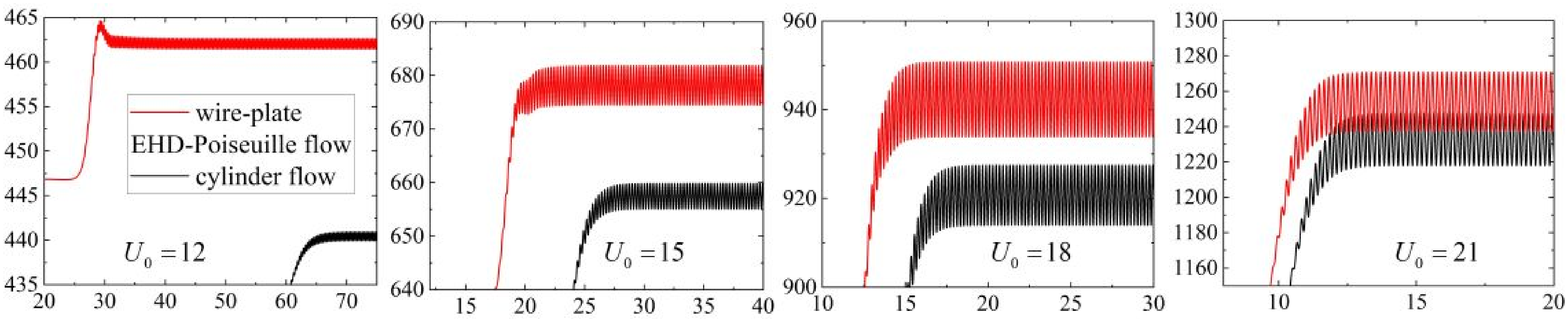}
		\put(-410,70){(d)}
		\put(-410,35){$ C_d $}
		\put(-45,-8){$ t $}	
		\put(-147,-8){$ t $}
		\put(-247,-8){$ t $}	
		\put(-347,-8){$ t $}				
		\label{fig.p3cl}
	\end{minipage}
	\caption{Comparision for time evolution of nonlinear results between cylinder wake flow (black lines) and wire-plate EHD-Poiseuille flow (red lines) at different $ U_0 $. (a) $ U_x $ at the sampling point (10,0); (b) $ U_y $ at the sampling point (10,0) ; (c) $ C_l $ for the wire; (d) $ C_d $ for the wire.}
	\label{fig.p3non2}
\end{figure}

Figure \ref{fig.p3non2} compares the time evolution of $ U_x $, $ U_y $ at a sampling point (10,0) (shown in figure \ref{fig.U12non}) and $ C_d $ and $ C_l $ for the wire with and without the EHD effect  with $ U_0=12, 15,18, 21 $ (the corresponding hydrodynamic Reynolds number $ Re^W=28.8,36,43.2,50.4 $, respectively). First, in all panels, one can observe that the EHD effect has brought forward the start of the vortex shedding. The transient phase is due to the fact that we use a zero velocity field as the initial condition. Numerical simulations of cylinder wake subjected to a DC electric field have been  performed by \cite{barz2018electrokinetic}. The electrodes are placed around the cylinder at different angles, and there is a small electrokinetic velocity at the cylinder surface. Their results showed that a vertical equivalent electrostatic force component reduces the time required for a stable vortex to disengage, which is similar to our results in wire-plate EHD-Poiseuille flow. Panel (a) shows that in the vortex shedding phase, the presence of charges slightly increases the oscillatory streamwise velocities. In panel (b), we find that the EHD effect decreases the oscillation amplitude of the $y$-velocity at $ U_0=12 $, but it has little effect on the $y$-velocity at larger $ U_0 $. Panel (c) shows the EHD effect on the lift coefficient $C_l$, where the amplitude of oscillating $C_l$ in the Newtonian cylinder wake is slightly smaller than that in the EHD case. This means that the EHD effect can increase the maximum lift coefficient in the wake flow. Moreover, the values of the drag coefficient $ C_d $ in panel (d) increase significantly due to the imposed EHD effect. As the cross-flow intensity increases, the boosting effect become weaker (from left to right).

\begin{figure}
	\centering
	\begin{minipage}[h]{0.4\linewidth}
		\centering
		\includegraphics[height=4.2cm]{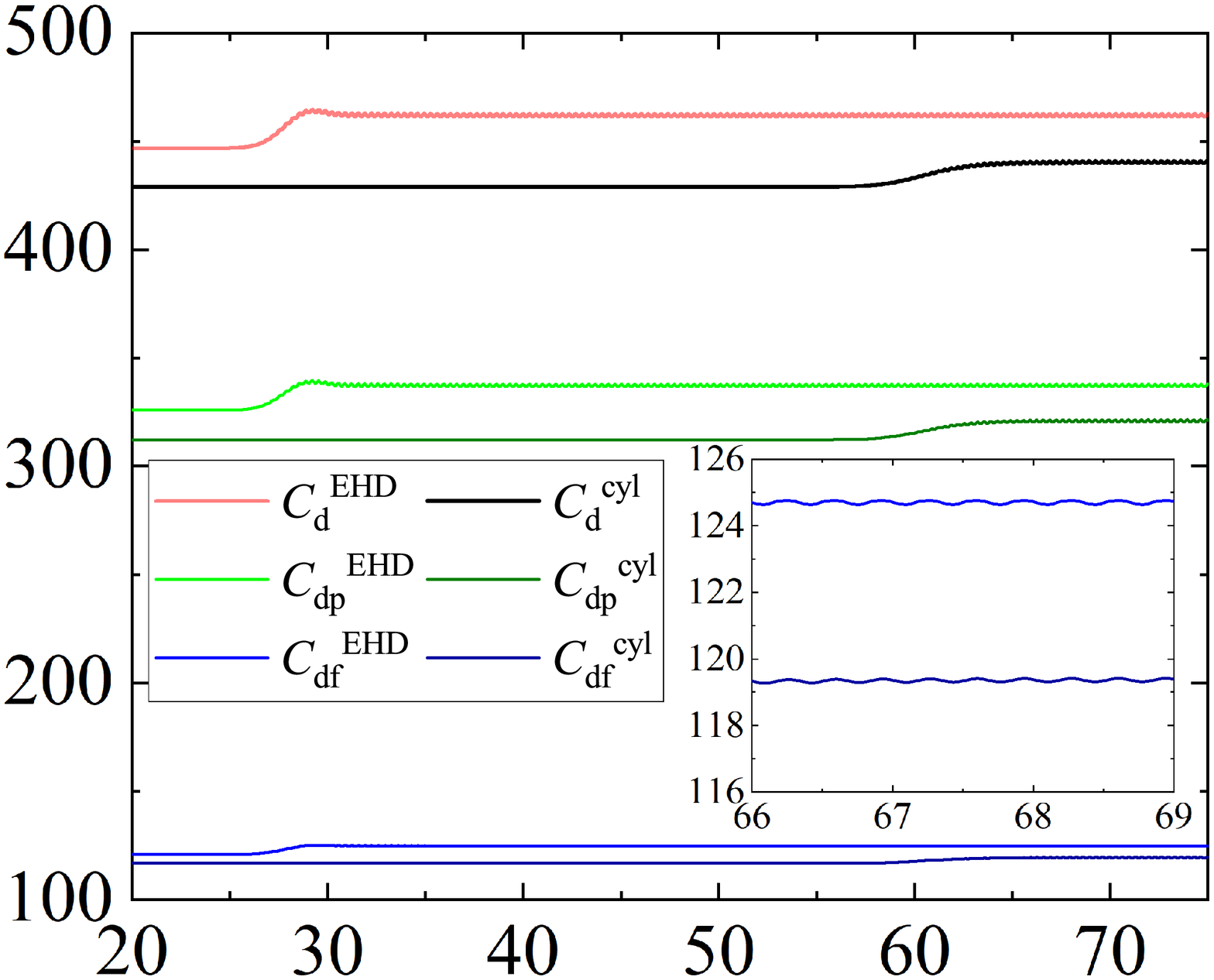}
		\put(-165,115){(a)}
		\put(-75,-5){$ t $}		
		\label{fig.U12cd}
	\end{minipage}	
	\hspace{20pt}
	\begin{minipage}[h]{0.4\linewidth}
		\centering
		\includegraphics[height=4.2cm]{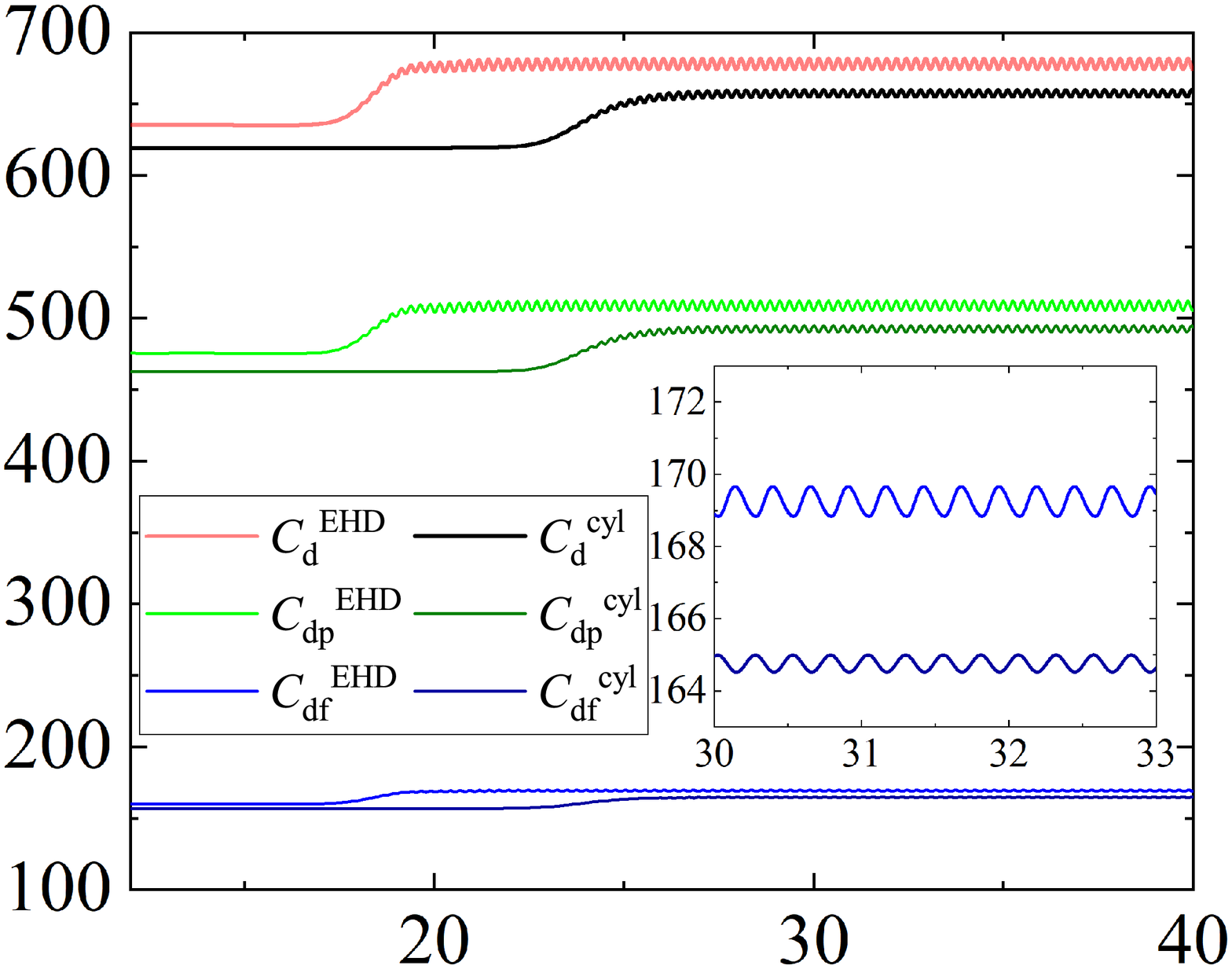}
		\put(-165,110){(b)}
		\put(-75,-5){$ t $}			
		\label{fig.U15cd}
	\end{minipage}
	\hspace{20pt}
    \begin{minipage}[h]{0.4\linewidth}
	\centering
	\includegraphics[height=4.2cm]{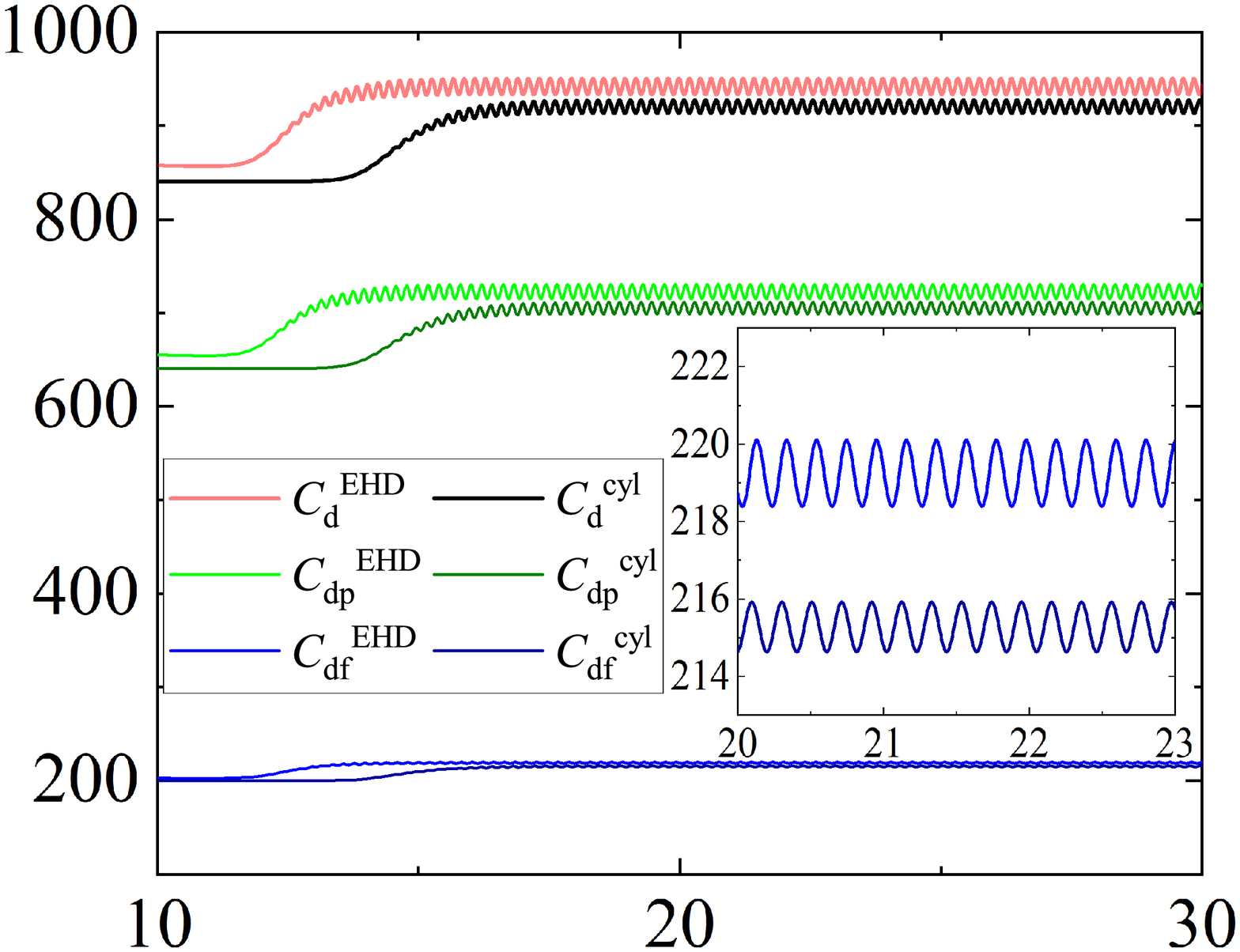}
	\put(-170,115){(c)}
	\put(-75,-5){$ t $}		
	\label{fig.U18cd}
    \end{minipage}
	\hspace{20pt}
    \begin{minipage}[h]{0.4\linewidth}
	\centering
	\includegraphics[height=4.2cm]{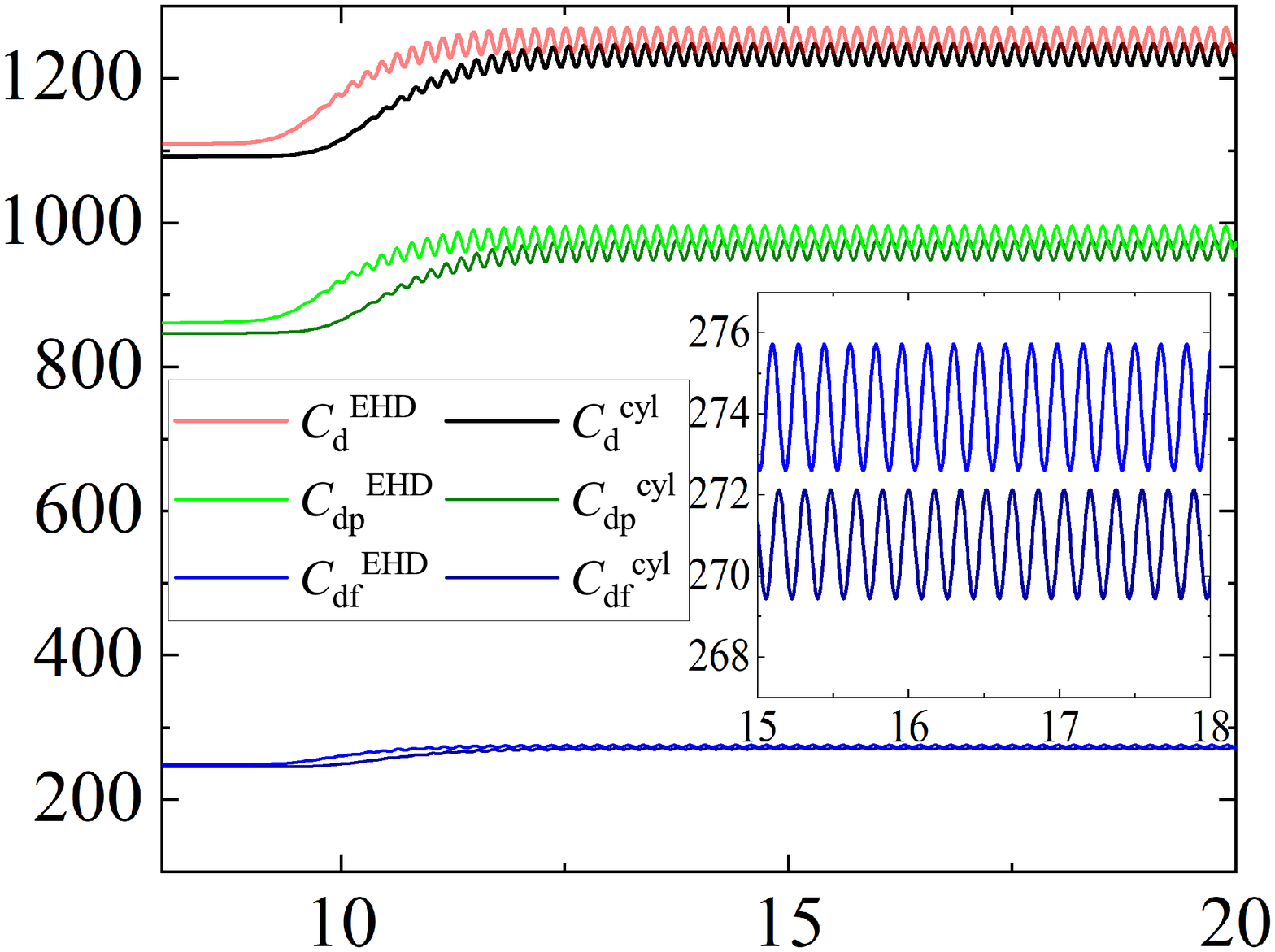}
	\put(-175,115){(d)}
	\put(-75,-5){$ t $}		
	\label{fig.U21cd}
    \end{minipage}
	\caption{Comparision for time evolution of drag coefficient $ C_d $ and its decomposition of pressure drag coefficient $ C_{dp} $ and friction drag coefficient $ C_{df} $ between cylinder wake flow (with the superscript 'cyl') and wire-plate EHD-Poiseuille flow (with the superscript 'EHD') at $ Re^E=2.4 $ at (a) $ U_0=12 $, (b) $ U_0=15 $, (c) $ U_0=18 $, (d) $ U_0=21 $.}
	\label{fig.cd}
\end{figure}

\begin{figure}
	\centering
	\includegraphics[width = 0.4\textwidth]{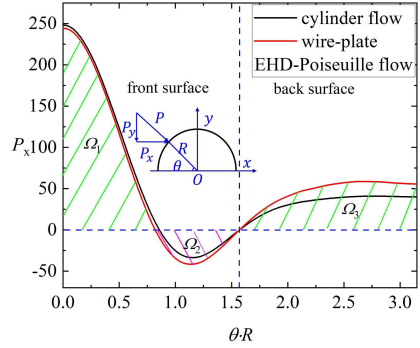}
	\caption{The streamwise pressure distribution with respect to the arc length acting on the upper half wire for cylinder wake flow and wire-plate EHD-Poiseuille flow at $ Re^E=2.4 $ and $ U_0=12 $. $ \Omega_1-\Omega_3 $ denote the area of the shaded parts.}\label{fig.wireP}
\end{figure}

\begin{figure}
	\centering
	\begin{minipage}[h]{0.4\linewidth}
		\centering
		\includegraphics[height=4cm]{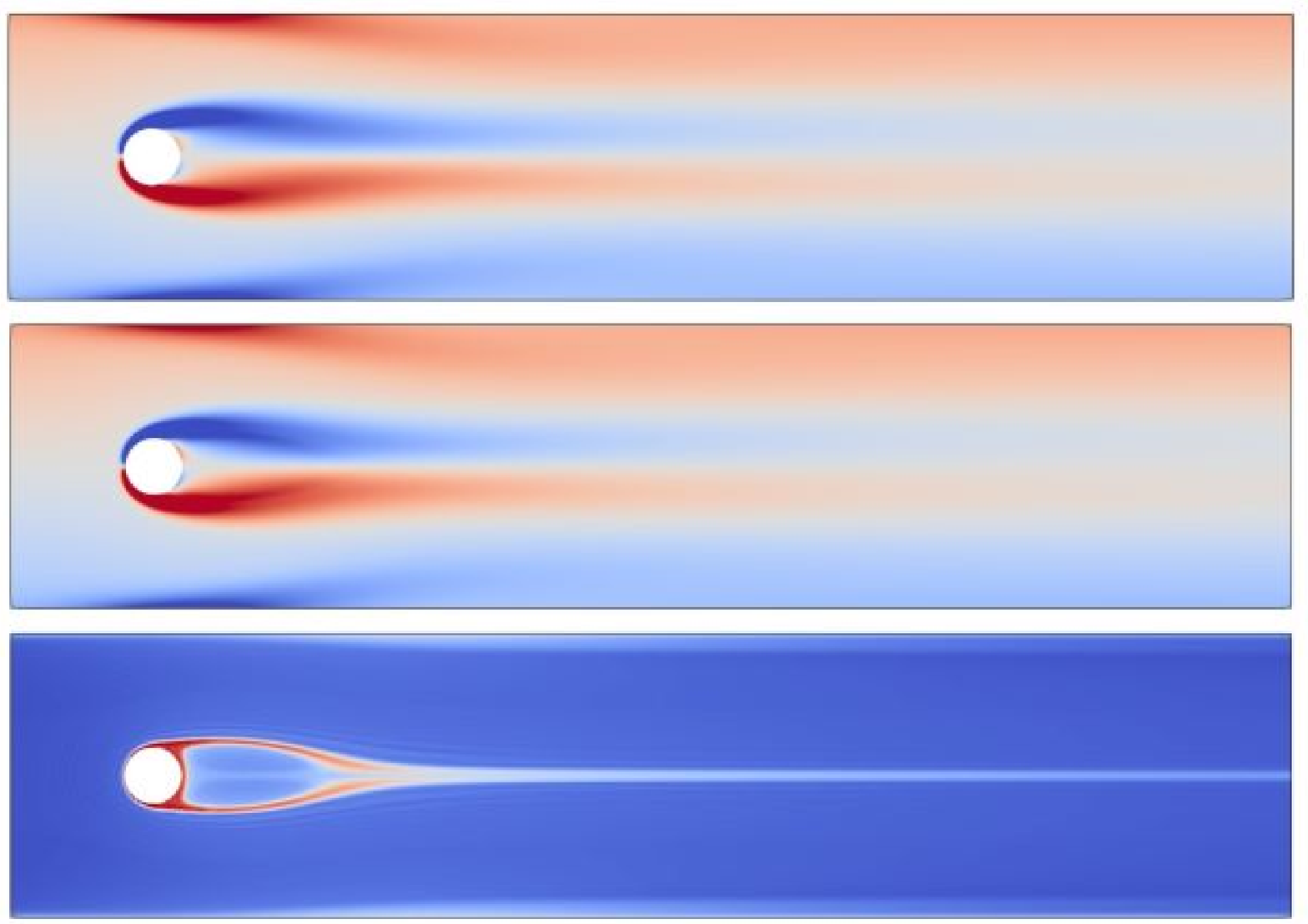}
		\put(-175,110){(a)}    
		\put(-175,65){(c)} 
		\put(-175,30){(e)} 		   
        \put(-110,120){SFD base flow}		
		\label{fig.U12sfd}
	\end{minipage}
	\hspace{20pt}
	\begin{minipage}[h]{0.4\linewidth}
		\centering
		\includegraphics[height=4cm]{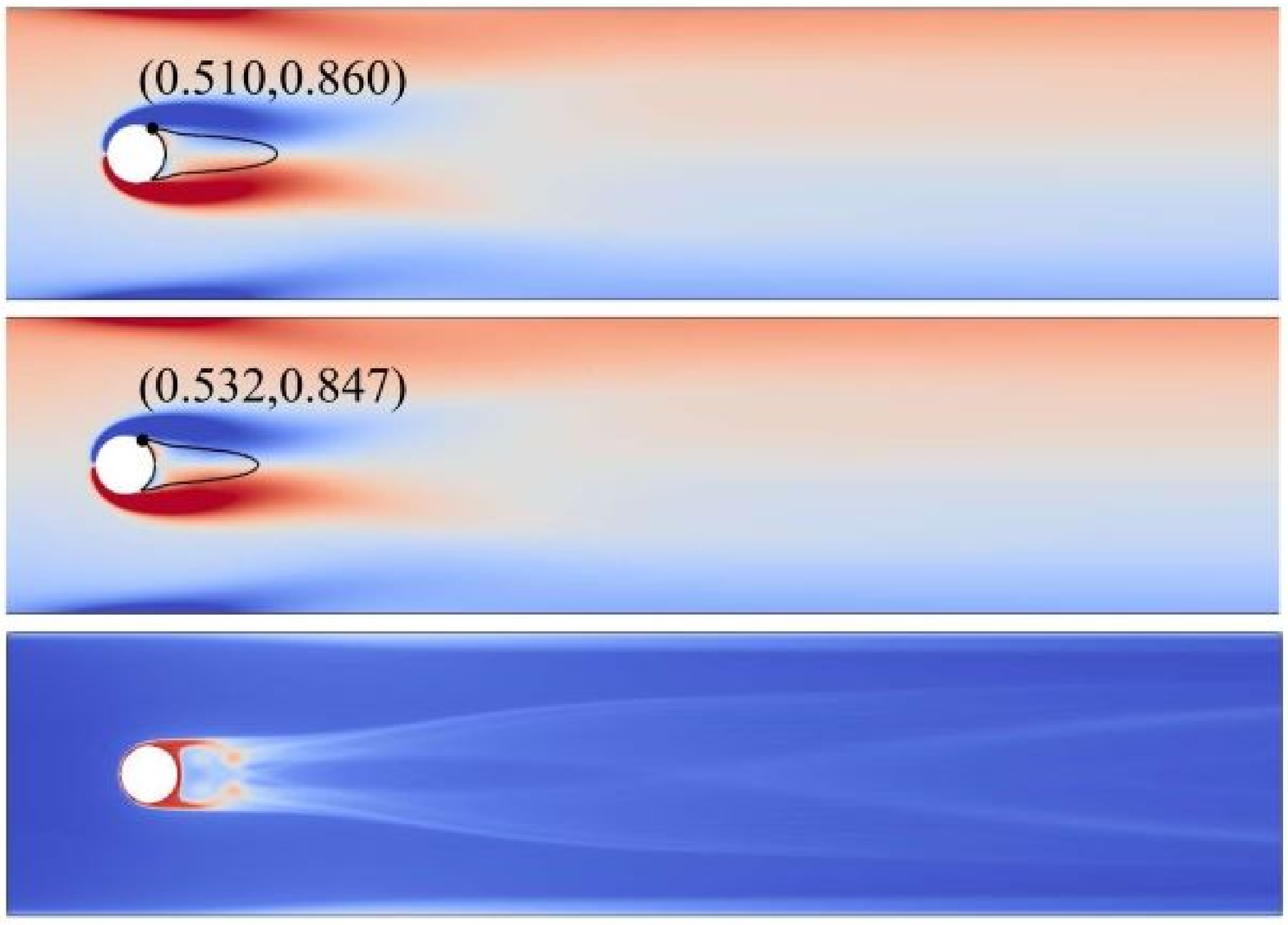}
		\put(-175,110){(b)}	
		\put(-175,65){(d)} 	
		\put(-175,30){(f)} 	
		\put(-110,120){Time-mean flow}
		\label{fig.U12mean}
	\end{minipage}
	\caption{SFD base flow (left) and mean flow (right) for cylinder wake flow and wire-plate EHD-Poiseuille flow at $ U_0=12 $. From top to bottom, they are (a)(b) vorticity of cylinder wake flow;(c)(d) vorticity of wire-plate EHD-Poiseuille flow; (e)(f) positive charge density of wire-plate EHD-Poiseuille flow. The black solid lines are the outlines of the recirculation zones and the black dots are the separation points.}
	\label{fig.p3non3}
\end{figure}

We would like to investigate further how the electric field influences the drag in the EHD wake flow. We check the two parts of the drag coefficient, i.e. pressure drag coefficient $ C_{dp} $ and the friction drag coefficient $ C_{df} $ separately for both cylinder wake flow and wire-plate EHD-Poiseuille flow, as shown in figure \ref{fig.cd}. It can be observed from the four panels that within the parameters we choose, $ C_{dp} $ are all higher than $ C_{df} $, indicating that pressure contributes more to the drag forces. In addition, from every panel, we find that EHD effect can increase both $ C_{dp} $ and $ C_{df} $. However, the increase of $ C_{df} $ by the EHD effect is small in these four cases, and the increased value decreases with increasing $ U_0 $. On the other hand, the enhancement of $ C_{dp} $ is more significant, which indicates that the EHD effect mainly boosts the pressure drag, even though the enhancement also becomes weaker as $ U_0 $ increases. On the other hand, when examining the recirculation zone (displayed in the mean flow fields at $ U_0=12 $ as shown in panels (b) and (d) of figure \ref{fig.p3non3}), we find that the recirculation zone becomes smaller and the separation points move aft when we consider the effect of the electric field. In order to further explore the reason for the increase of $ C_{dp} $ by the EHD effect, we plot the distribution of the projection of the pressure on the $x$-axis (denoted as $P_x$ in figure \ref{fig.wireP}) acting on the upper half wire with respect to the arc length $ \theta\bcdot R $, as shown in figure \ref{fig.wireP}. Taking the wire-plate EHD-Poiseuille flow as an example, the drag force due to the pressure can be regarded as the net area enclosed by the red line and the $x$-axis, that is $ \Omega_1+\Omega_3-\Omega_2 $.  Therefore, we can find that the increase of the $ C_{dp} $ is mainly because the charge injection increases the pressure on the back face of the wire, though it shrinks the recirculation zone.

\begin{figure}
	\centering
	\begin{minipage}[h]{0.4\linewidth}
		\centering
		\includegraphics[height=4.5cm]{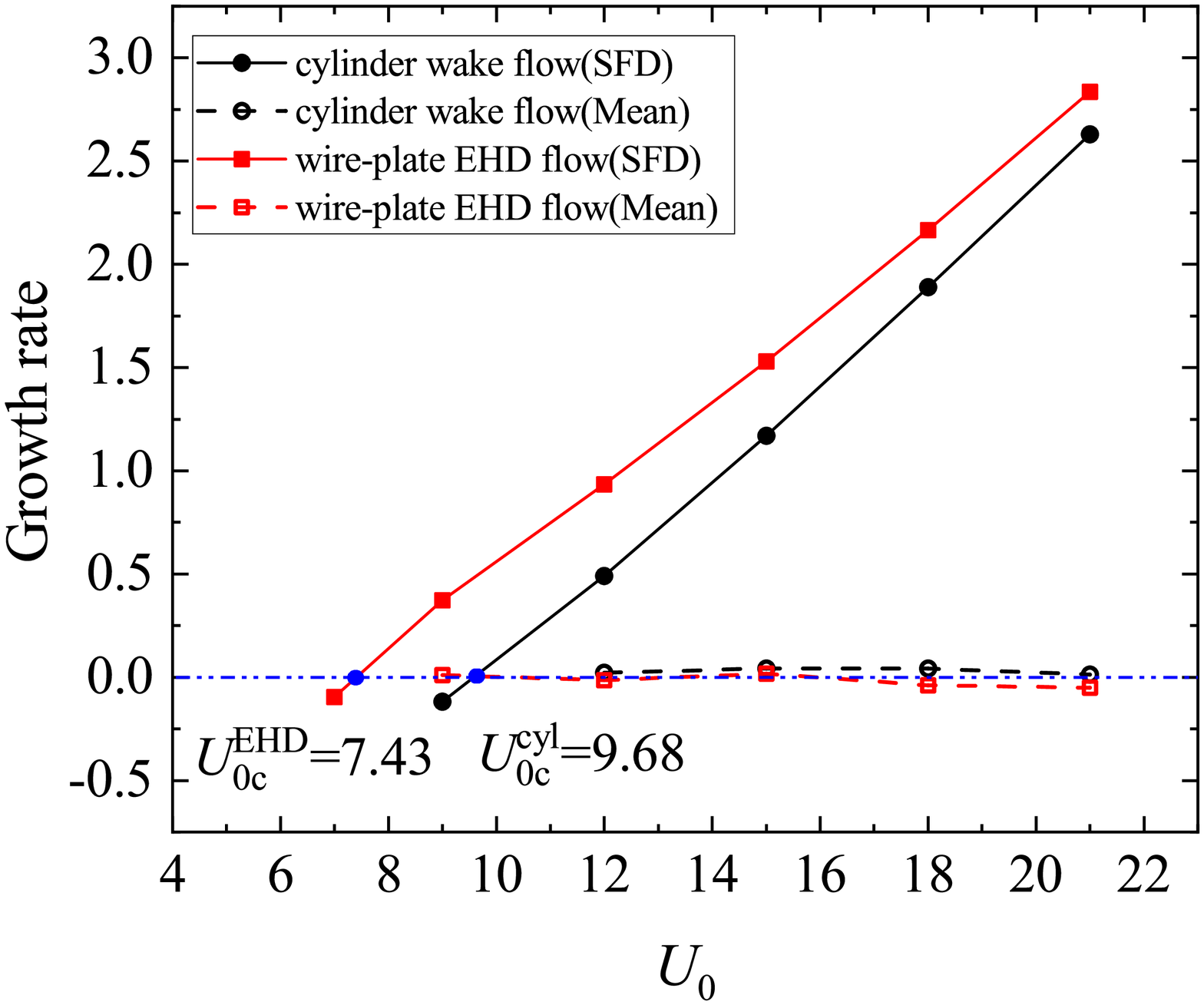}
		\put(-160,120){(a)}		
		\label{fig.p3gr}
	\end{minipage}
	\hspace{20pt}
	\begin{minipage}[h]{0.4\linewidth}
		\centering
		\includegraphics[height=4.5cm]{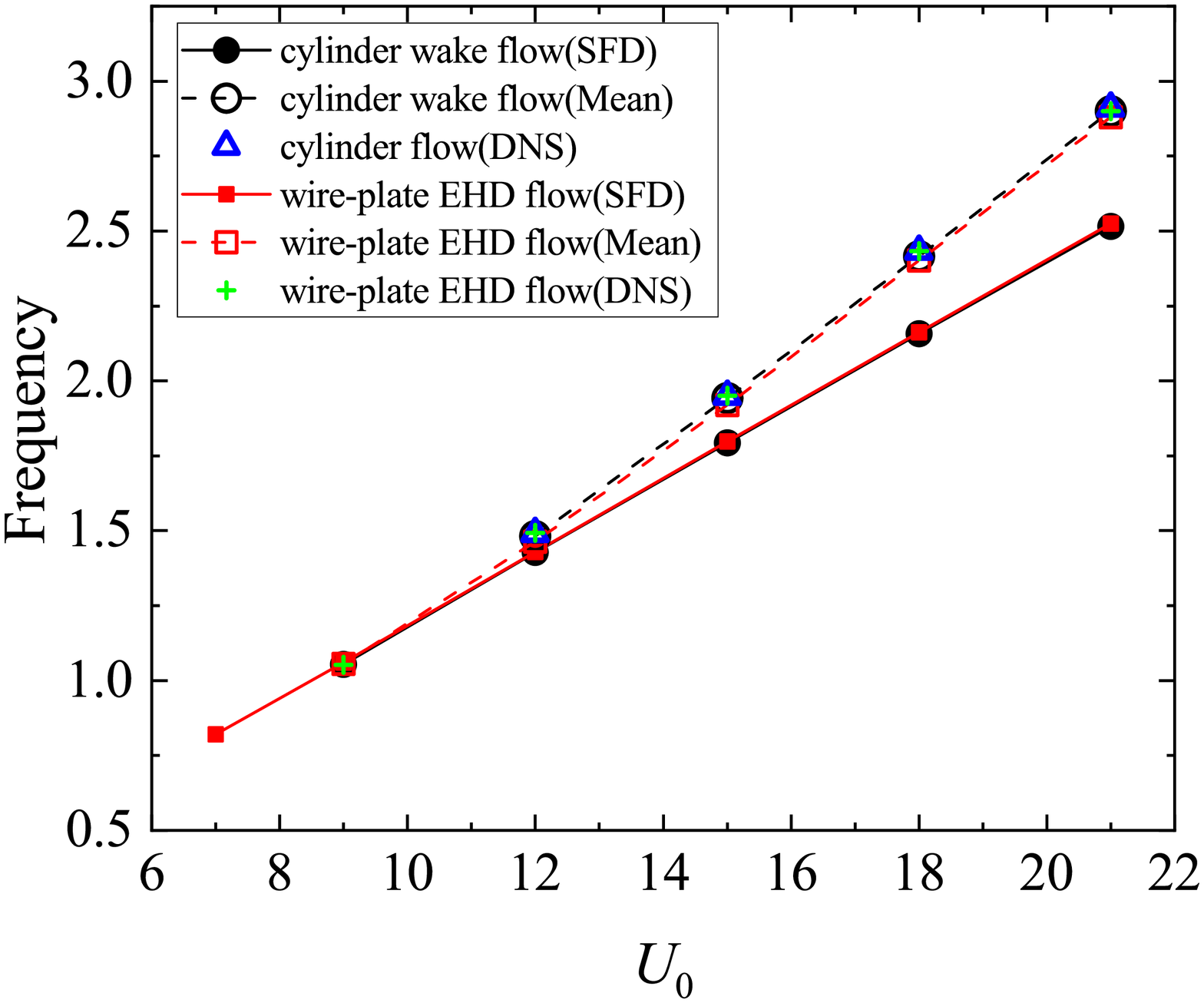}
		\put(-160,120){(b)}		
		\label{fig.p3fr}
	\end{minipage}
	\caption{Growth rates and frequencies of cylinder wake flow and wire-plate EHD-Poiseuille flow at different $ U_0 $. }
	\label{fig.p3sta}
\end{figure}

\subsubsection{Global stability analysis of the EHD flow}
It has been found that for the cylinder wake \citep{barkley2006linear}, the nonlinear oscillating frequencies obtained from DNS are close to the frequency obtained from a linear stability analysis based on the time-mean flow field. In addition, the growth rate is virtually zero. This property is called real-zero imaginary-frequency (RZIF) property \citep{turton2015prediction}. It also exists in travelling waves of thermosolutal convection \citep{turton2015prediction}, and the ribbons and spirals of counter-rotating Taylor-Couette flow \citep{bengana2019spirals}. In order to estimate whether the wire-plate EHD flow with a strong cross-flow has the RZIF property, we use both the time-mean flow and SFD base flow to carry out the global linear stability analysis in this section. 

\begin{figure}
	\centering
	\subfigure{
	\begin{minipage}[h]{0.4\linewidth}
		\centering
		\includegraphics[height=2cm]{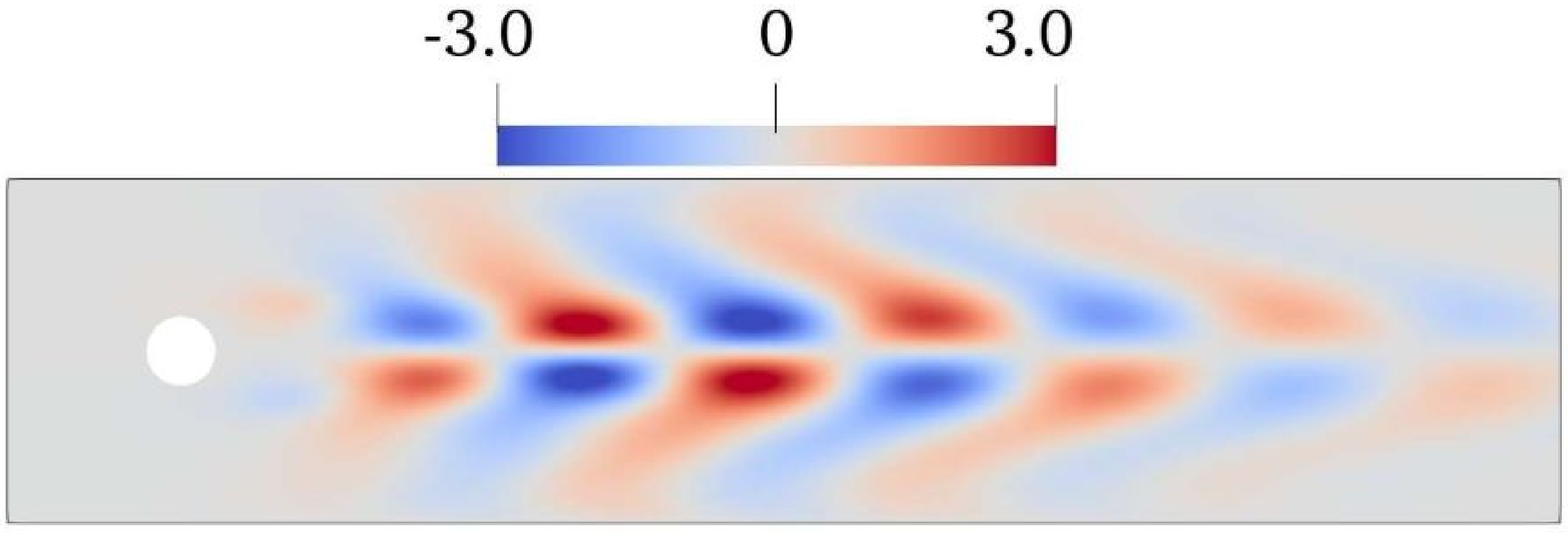}
		\put(-170,45){(a)}
		\put(-105,60){$ x $-velocity}
		\put(-55,75){Newtonian cylindrical wake flow:}					
		\label{fig.U12negvx}
	\end{minipage}}
	\hspace{20pt}
	\vspace{20pt}	
	\subfigure{
	\begin{minipage}[h]{0.4\linewidth}
		\centering
		\includegraphics[height=2.05cm]{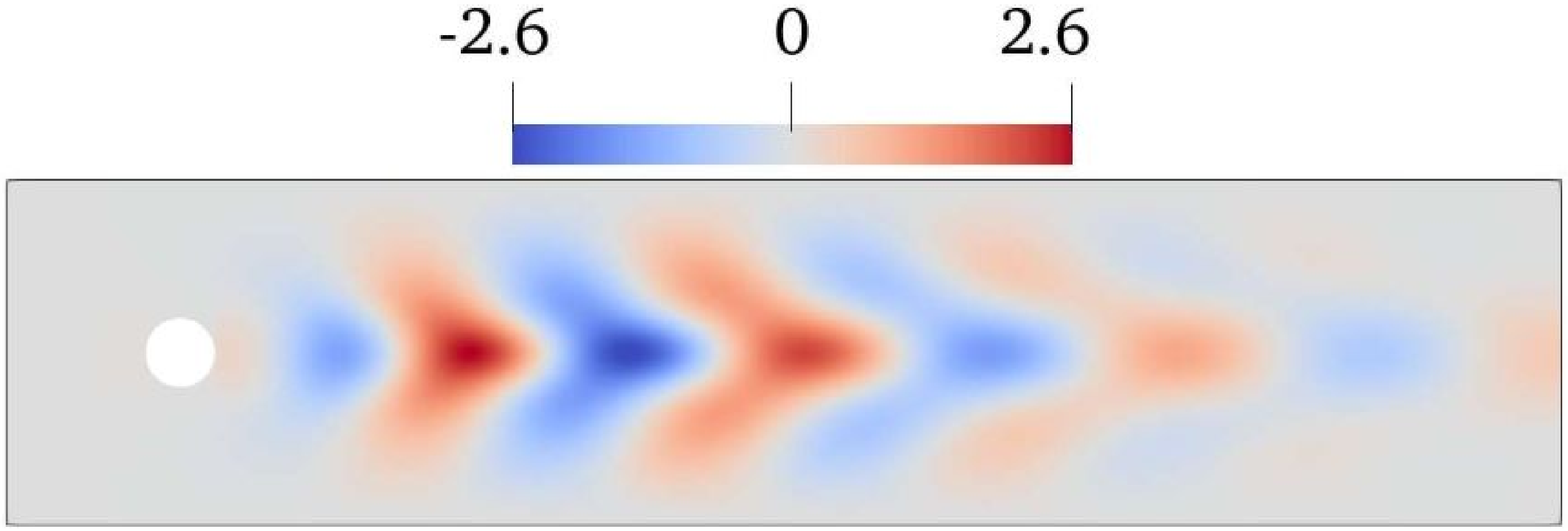}
		\put(-170,45){(b)}
		\put(-100,60){$ y $-velocity}
		\put(-30,80){}	
		\label{fig.U12negvy}
	\end{minipage}}
	\hspace{20pt}	
	\subfigure{
	\begin{minipage}[h]{0.4\linewidth}
		\centering
		\includegraphics[height=2cm]{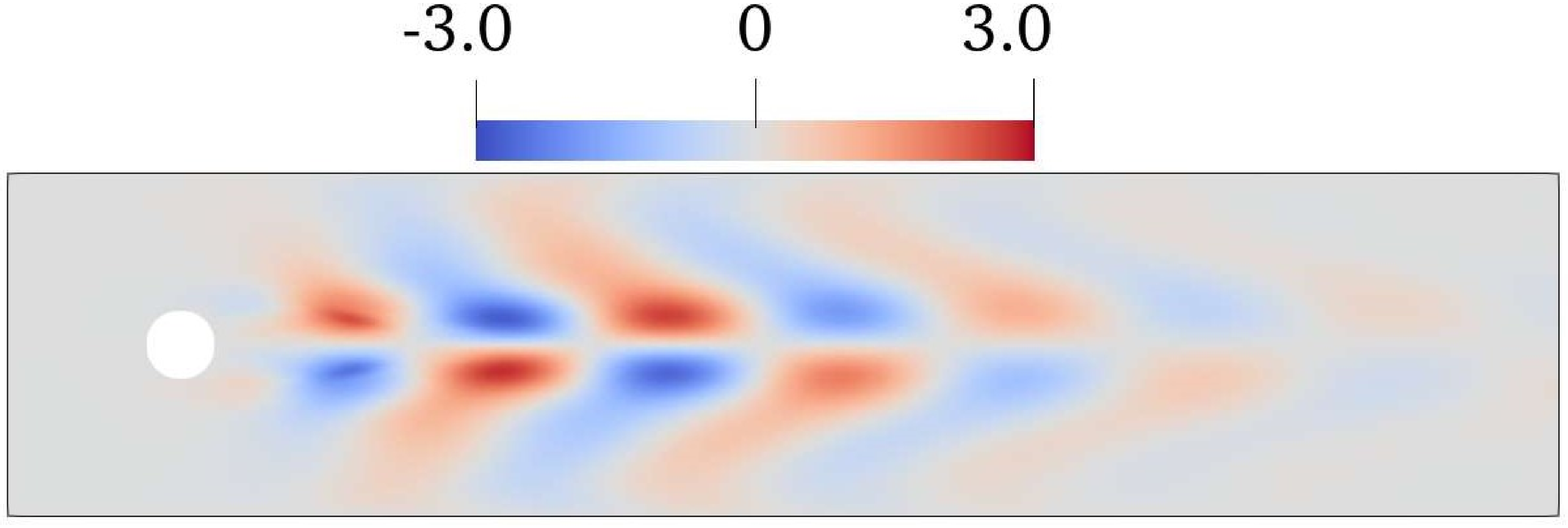}
		\put(-170,45){(c)}
		\put(-105,60){$ x $-velocity}
		\put(-50,75){Wire-plate EHD-Poiseuille flow:}					
		\label{fig.U12egvx}
	\end{minipage}}
	\hspace{20pt}
	\subfigure{
	\begin{minipage}[h]{0.4\linewidth}
		\centering
		\includegraphics[height=2cm]{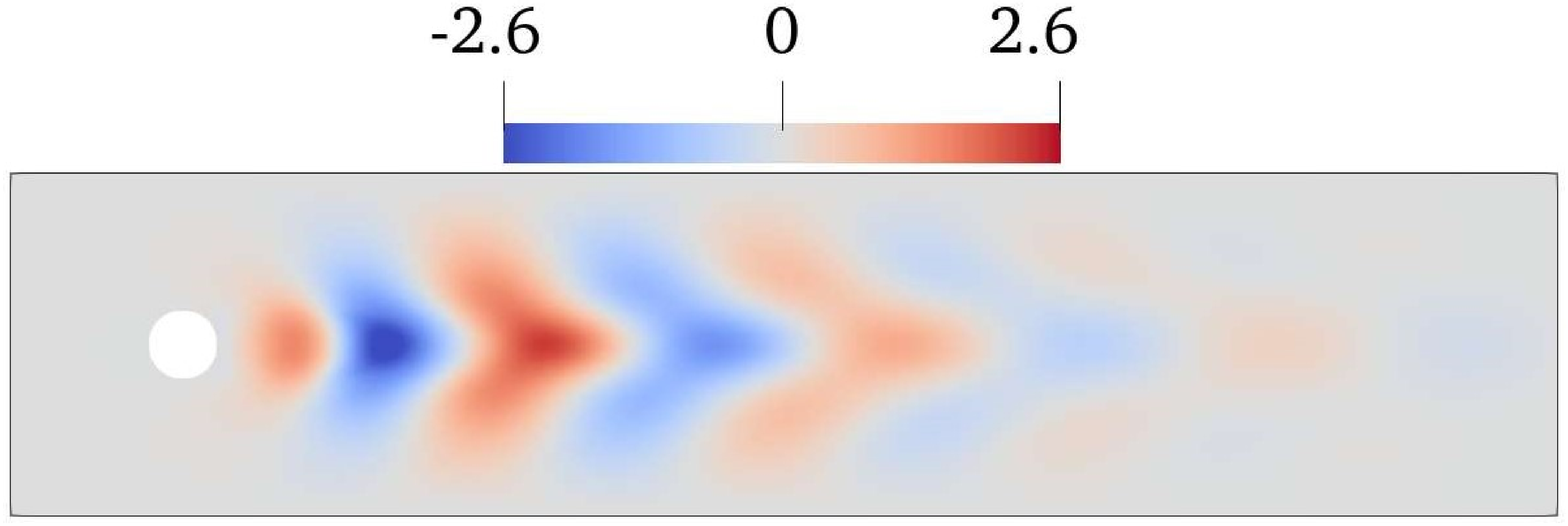}
		\put(-170,45){(d)}
		\put(-105,60){$ y $-velocity}
		\put(-30,80){}					
		\label{fig.U12egvy}
	\end{minipage}}
	\hspace{20pt}
	\subfigure{
	\begin{minipage}[h]{0.4\linewidth}
		\centering
		\includegraphics[height=2cm]{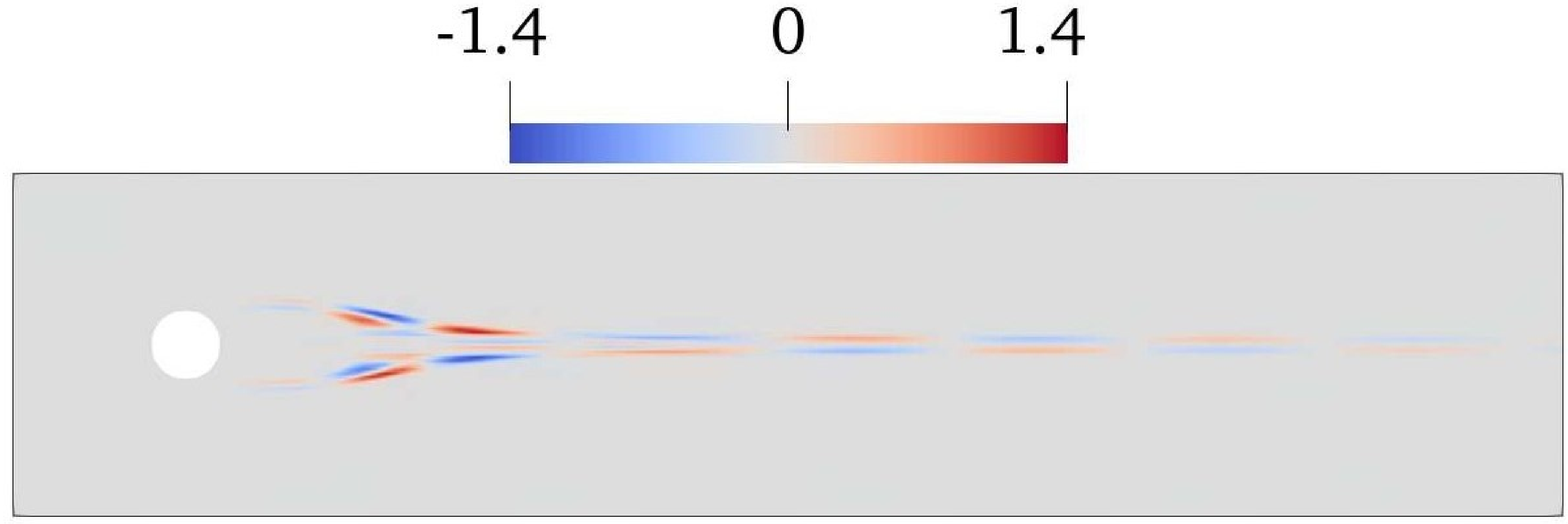}
		\put(-170,45){(e)}
		\put(-170,60){positive charge density for SFD base flow}
		\label{fig.U12egq1}
	\end{minipage}}
	\hspace{20pt}
	\subfigure{
	\begin{minipage}[h]{0.4\linewidth}
		\centering
		\includegraphics[height=2cm]{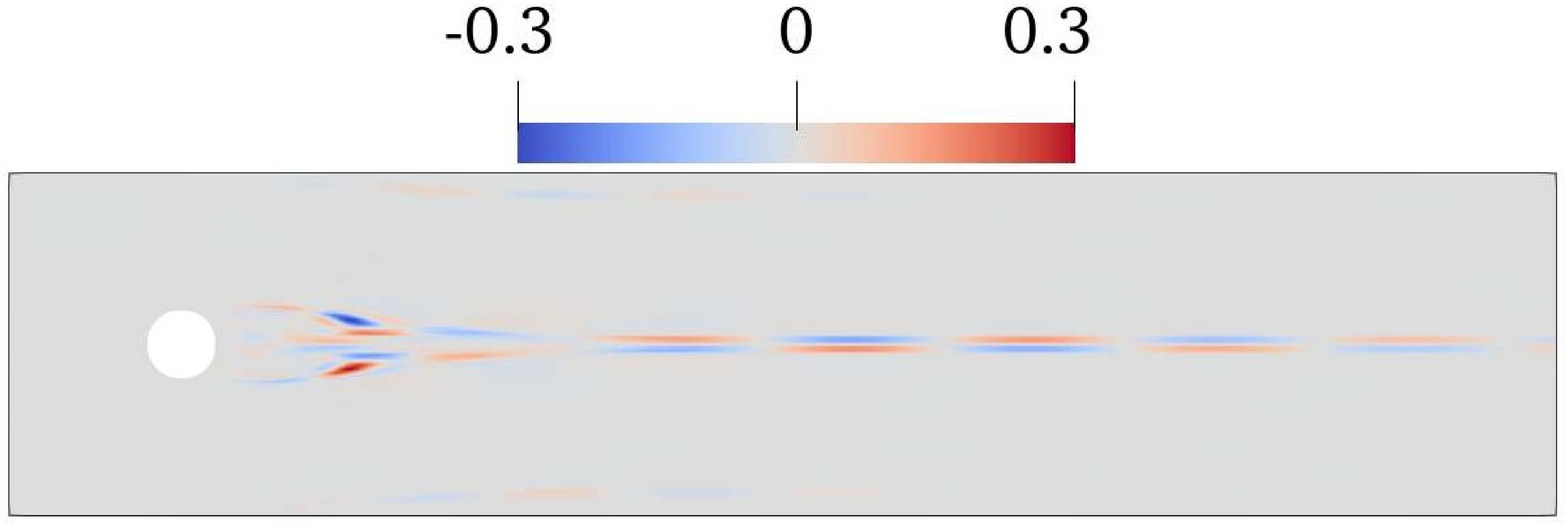}
		\put(-170,45){(f)}
		\put(-170,60){negative charge density for SFD base flow}
		\label{fig.U12egq2}
	\end{minipage}}
	\hspace{20pt}
	\subfigure{
	\begin{minipage}[h]{0.4\linewidth}
		\centering
		\includegraphics[height=2cm]{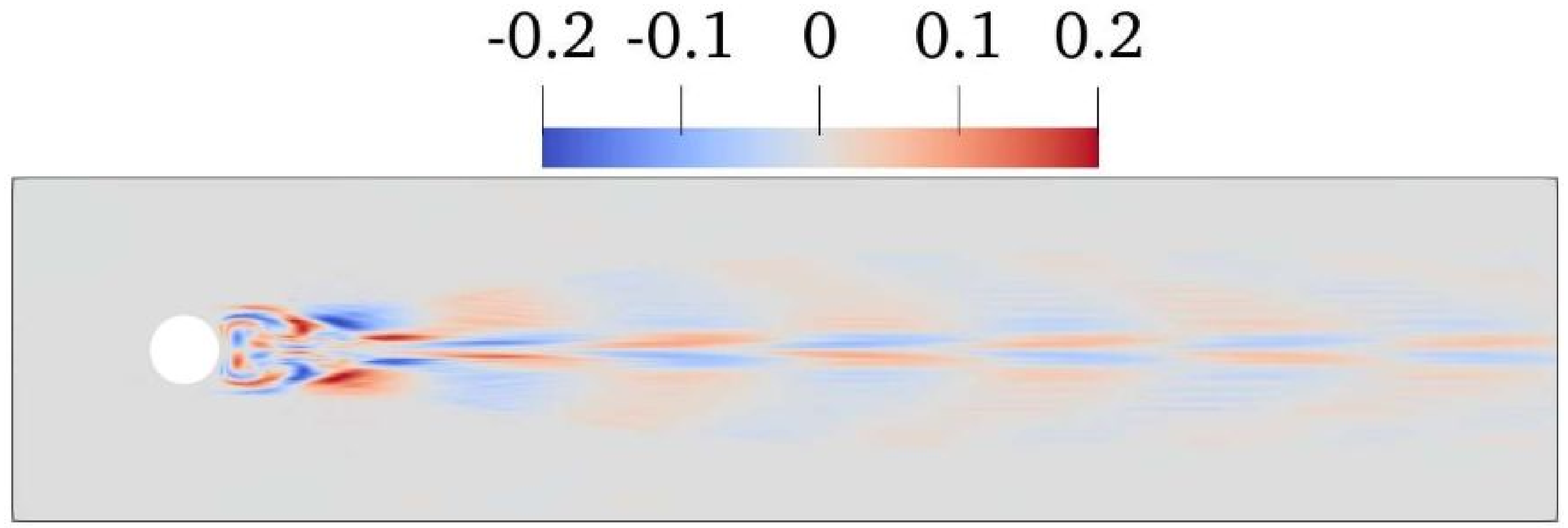}
		\put(-170,45){(g)}
		\put(-160,60){positive charge density for mean flow}		
		\label{fig.meaneq1}
	\end{minipage}}
	\hspace{20pt}
	\subfigure{
	\begin{minipage}[h]{0.4\linewidth}
		\centering
		\includegraphics[height=2cm]{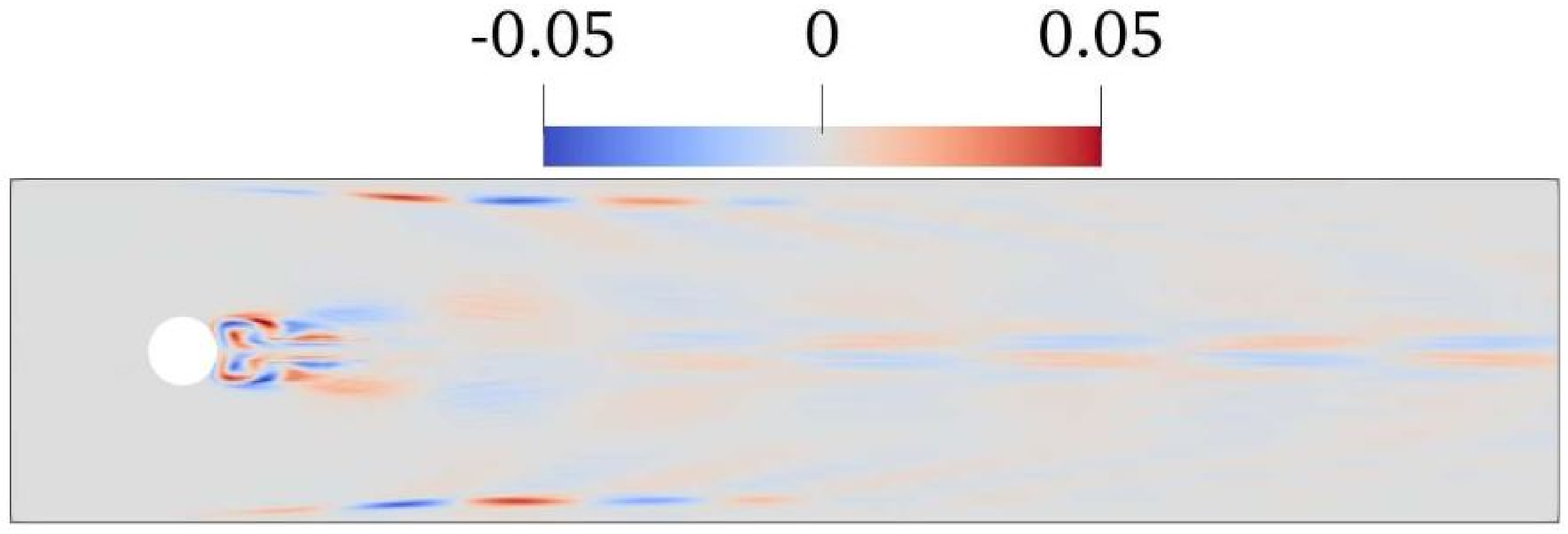}
		\put(-170,45){(h)}
		\put(-160,60){negative charge density for mean flow}			
		\label{fig.meaneq2}
	\end{minipage}}
	\caption{The leading eigenvectors of cylinder wake flow ((a),(b)), and wire-plate EHD-Poiseuille flow based on SFD base flow((c),(d),(e),(f)) and mean flow ((g), (h)) at $ U_0=12 $ for (a)(c) $ x $-velocity; (b)(d) $ y $-velocity; (e)(g) positive charge density; (f)(h) negative charge density.}
	\label{fig.eigenFU12}
\end{figure}  

Figure \ref{fig.p3non3} shows the SFD base flow (left panels) and time-mean flow (right panels) for the Newtonian cylindrical wake and the wire-plate EHD flow at $ U_0=12 $. It can be found from panels (a) and (c) that the vorticity fields of the SFD base flow in the two cases resemble each other. In addition, the positive charge distribution of the SFD base flow (panel (e)) shows a similar pattern to the steady flow as shown in figure \ref{fig.base3}. The vorticity field of time-mean fields of the cylinder wake and the wire-plate EHD-Poiseuille flow are displayed in panels (b) and (d), respectively. We can see that the mean flows for the two types of flows are similar, and display a different pattern from SFD base flows (panels a and c), where the vorticity concentrates only near the wire. Moreover, from panel (f), we find that the concentration of positive species is also near the wire. It is noted that the expanded structure in panel (f) is due to the vortex shedding bringing the positive charges to these downstream regions. Once these positive charges are time-averaged, one can see such a pattern.

Figure \ref{fig.p3sta} presents the growth rates and frequencies of Newtonian cylinder wake flow and wire-plate EHD-Poiseuille flow at different $ U_0 $ based on SFD base flow and mean flow. The growth rates in panel (a) resulting from the SFD base flow of the cylinder wake (black solid line with circle symbols) and wire-plate EHD-Poiseuille flow (red solid line with square symbols) illustrate that although the presence of the charges has little effect on the flow pattern, it boosts the instability of the system. We note that the instability threshold for the flow without the EHD effect is about $U_{0c}=9.68$, and it matches with the critical value for a flow past a confined cylinder ($ U_{0c}=9.63 $) obtained by \cite{sahin2004numerical}. The threshold reduces to 7.43 with the EHD effect. In addition, the frequencies of the two flows are close (panel (b), black solid line with circle symbols for the cylinder wake flow and red solid line with square symbols for the wire-plate EHD-Poiseuille flow). On the other hand, panel (a) also shows that the growth rates of time-mean flow for cylinder wake (black dash line with circle symbols) and wire-plate EHD-Poiseuille flow (red dash line with square symbols) are both close to zero, meaning that time-mean flows of the two flows are marginally stable. Panel (b) also indicates that the frequencies based on the time-mean flows for the two cases (black dash line with circle symbols for the cylinder wake flow and red dash line with square symbols for the wire-plate EHD-Poiseuille flow) are both close to nonlinear frequencies (blue $ \times $ symbols). These results indicate that the RZIF property also exists in the wire-plate EHD-Poiseuille flow within the parameter range we considered.

Finally, the global modes in the EHD cylindrical wake flow are shown. We take $ U_0=12 $ as an example in figure \ref{fig.eigenFU12} to display the leading eigenmodes for both Newtonian cylinder wake flow and wire-plate EHD-Poiseuille flow. Panels (a) and (b) are the eigenvectors of $ x $-velocity and $ y $-velocity for the Newtonian cylinder wake based on SFD base flow. Panels (c) and (d) depict the eigenmodes of $ x $-velocity and $ y $-velocity for wire-plate EHD-Poiseuille flow subjected to the SFD base flow, and we notice that they bear resemblance to those of the cylinder wake flow. Panels (e) and (f) show the eigenvectors of positive and negative charge density, respectively. It can be found that they are all antisymmetric with respect to the horizontal central line $ y=0 $. As for the eigenmodes of the mean flow, we find that the distribution of velocity perturbations shows a similar pattern to those of SFD base flow, so we do not present them here. However, the eigenvectors of positive and negative species show different patterns from those of SFD base flow, and they are exhibited in panels (g) and (h). Loosely speaking, the oscillatory behavior of the charge fields causes the difference seen between the linear stability analyses based on the SFD and time-mean flows. The perturbations of positive and negative species are also antisymmetric to $ y=0 $, but concentrate near the wire.

In conclusion, since the flow is inertia-dominant when the cross-flow is strong, both the flow pattern and the leading eigenmodes are similar to the conventional flow past a cylinder. However, charge injection from the wire electrode can significantly advance the onset of the vortex shedding, and increase the drag coefficient and the instability of the linear system. It is also reasonable to observe that the modification by the EHD effect becomes smaller as the cross-flow becomes stronger. In addition, we find that the wire-plate EHD flow with a strong cross-flow possesses the RZIF property within the parameter range.

\section{Conclusions} \label{Conclusions}

In this work, we conducted direct numerical simulations and linear global stability analysis of the 2-D EHD flow subjected to a Poiseuille cross-flow in a wire-plate configuration. For the mechanisms of the generation of space charges, both conduction and injection of charges have been considered. In this wire-plate EHD-Poiseuille problem, the process of  dissociation and recombination of the species cannot be ignored since the recombination time is close to the transit times and the convective time. Therefore, the consideration of the conduction mechanism makes our results more realistic and meaningful. The flow is characterised by many physical parameters. In this work, we have focused on varying the intensity of the cross-flow to organise our discussions. Three flow patterns have been investigated, namely, wire-plate EHD flow without a cross-flow, with a weak cross-flow and with a strong cross-flow. In the no cross-flow case, the electroconvection in the wire-plate configuration was studied. The weak cross-flow case was investigated because it can be related to the flow pattern in ESP. With the case of a strong cross-flow, we studied the EHD effect on the cylindrical wake flow.

We first investigated the wire-plate EHD flow without a cross-flow. According to the experiment of \cite{mccluskey}, only when $ Re^E\textgreater 0.23 $, charge injection occurs. Therefore, taking it as the borderline, the flow phenomena are divided into two regimes, namely, conduction regime ($ Re^E\textless0.23, C_I=0 $) and injection regime ($ Re^E\textgreater0.23, C_I=0.2 $), and they are studied  separately. In the conduction regime, the charges are generated everywhere in the domain, and their movement is mainly caused by electric drift and flow convection. The linear stability analysis shows that the growth rate of the linear flow increases with the increase of $ Re^E $, and the frequency is zero. In the injection regime, the injected charges are issued from the wire and convected to the plate electrodes vertically due to the Coulomb force. The oscillatory instability can be predicted well by a global stability analysis of the steady flow. The growth rate and frequency both increase with increasing $ Re^E $. Besides, the leading eigenvectors show different patterns in the conduction regime and in the injection regime, demonstrating their different influences on the flow. 


Then, the case with a weak cross-flow was studied. Due to the action of the cross-flow, the charge density plume tilts downstream. Since the secondary EHD flow can help the particles in the dusty gas settle on the collecting plate, this flow pattern is commonly seen in ESP. We performed a detailed linear stability analysis of this case. The results indicate that at small $ Re^E $, the flow stays steady. When $ Re^E $ exceeds a certain critical value, the flow state will transition from stable to time-periodic oscillation. A violent flapping of the charge density beams can be observed in the oscillation state. Furthermore, we explore the effect of the cross-flow velocity $U_0$. It is found that a larger $U_0$ increases the growth rate of the perturbation within the chosen parameter range. Moreover, the eigenvectors are presented, which are similar to those in wire-plate EHD flow without cross-flow, but the flow structures tilt downstream. Besides, we can learn from the results that in order to gain a good performance in ESP, it is better to adopt a lower voltage and a relatively weak cross-flow to not cause flow oscillation or instability, otherwise, the flapping of the flow structures on the collecting plates will cause the charged dusts in ESP to bounce back. Furthermore, in the case of the weak cross-flow, we found that the EHD flow reduces the drag on the wire, especially the shear forces on the wire surface, which may imply that the EHD effect results in a large passing-through rate of the fluids in the channel, consistent with \cite{soldati1998turbulence}'s results on the turbulent flow in the wire-plate ESP. 

Finally, the wire-plate EHD flow with a strong cross-flow has been investigated. In this inertia-dominant case, the flow can be compared to a confined Newtonian cylinder wake flow. The aim of studying this case is to examine how the EHD modifies the wake flow and its stability. The nonlinear simulations show that the EHD effect can bring forward the vortex shedding behavior. In addition, we observed that $ C_d $ increases significantly due to the presence of the EHD effect. The decomposition analysis of the drag coefficient and the distribution of the pressure around the wire/cylinder show that the EHD effect mainly increases the pressure force acting on the aft side of the wire/cylinder. However, interestingly, the velocity field is hardly changed by the EHD, indicating the inertia-dominant characteristic of this flow pattern. From the perspective of flow stability analysis, we have used both SFD base flow and time-mean flow as the base flow. The SFD results showed that EHD flow significantly boosts the linear growth rate, although the effect decreases with the increasing intensity of the cross-flow. In addition, the critical $ U_0 $ for the onset of vortex shedding is reduced when considering the EHD effect. Furthermore, the time-mean stability analysis indicated that both flow types with and without the EHD effect are marginally stable. The frequencies obtained by the linear stability analysis based on the time-mean flow are also closer to the nonlinear frequencies than those of steady base flow, especially at large $ U_0 $. Therefore, the real-zero imaginary-frequency (RZIF) property exists in the wire-plate EHD-Poiseuille flow as well. 

In the end, we mention the limitation of this work. First, we used an experimentally determined borderline \citep{mccluskey} between the conduction and injection mechanisms for the charge generation. This specific threshold value may be different in other flow conditions. Second, the blockage ratio of the wire diameter to the channel height is fixed at 0.2. It would be interesting to study the wire-plate EHD-Poiseuille flow with different blockage ratios. Additionally, the working medium in our work is assumed to be liquid, which may not be directly translated to the ESP, even though an analogy of the flow with particles in ESP and fluids in EHD has been made by \cite{atten1987electrohydrodynamic}. The gas flow in the wire-plate EHD flow could be investigated as part of a large parametric study in the future.

Declaration of Interests. The authors report no conflict of interest.

\begin{acknowledgments} 
The financial support from the Ministry of Education, Singapore is acknowledged (the WBS No. R-265-000-689-114). X. H. is supported by a doctoral research scholarship from the National University of Singapore and a scholarship from the China Scholarship Council. P. A. V. acknowledges the grant PGC2018-099217-B-I funded by MCIN/ AEI/10.13039/50110001103. We thank all the reviewers for the stimulating comments.
\end{acknowledgments}

\begin{appendix}
\section{Validation and comparison with experiment}\label{valinon}
In this section, we verify our nonlinear solver of wire-plate EHD-Poiseuille flow by comparing our results to the experimental results in \cite{mccluskey}. \cite{mccluskey} have performed an experimental study of the wire-plate EHD flow subjected to a laminar Poiseuille flow. They mounted a wire with a diameter of $ D^*=0.1 $mm in the center between two parallel plates apart from each other by $ 3.5 $mm. Since in this experiment the aspect ratio of the cylinder length to the diameter is 300, it is believed that there was no influence from the sidewalls on the central part of the flow. Therefore, we can make a comparison of our 2-D simulations with the experimental flow in the center part. The liquid used in the experiment was Benzyl Neocaprate (BNC), and its physical properties are: dynamic viscosity $ \mu^*=6\times 10^{-3}\mathrm{Pa\bcdot s} $, mass density $ \rho^*=957\mathrm{Kg/m^3} $, relative permittivity $ \epsilon_r^*=3.8 $, electrical conduction $ \sigma^*=10^{-9} \mathrm{S/m} $. According to the Walden rule \citep{castellanos1998electrohydrodynamics}, we can obtain that the ionic mobility $ K_+^*=K_-^*=5\times10^{-9} \mathrm{m^2/(V\bcdot s)}. $ Therefore, the values of non-dimensional parameters corresponding to $ \Phi_0^*=3 $kV can be calculated as: $ C_0=3, M=37, K_r=1,  Re^E=2.4, O_s=8.6, \lambda=0.02857 $, and $ \alpha=8\times10^{-6}$. In our numerical simulations, we found that when using the charge diffusion coefficient $\alpha=8\times10^{-6}$, it is difficult for the numerical simulations to converge. Thus, we will use $\alpha=0.001$ in most of the cases to be discussed. As to be shown in figure \ref{fig.fig83}, the charge diffusion effect has a small effect on the charge distribution in the bulk flow. In addition, in order to determine the value of $ C_I=Q_0^*R^{*2}/(\varepsilon^*\phi_0^*) $, an estimation of $ Q_0 $ should be given, which is hard to measure directly in the experiment. Here we assume $ Q_0^*=1\mathrm{C/m^2} $, and we will have $ C_I=0.025 $.

\subsection{Wire-plate EHD flow without cross-flow}

\begin{figure}
	\centering
	\includegraphics[height=4cm,width=6cm]{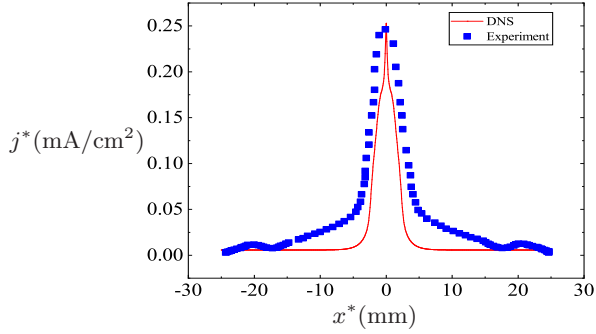}
	\put(-222,58){$ j^*(\mathrm{mA/cm^2}) $}
	\put(-100,-8){$ x^*(\mathrm{mm}) $}
	\caption{Comparison of current density distribution along the plate electrodes on top plate for
		$ \Phi_0^*=3 $kV without cross-flow between our results (DNS) and figure 5 in \cite{mccluskey} (experiment).}
	\label{fig.fig5}
\end{figure}

We first validate our simulation results without cross-flow. The geometry is symmetric with respect to $ x=0 $, that is $ \Lambda_1=\Lambda_2=500 $. The comparison between the current density obtained by our numerical simulations and the experiment by \cite{mccluskey} is displayed in figure \ref{fig.fig5}. We can see that the numerical results show the same trend as the experimental results, although the current density decreases faster away from the center part than the experimental result. This discrepancy may be attributed to the inevitable three-dimensional effects in the experiments. A similar phenomenon of comparison between numerical simulation and experimental work has also been observed in other studies of corona discharge in point-plane configuration \citep{zhang2007numerical,adamiak2004simulation,zhang2009stationary}.

\subsection{Wire-plate EHD flow with cross-flow}

Now we present our numerical simulations of the EHD flow subjected to a cross-flow and compare the results to the experimental results obtained by \citep{mccluskey}. In their experiment, the diameter of the wire is $ D^*=0.1 $mm, and the plate electrodes are 115mm long with the 25mm upstream length. In addition, the gap between two plates is 35mm. In order to be consistent with the experimental setting, the geometric parameters are selected as: $ \Lambda_1=500 $, $ \Lambda_2=1800 $, $\lambda=0.02857  $. The typical hydrodynamic Reynolds number in their experiment is $ Re=740 $ ($ Re=2\rho^* U_0^*L_y^*/\mu^* $) which is related to the height between two plates $ (2L_y^*) $ and the mean velocity ($ U_0^* $) of Poiseuille flow. So we can get that: $ Re^W=\rho^* U_0^*R^*/\mu^*=Re\times R^*/(2L_y^*)=10.4 $. Therefore, 
\begin{equation}
	U_0=\frac{Re^W}{Re^E}=\frac{10.4}{2.4}=4.3.
\end{equation}

\begin{figure}
	\centering
	\begin{minipage}[h]{0.4\textwidth}
		\centering
		\includegraphics[height=4cm]{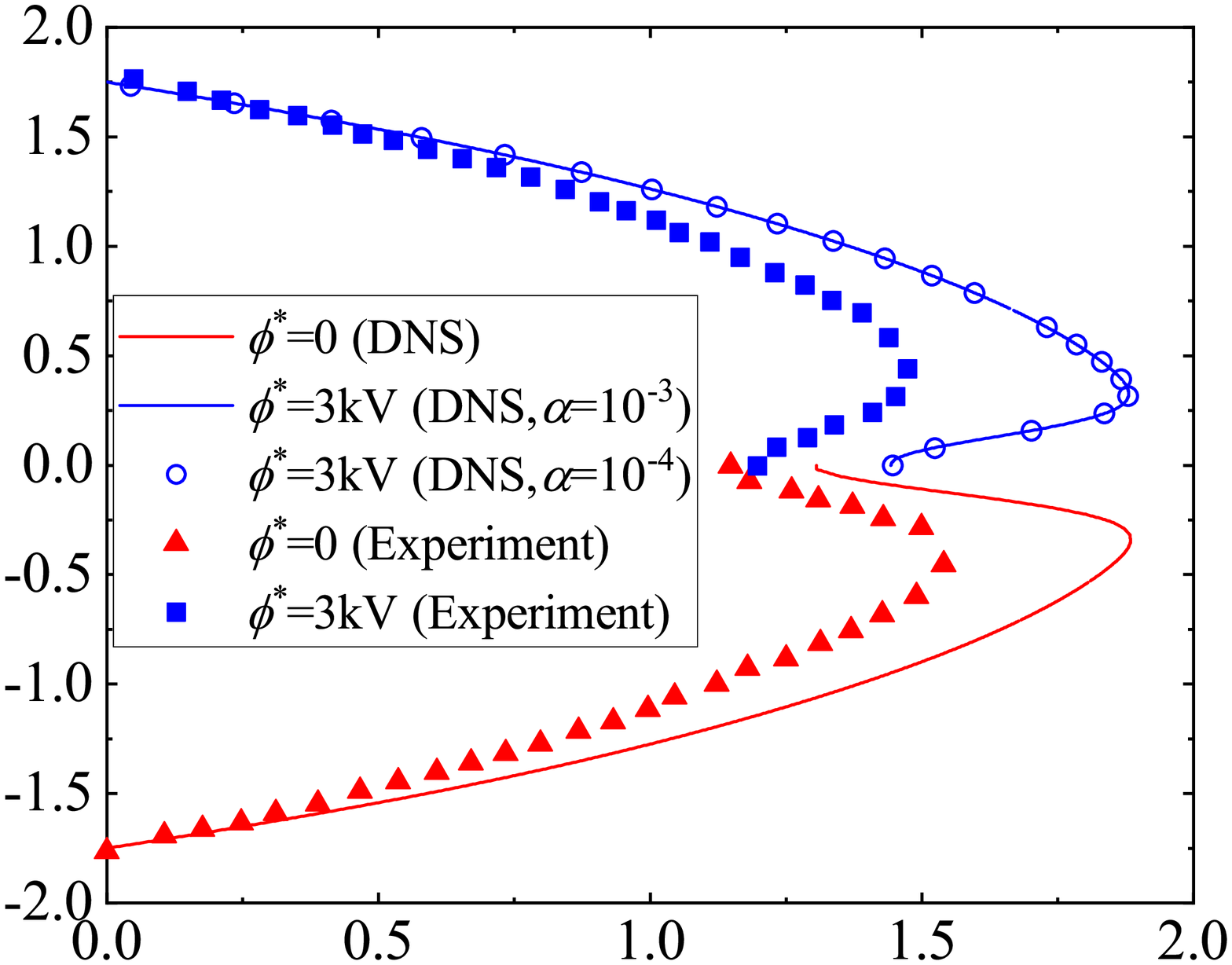}
		\put(-160,95){(a)}
		\put(-172,55){$ y^*(\mathrm{mm}) $}
		\put(-90,-10){$ |U^*|(\mathrm{m/s}) $}		
		\label{fig.20phiUin}
	\end{minipage}
	\hspace{20pt}
	\begin{minipage}[h]{0.4\textwidth}
		\centering
		\includegraphics[height=4cm]{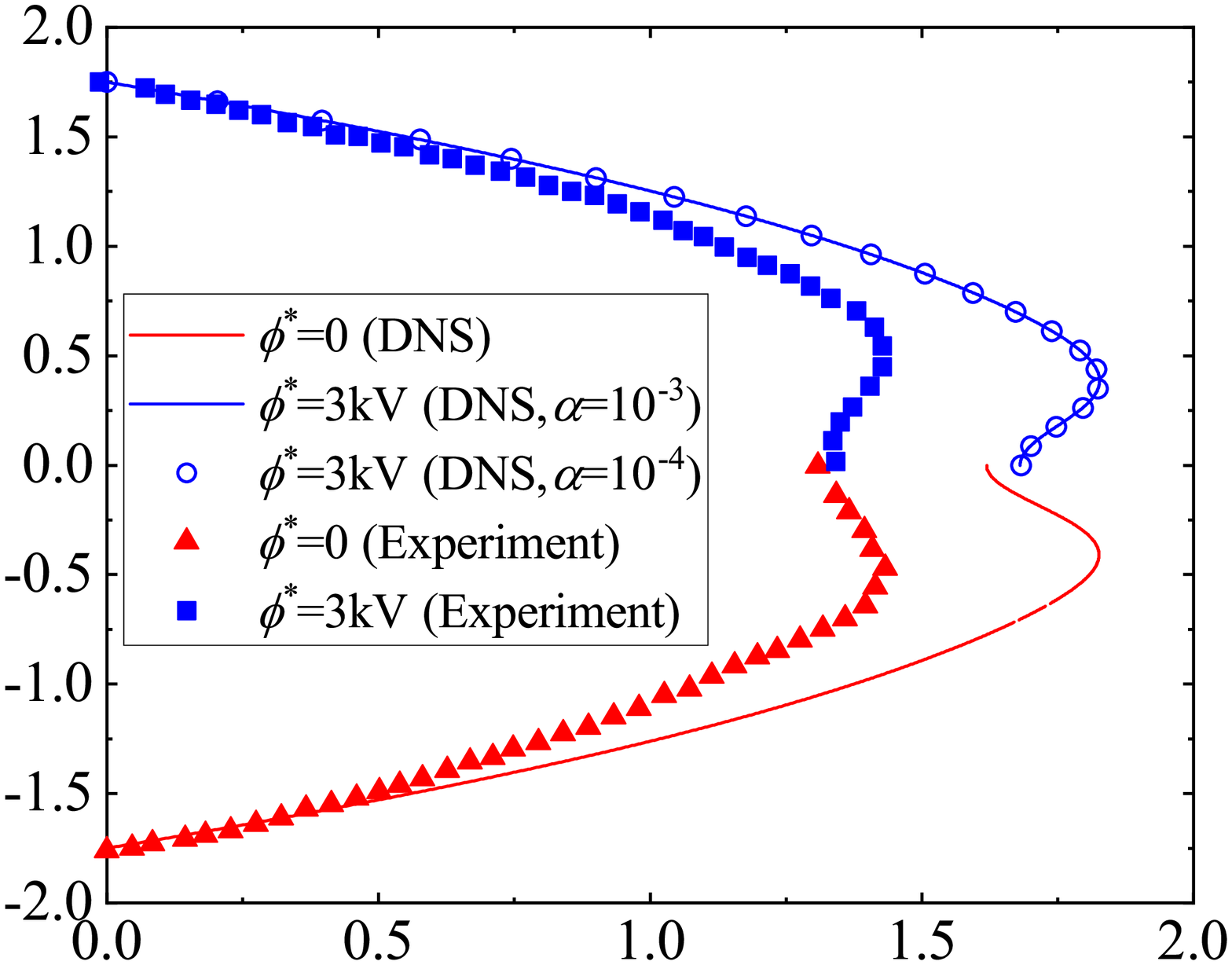}
		\put(-160,95){(b)}
		\put(-172,55){$ y^*(\mathrm{mm}) $}
		\put(-90,-10){$ |U^*|(\mathrm{m/s}) $}	
		\label{fig.60phiUin}
	\end{minipage}
	\hspace{20pt}
	\begin{minipage}[h]{0.4\textwidth}
		\centering
		\includegraphics[height=4cm]{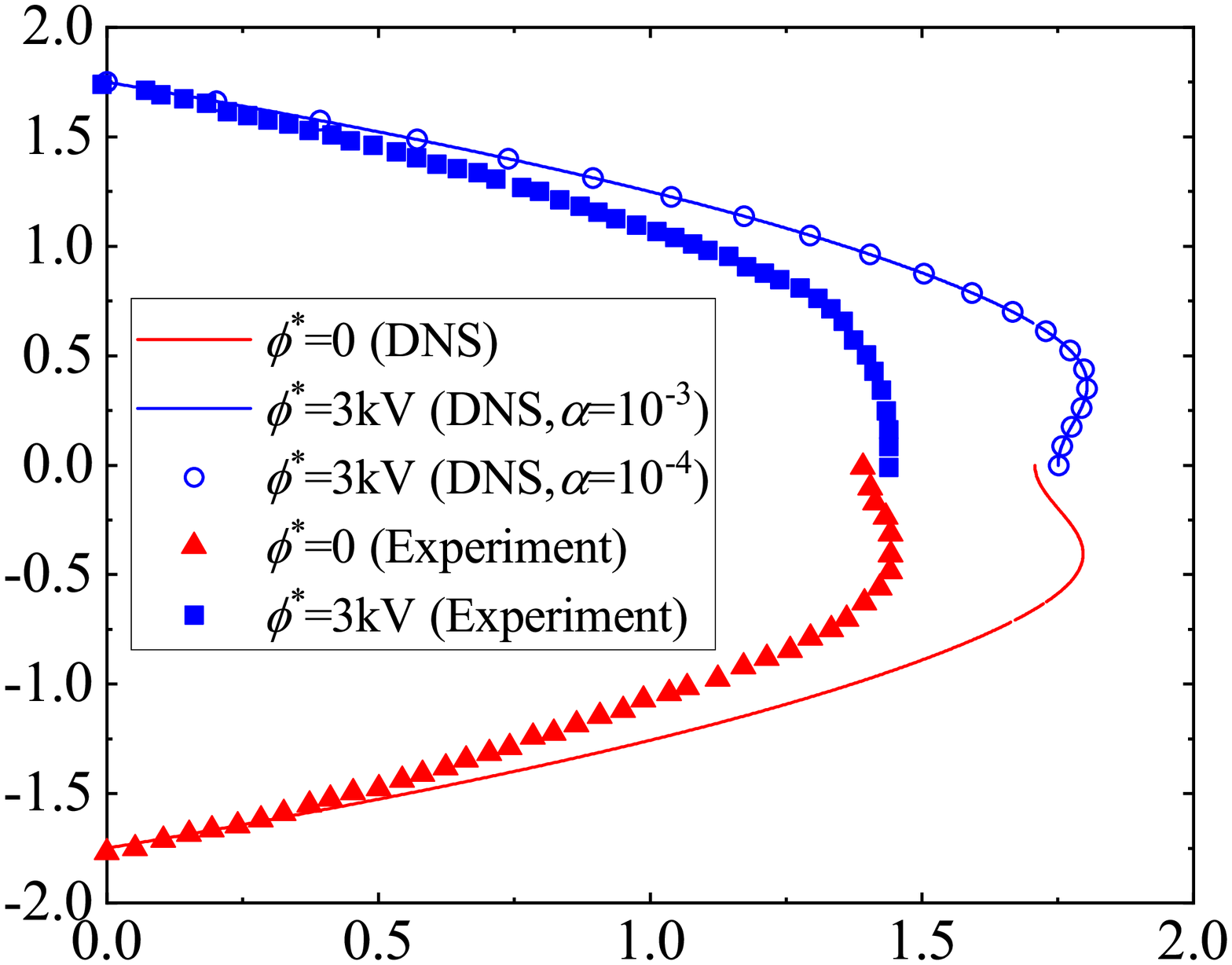}
		\put(-160,95){(c)}
		\put(-172,55){$ y^*(\mathrm{mm}) $}
		\put(-90,-10){$ |U^*|(\mathrm{m/s}) $}			
		\label{fig.100phiUin}
	\end{minipage}
	\hspace{20pt}
	\begin{minipage}[h]{0.4\textwidth}
		\centering
		\includegraphics[height=4cm]{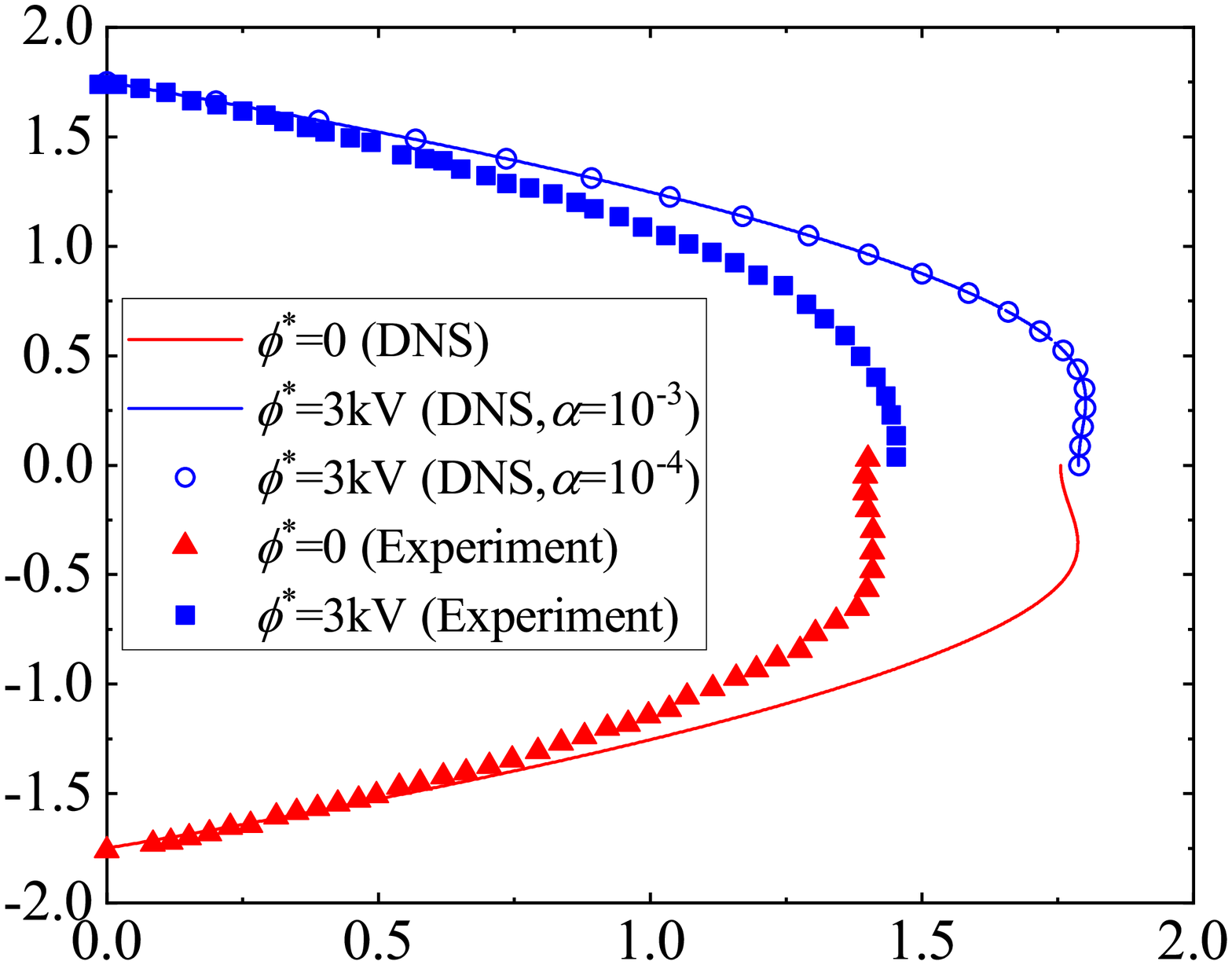}
		\put(-160,95){(d)}
		\put(-172,55){$ y^*(\mathrm{mm}) $}
		\put(-90,-10){$ |U^*|(\mathrm{m/s}) $}			
		\label{fig.140phiUin}
	\end{minipage}
	\caption{Comparison of velocity profiles for $ Re = 740 $ and $ \Phi_0^* = 0 $ kV (upper side) and $ \Phi_0^* = 3 $ kV (lower side) for different downstream distances. (a) $ x = 20D^* $; (b) $ 60D^* $; (c) $ 100D^* $; (d) $ 140D^* $. The experimental data are extracted from \cite{mccluskey}.}
	\label{fig.fig83} 
\end{figure}

It has been clarified in the experiment that the profile upstream the wire was a laminar Poiseuille flow. Therefore, in the simulation, we choose as the initial condition of the $x$-velocity the Poiseuille flow profile, i.e. $ f(y)=\frac{3}{2}U_0(1-y^2/L_y^2) $. In addition, the initial charge density is $ N_+=N_-=1 $. Figure \ref{fig.fig83} compares the velocity profiles at different downstream distances ($ D^* $ is the dimensional diameter of the wire) without and with an electric field, i.e., $\phi^*=0,3$kV. It can be seen that in the regions close to the walls, the velocity profiles in the simulations are all close to the experimental results; nevertheless, our results are persistently larger than the experimental results in the bulk region. The reason is unknown to us. 
What we have done to further verify our numerical results was to compare the case without the electric field to other more recent numerical/experimental works. In the following, we will perform a two-dimensional simulation of the classical Newtonian cylindrical wake flow and compare our velocity profiles with the results in the literature.

\begin{figure}
	\centering
	\begin{minipage}[h]{0.4\textwidth}
		\centering
		\includegraphics[height=4.2cm]{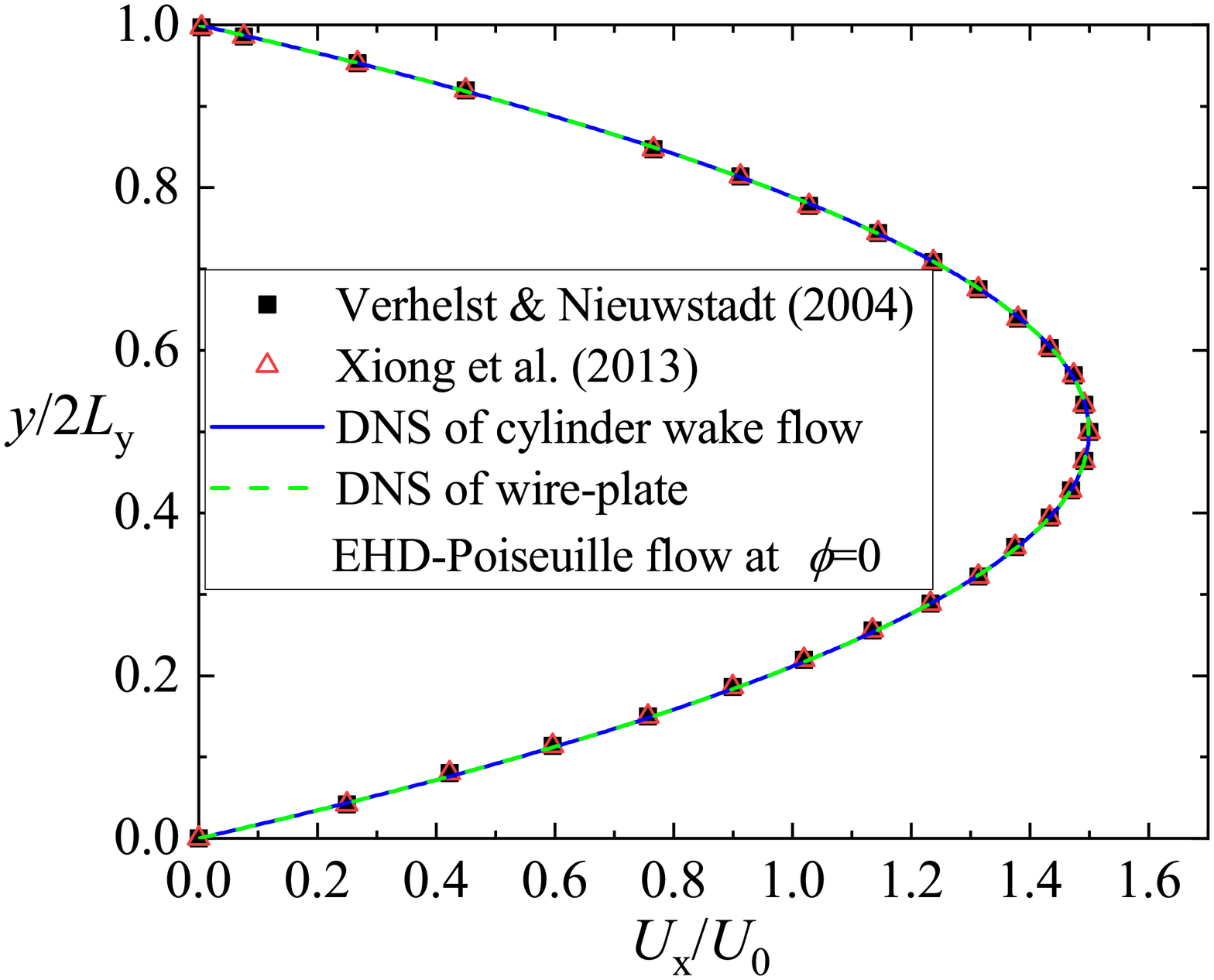}
		\put(-155,115){(a)}
		\label{fig.x1}
	\end{minipage}
	\hspace{20pt}
	\begin{minipage}[h]{0.4\textwidth}
		\centering
		\includegraphics[height=4.2cm]{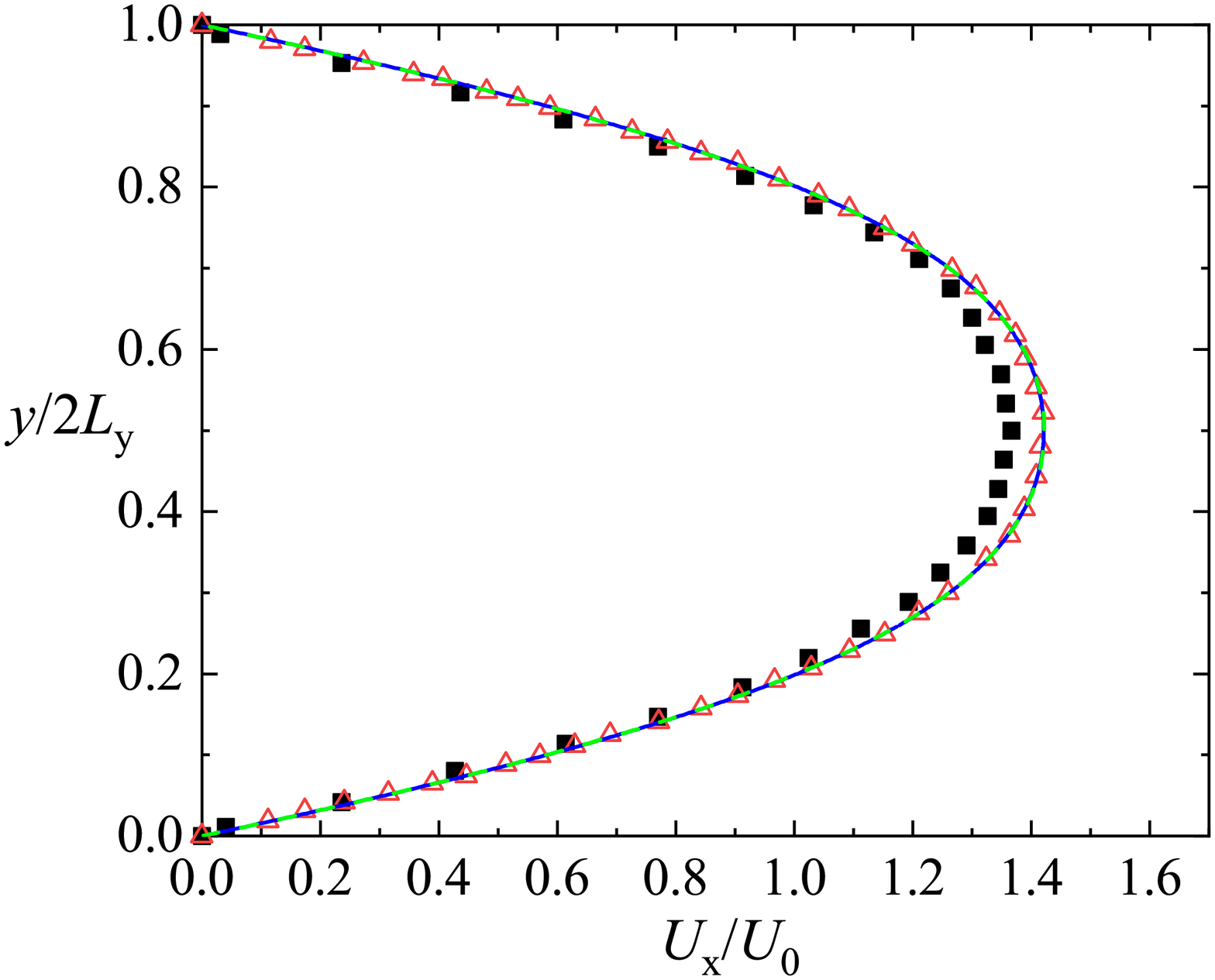}
		\put(-155,115){(b)}
		\label{fig.x2}
	\end{minipage}
	\hspace{20pt}
	\begin{minipage}[h]{0.4\textwidth}
		\centering
		\includegraphics[height=4.2cm]{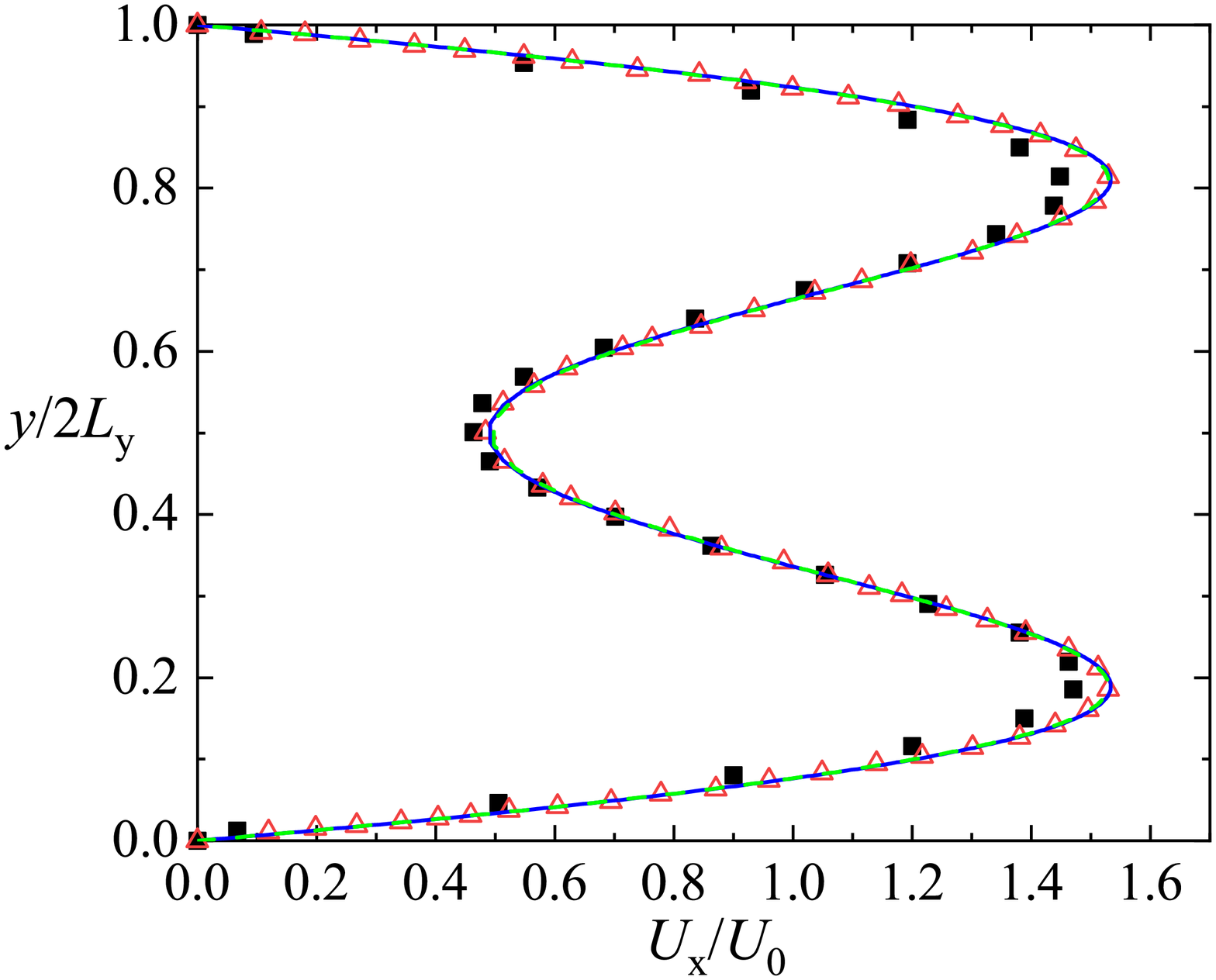}
		\put(-155,115){(c)}
		\label{fig.x3}
	\end{minipage}
	\hspace{20pt}
	\begin{minipage}[h]{0.4\textwidth}
		\centering
		\includegraphics[height=4.2cm]{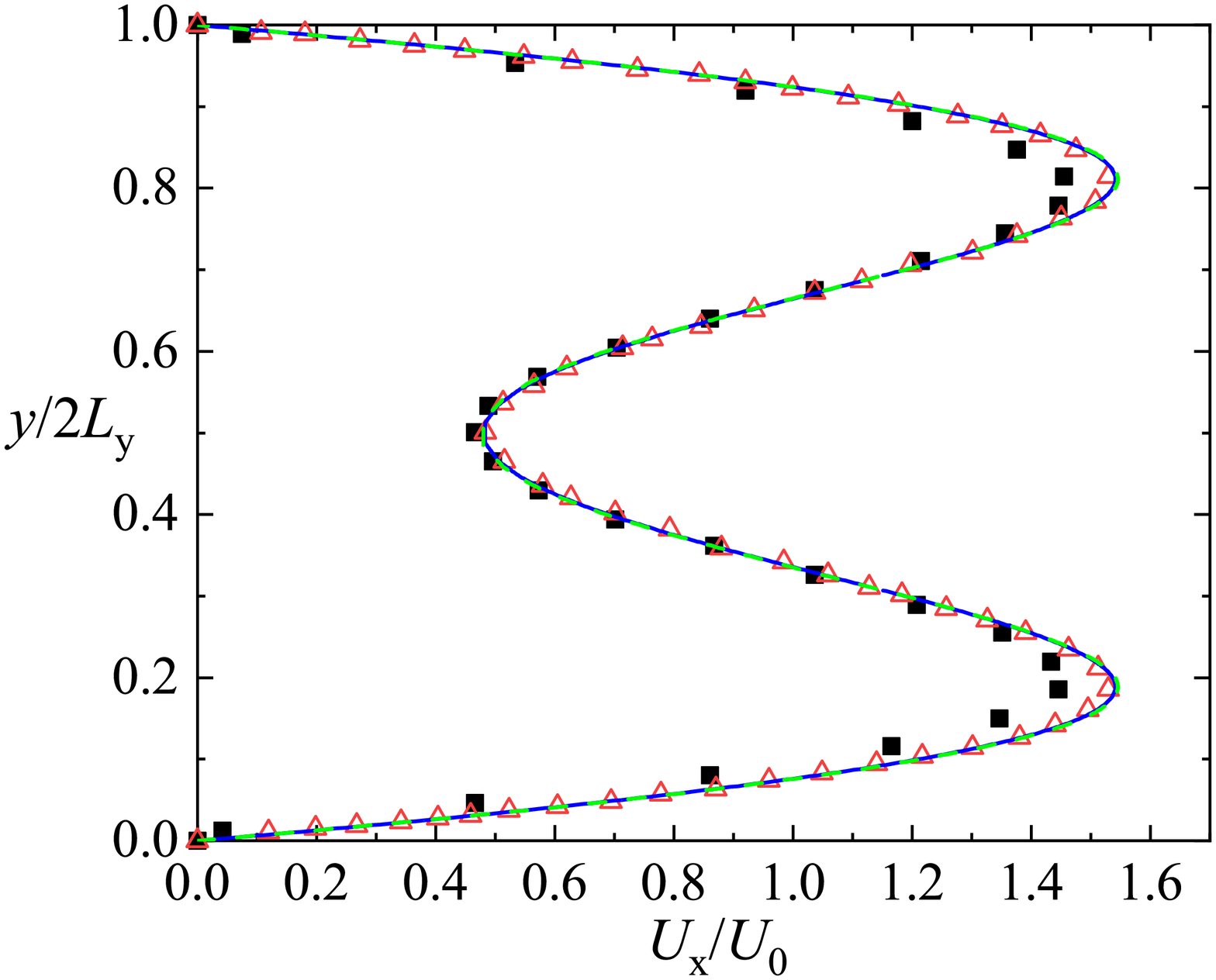}
		\put(-155,115){(d)}
		\label{fig.x4}
	\end{minipage}
	\hspace{20pt}
	\begin{minipage}[h]{0.4\textwidth}
		\centering
		\includegraphics[height=4.2cm]{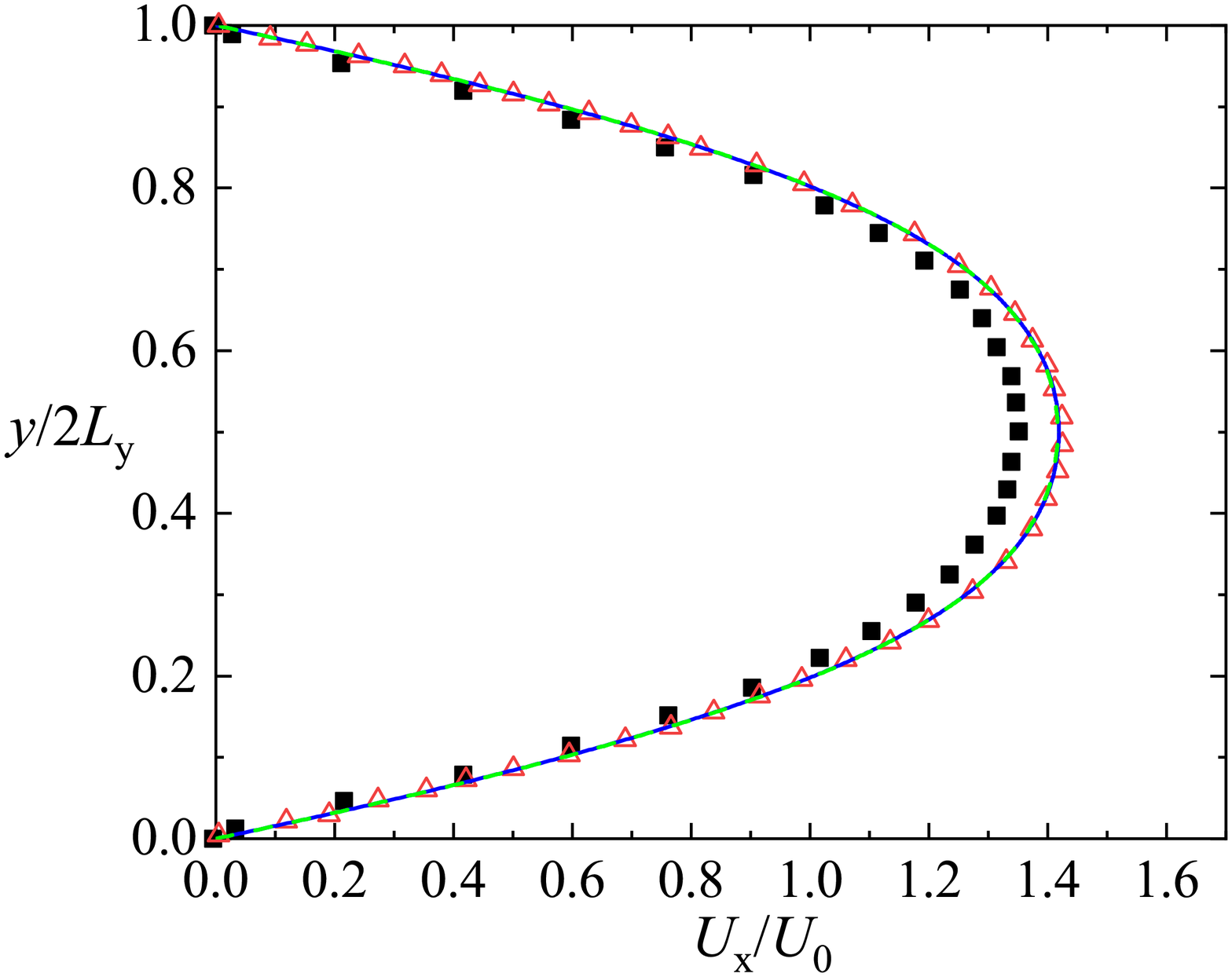}
		\put(-155,115){(e)}
		\label{fig.x5}
	\end{minipage}
	\caption{Comparison of streamwise velocity profiles for different downstream distances. (a) $ x = -21R $; (b) $ -3R $; (c) $ -1.5R $; (d) $ 1.5R $; (e) $ 3R $. }
	\label{fig.cylinder}
\end{figure}

\cite{verhelst2004visco} performed an experimental study for the cylindrical wake flow in a rectangular channel with a width of 20mm and a height of 160mm. In addition, the cylinder diameter is 10mm and then $ Re=\frac{U_{max}^*R^*}{\nu^*}=0.23 $. Since the Reynolds number in the experiment is low, the 3-D effect is weak and we can make a comparison of our 2-D simulations with theirs. \cite{xiong2013numerical} carried out the 2-D numerical simulation for the same cylinder wake flow as the experimental work. The channel height is normalised as 1, the cylinder diameter is 0.5, and $ Re=\frac{U_{0}^*D^*}{\nu^*}=0.307 $. In order to validate our nonlinear solver, we perform a 2-D numerical simulation of a flow past a cylinder calculated by Nek5000. The parameters are the same as those in \cite{xiong2013numerical}. Then we also calculate the results using the equations of wire-plate EHD-Poiesuille flow at $ \phi = 0 $ kV, which is also simulated by Nek5000. The parameters are that channel height is 4, cylinder diameter is 2, $ Re^E=2.4 $, $ Re^W=\frac{U_{0}^*R^*}{\nu^*}=0.1535 $ (which is equivalent to the above two Reynolds numbers), therefore:
\begin{equation}
	U_x(y)=\frac{3}{2}\frac{Re^W}{Re^E}(1-y^2/L_y^2)=\frac{3}{2}\times0.064\times(1-y^2/L_y^2).
\end{equation}
From figure \ref{fig.cylinder},  we can see that the three 2-D numerical simulations are in good agreement, and their values are slightly larger than the 3-D experimental results, which is due to the inevitable three-dimensional effects and the presence of the wall in reality \citep{xiong2013numerical}. According to these results, we consider that our nonlinear code is verified.

\section{Validation of linear results}\label{valili}

\begin{figure}
	\centering
	\includegraphics[height=4.5cm]{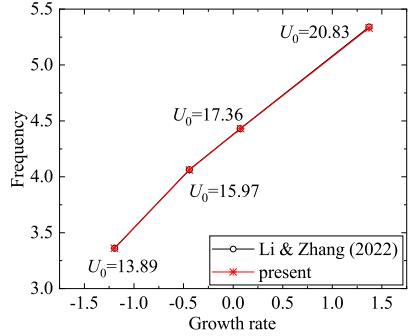}
	\caption{Eigenvalues of the leading eigenmode at different cross-flow velocity $ U_0 $ at $ \lambda=0.5 $. }
	\label{fig.valiL}
\end{figure}

Next, we validate the solution for the linearized equations. Figure \ref{fig.valiL} compares eigenvalues of the leading eigenmode of wire-plate EHD-Poiseuillle flow at $ \Phi=0 $ (without electric field) obtained by the Arnoldi method with the results of the global stability analysis in \cite{li2022reinforcement} for the confined wake flow. The parameters in \cite{li2022reinforcement} are: $ \lambda=0.5 $ and $ Re=100, 115, 125, 150 $, respectively, where $ Re=U_{max}^*D^*/\nu^* $. Our parameters are consistent with those in \cite{li2022reinforcement} after proper transformation: $ C_0=0.0, \lambda=0.5, Re^E=2.4 $, and $ U_0=13.89, 15.97, 17.36, 20.83 $, respectively. In addition, the charge injection from the wire $ C_I/(\lambda^2 C_0)  $ in \ref{bc2d14} needs to be set to zero since $ C_0 $ is in the denominator.  The results in figure \ref{fig.valiL} show that the two results match well with each other. Additionally, we compare the growth rate at different $ U_0 $ and $ Re^E=2.4 $ obtained by the power method and the Arnoldi method, as shown in table \ref{table:gr}. The parameters are $ C_0=3, M=37, K_r=1,  C_I=0.2, O_s=8.6, \lambda=0.2$ and $ \alpha=0.001 $. For all the cases, the relative error does not exceed $ 2.5\% $. Therefore, we consider that our linear solver is also accurate and reliable. 


\begin{table}
	\def~{\hphantom{0}}
	\begin{center}
		\begin{tabular}{l c c c c c c}							
			$ U_0 $ & 0  & 0.3 & 0.3 & 15 & 18 &21\\ 
			$ Re^E $ & 0.8  & 4.5 & 5 & 2.4 & 2.4 &2.4\\ 			
			growth rate (Power)  & ~-0.0074~ & ~0.00425~&~0.0281~ &~1.527~&~2.166~ &~2.837~\\
			growth rate (Arnoldi) & ~-0.0074~ & ~0.00415~& ~0.0279~ &~1.529~&~2.166~ &~2.836~\\			
		\end{tabular}
		\caption{Comparison of growth rate obtained by Power method and Arnoldi method for wire-plate EHD-Poiseuille flow at different $ U_0 $ and $ Re^E $. }
		\label{table:gr}
	\end{center}
\end{table}

\section{Grid independence verification}\label{valimesh}

\begin{figure}
	\centering
	\includegraphics[height=2.5cm]{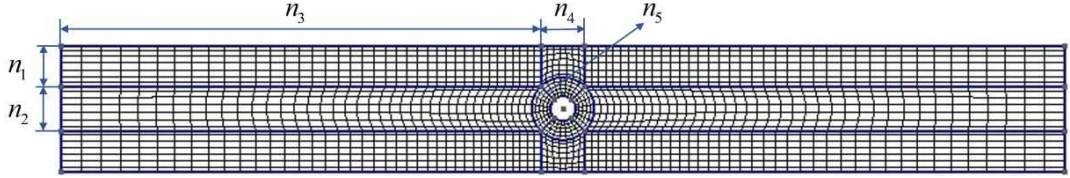}
	\caption{Distribution of the Legendre spectral elements for the wire-plate EHD-Poiseuille problem.}
	\label{fig.mesh}
\end{figure}

We then verify the independence of our numerical results with respect to the grid resolution. In this work, we choose the domain size $ \Lambda_1=\Lambda_2=40 $, and $ \lambda=0.2 $. The distribution of the spectral elements in the computation domain is shown in figure \ref{fig.mesh}. The region is divided into several parts to ease the discretization, and a circular region with a radius of 2.5 refines the mesh near the wire. It is noted that the spectral elements mesh depends on two grid levels, one being the number of spectral elements $ N_e $ and the other being the polynomial order within each element $ N $. The notations for the numbers of elements in each part is labeled in figure \ref{fig.mesh}.

\begin{table}
	\begin{center}
		\def~{\hphantom{0}}
		\begin{tabular}{l c c c c c c c c}
			Mesh & $ n_1 $ & $ n_2 $ & $ n_3 $ & $ n_4 $ &$ n_5 $ & Element & $ \bar{|U|}_{max} $ & Relative error to M5\\
			M1 & ~5~ & ~5~ & ~30~&  ~5~&  ~3~& ~760~ & ~42.710~ & $ 0.47 \% $  \\
			M2  & ~6~ & ~6~&~35~ & ~5~  & ~3~& ~1130~  &~42.715~ &$ 0.46 \% $ \\
			M3 &  ~8~& ~8~ & ~40~ & ~8~ & ~6~ & ~1876~ & ~42.866~ &$ 0.11 \% $ \\
			M4 & ~9~ & ~9~ & ~45~&  ~9~ & ~6~ & ~2400~ &  ~42.908~ &  $ 0.009 \% $\\
			M5 & ~12~ &~12~ &~45~ & ~12~ & ~6~ & ~3366~  &~42.912~ &   \\
		\end{tabular}
		\caption{ Grid independence validation at $ U_0=21, Re^E=2.4 $.}
		\label{table:mesh}
	\end{center}
\end{table}

\begin{figure}
	\centering
	\includegraphics[height=4.2cm]{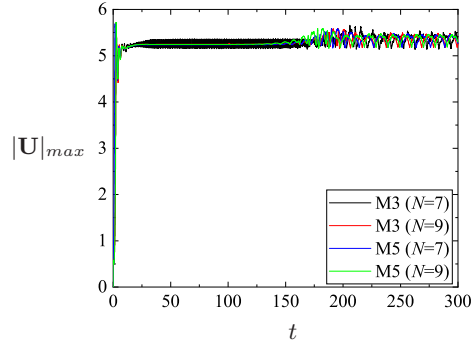}
	\put(-175,60){$ |\Ub|_{max} $}
	\put(-70,-10){$ t $}
	\caption{Time evolution of maximum velocity magnitude at different mesh at  $U_0=0.3, Re^E=6 $. }
	\label{fig.meshO10}
\end{figure}

We first take $ U_0=21, Re^E=2.4 $ which is the maximum $ U_0 $ we study in this work for the validation. The other parameter are $ C_0=3, M=37, K_r=1,  C_I=0.2, O_s=8.6, \lambda=0.2 $ and $ \alpha=0.001 $. Here we fix the polynomial order $ N=7 $  \citep{Paul2008} and change the number of elements to study the mesh convergence. The time-averaged maximum velocity norm $ |U|_{max} $ is shown in table \ref{table:mesh} for different meshes ranging from a coarse grid $M_1$ with 760 elements to a refined grid $M_5$ with 3366 elements. It can be seen that the relative errors of M3 and M4 to M5 are both less than $ 0.12\% $, indicating that results converge at these mesh sizes. M3 is adopted for most cases considering both the accuracy and the computational time.

In the case of wire-plate EHD flow with a weak cross-flow, we adopt the value of $ Re^E $ up to 6. In addition, it is found that the oscillation behavior is sensitive to the mesh. Therefore, we also perform a grid independence test for $ U_0=0.3, Re^E=6 $. The other parameter are $ C_0=3, M=37, K_r=1,  C_I=0.2, O_s=8.6, \lambda=0.2$ and $ \alpha=0.001 $. We compare the evolution of $ |\Ub|_{max} $ for four sets of mesh: M3 described in table \ref{table:mesh} with polynomial order $ N=7 $, M3 with $ N=9 $, M5 with $ N=7 $ and M5 with $ N=10 $, as shown in figure \ref{fig.meshO10}. We find from the figure that the results from the latter three meshes are close to each other. Therefore, we take M3 with $ N=9 $ for the calculation of wire-plate EHD flow with a weak cross-flow.

\end{appendix}

\bibliographystyle{jfm}
\bibliography{ref}

\end{document}